\documentclass{article}

\usepackage[english]{babel}
\usepackage[letterpaper,top=2cm,bottom=2cm,left=3cm,right=3cm,marginparwidth=1.75cm]{geometry}

\usepackage{amsmath,amssymb,amsfonts}
\usepackage{amsthm}
\usepackage{tabularx}
\usepackage{multirow} 
\newcolumntype{Y}{>{\centering\arraybackslash}X}
\usepackage{float}
\theoremstyle{remark}
\newtheorem{Remark}{Remark}
\usepackage{graphicx}
\usepackage[colorlinks=true, allcolors=blue]{hyperref}
\usepackage{caption}
\title{Your Paper}

\title{Varying Gravity from a Modified Fractional Model:\\
Observational Constraints and Slow--Fast Dynamics}

\author{
Rami Ahmad El-Nabulsi\textsuperscript{1,2,3} \and
Genly Leon\textsuperscript{4,5,6} \and
Esteban González\textsuperscript{7} \and
Kevin Marroquín\textsuperscript{4,8}
}

\date{
\textsuperscript{1} Center of Excellence in Quantum Technology, Faculty of Engineering, Chiang Mai University, Chiang Mai 50200, Thailand; \texttt{nabulsiahmadrami@yahoo.fr}\\
\textsuperscript{2} Quantum-Atom Optics Laboratory and Research Center for Quantum Technology, Faculty of Science, Chiang Mai University, Chiang Mai 50200, Thailand\\
\textsuperscript{3} Department of Optical Networks CESNET, Generála Píky 430/26, Prague, Czech Republic; \texttt{el-nabulsi@cesnet.cz}\\
\textsuperscript{4} Departamento de Matemáticas, Universidad Católica del Norte, Avenida Angamos 0610, Casilla 1280, Antofagasta, Chile; \texttt{genly.leon@ucn.cl}, \texttt{kevin.marroquin@alumnos.ucn.cl}\\
\textsuperscript{5} Centre for Space Research, North-West University, Potchefstroom 2520, South Africa\\
\textsuperscript{6} Institute of Systems Science, Durban University of Technology, Durban 4000, South Africa; \texttt{genlyl@dut.ac.za}\\
\textsuperscript{7} Departamento de Física, Universidad Católica del Norte, Avenida Angamos 0610, Casilla 1280, 1270709 Antofagasta, Chile; \texttt{esteban.gonzalez@ucn.cl}\\
\textsuperscript{8} Facultad de Matemáticas, Pontificia Universidad Católica de Chile, Edificio Rolando Chuaqui, Campus San Joaquín, Avda. Vicuña Mackenna 4860, Macul, Santiago, Chile; \texttt{kmarroquinv@estudiante.uc.cl}\\[1em]
\textbf{Corresponding author:} Genly Leon (\texttt{genly.leon@ucn.cl})
}

\begin{document}
\maketitle

\begin{abstract}We investigate a fractional gravity model in which both the Hubble parameter and the gravitational constant evolve dynamically due to fractional renormalization--group effects. The model incorporates a scalar field coupled to 
a time-varying $G$, generating nonlocal corrections characteristic of fractional--action cosmology. Analytical and numerical solutions reveal oscillatory regimes, cyclic phases, and rapid variations with implications for BBN and early-universe evolution. A robust numerical framework is developed to integrate the regularized system and compare the resulting $H(z)$ evolution with observational data from the Hubble parameter, baryon acoustic oscillations, type Ia supernovae, gravitational lensing, and black hole shadows, thereby enabling a consistent reconstruction of cosmographic quantities.
A Bayesian analysis shows that the Fractional model with $\mu=0$ is the only statistically viable variant. The inferred Hubble parameter is stable across models ($h\simeq 0.72$), while the fractional parameters are significantly better constrained in the $\mu=0$ case ($\alpha=1.20^{+0.25}_{-0.14}$, $\zeta=0.43^{+0.39}_{-0.29}$). The dynamical sector yields $m=30.8^{+28.0}_{-20.9}$ and $\Gamma=108.3\pm1.1$, leading to a positive discriminant and a well-determined relaxation timescale $\tau_{\rm rel}\simeq 9$ Gyr, confirming an overdamped regime. Although the $\mu=0$ model attains a slightly lower $\chi^2_{\min}$ than $\Lambda$CDM, the BIC strongly favors $\Lambda$CDM due to its smaller parameter space. Overall, the model reproduces late-time acceleration and mimics $\Lambda$CDM while introducing distinctive cosmographic signatures. The dynamical systems analysis clarifies the stability structure and parameter dependence, indicating that fractional nonlocal corrections may offer new pathways toward addressing 
the $H_0$ and $S_8$.
\end{abstract}

\section{Introduction}

The Friedman-Robertson-Walker (FRW) standard model of cosmology is considered today the most successful model for describing our homogeneous, isotropic, and flat Universe, favored by observations of the cosmic microwave background (CMB). The Universe began with a very hot, dense Big Bang and has been expanding ever since. Today, it is undergoing a phase of accelerated expansion, as confirmed by recent astronomical measurements, including WMAP data, along with the latest distance measurements from baryon acoustic oscillations (BAO) in the distribution of galaxies and the Hubble constant $H_0$ measurement \cite{1}. 

General relativity (GR) correctly describes both the geometry and the physics of the Universe. However, the theory is somewhat complicated and, beyond spatially curved universes, which are solutions of Einstein's field equations, is mathematically intricate. On the other hand, the Newtonian theory (NT) is, in various ways, far simpler, since the Friedmann equation derived from GR can also be derived within the NT under certain assumptions \cite{2}. In general, the NT adequately describes the large-scale cosmos after the decoupling of matter and radiation. Therefore, NT is not considered a universal cosmological theory for describing the Universe from its early epoch to its late-time accelerated expansion epoch. Besides, it requires the use of Birkhoff's theorem, which is based on GR, although NT provides some intuitive insights. On the other hand, even the FRW model by GR suffered from several problems. 

One of these problems is the poor observational constraints on the equation of state (EoS), which is why the literature contains many dissimilar phenomenological EoS models that nevertheless agree well with the coarse features of the present cosmological epoch of the Universe. 

Notably, standard cosmology also suffers from additional serious problems, such as the so-called Cosmological Constant Problem (CCP), in which the observed value of the cosmological constant differs by 60-120 orders of magnitude from the value predicted by particle physics \cite{3}. 

The coincidence problem also represents another serious problem in standard cosmology, which requires that dark matter (DM) and dark energy (DE) densities, at the present epoch of the~ Universe's evolution, are of the same order of magnitude, leading to their analogous effects on the dynamics of the cosmos \cite{4}. 

One of the best-fit cosmological models that describes the Universe's observed evolution is the $\Lambda$CDM model \cite{5}. However, the $\Lambda$CDM model failed to explain the fine-tuning and coincidence problems. Another particularly challenging problem in contemporary cosmology is the Hubble tension, which is the discrepancy between the current rate of expansion of the Universe and what cosmological theories predict based on our observations of the CMB in the early Universe and the Universe's initial conditions. 

Recent Astronomical measurements of the Hubble parameter at present are based on recent astronomical observations of Type Ia~supernovae~ ~(SNe) by SH0ES (Supernovae and $H_0$ for the Equation of State of dark energy) collaboration, from Cepheids, and the Planck CMB is not consistent with the value reported by recent~CMBR~observations utilizing the Planck satellite and $\Lambda$CDM~by at least $3\sigma$ \cite{6}.
The present measurement has a 10\%  accuracy of the Hubble Space Telescope gives $H_0=72\pm 8 $ km $\times$ s $^{-1} \times $Mpc$^{-1}$ \cite{7} whereas SH0ES program gives $H_0=74.03\pm 1.42 $km$\times$ s $^{-1} \times $Mpc$^{-1}$ \cite{8,9}.

Recent analysis based on a Bayesian statistics approach and the maximum likelihood method estimates $H_0=72.2\pm 4 $ km$\times$ s $^{-1} \times $Mpc$^{-1}$ for the non-flat $\Lambda$CDM model and $H_0=68.3\pm 3.1 $ km$\times$ s $^{-1} \times $Mpc$^{-1}$ for the flat $\Lambda$CDM \cite{10,11}. Given that the Planck estimate of the Hubble parameter strongly depends on the assumption of the standard cosmological model. 

In the $\Lambda$CDM model, the Hubble tension may point to new physics involving the properties of DM or DE. Several cosmological models have been proposed in the literature to alleviate these problems, e.g., the recent cosmological model based on the Extended Uncertainty Principle (EUP) containing a minimal fundamental measurable momentum for describing large-scale quantum gravitational effects \cite{12}; the $\Lambda$CDM-NG (Newtonian Gravity) model which is free from the Hubble tension, and was introduced to move matter between the source terms to restore the source of gravitational potential to its Newtonian form, for a particular range of the gravitational potential of condensed
objects \cite{13}; using modified gravity such as the Chern-Simons theory \cite{14} and electromagnetic non-minimal
Maxwell-Gauss-Bonnet coupling cosmology \cite{15}, etc.

One plausible model introduced in the literature to alleviate the Hubble tension is based on fractional action cosmology (FAC) \cite{16,17,18,19,20,21} and has been explored in detail through various cosmological models in \cite{21,22,23,24,25,26,27,28,29,30,31,32,33,34,35,36,37,38,39,40,41,42,43,44,45,46,47}. 

FAC is based on the concept of the fractional action-like variational approach (FALVA), which generalizes the standard action $S= \int_{t_1}^{t_2}L(q, \dot{q}, t)dt$ to its fractional integral counterpart $S=\frac{1}{\Gamma(\alpha)}\int_{t_1}^{t_2}L(q, \dot{q}, t)(t-\tau)^{\alpha-1}d\tau$.
Here, $L$ is the Lagrangian of the physical system, $q$ is the generalized coordinate, $\dot{q}= dq/dt$, $\alpha$ is
a real or a complex fractional parameter, and $\Gamma(\alpha)= \int_{0}^{\infty} t^{\alpha-1}e^{-t}dt$ is the Euler gamma function. Besides, $\tau$ is the intrinsic time and $t$ is the observer time, with $t\neq \tau$.

This paper presents a new theory, the ``Fractional Renormalization-Group Improved Cosmology (FRGIC)" or ``Fractional Einstein Quantum Gravity" (FEG), motivated by the ``renormalization group (RG) induced quantum effects", with significant cosmological implications. Recall that RG is considered a standard tool in elementary particle physics, used to introduce the leading quantum corrections to the Born approximation of a scattering cross section.

FRGIC is motivated by several astronomical observations and phenomenological studies, which indicate that both the cosmological constant $\Lambda$ and the gravitational constant $G$ vary with space and time. In this framework, the RG-induced quantum effects approach plays a leading role, as it allows us to derive both the cosmological constant and the gravitational constant from an ultraviolet-attractive fixed point \cite{47,48,49,50}. 

This approach introduces an effective average action for Euclidean quantum gravity $\Gamma_k[g_{\mu\nu}]$ since it permits describing gravity at a typical distance scale $l=k^{-1}$, and resulting in an exact function, the RG equation for the \emph{k}-dependence of $\Gamma_k[g_{\mu\nu}]$.

By letting $k\propto t^{-1}$, $t$ being the cosmological time, $G(k)$ and $\Lambda(k)$ are then converted to dynamical quantities in the field equations, and hence, affecting several astrophysical scenarios \cite{51,52,53}. This approach has been considered in the literature and has been shown to be relevant to both the initial Planck era and the structure of black holes \cite{51,52,53}, as well as the late-time accelerated expansion of the universe \cite{54,55,56}. A scalar-field-dominated cosmology with variable $\Lambda=\Lambda(G)$ and $G$ has been studied based on RG in \cite{54,57}. 

In the present work, we will study FRGIC or FEG by deriving the evolution equation within the framework of the Friedmann-Lemaître-Robertson-Walker (FLRW) metric.

Using recent astronomical data, we will prove that our model is consistent with current observational constraints, providing a late-time accelerating expansion of the Universe and a potential solution to the Hubble tension. Moreover, we will prove that such a model yields additional terms in the Friedman equation, with important implications for cosmology and Big Bang Nucleosynthesis (BBN). In particular, we will prove the emergence of a time-varying/rapidly varying gravitational constant such that $\dot{G}/G\gg H$, $H$ being the Hubble parameter. 

We also examine the physical interpretation of the cosmological solutions, focusing on the effects of the fractional order of the action in FRGIC/FEG. Scalar fields are expected to play a significant role in cosmology at both the high- and low-energy limits, and they arise in unified theories. In the early Universe, the scalar field is the inflaton responsible for inflation \cite{58}.

Scalar fields drive the accelerated expansion of the universe and offer a plausible explanation for DE \cite{59}, and dark matter (DM) \cite{60}. Scalar fields have been widely considered in the literature and have proven relevant in cosmology, although their potential as major matter components in the evolution of the Universe remains poorly understood \cite{61,62,63,64,65,66,67,68,69,70}.

In this paper, we analyze the fractional cosmological framework with a time-varying gravitational constant by reformulating the field equations into autonomous structures. \S\ref{Sec.2} introduces the fractional model and the dimensionless equations that define the dynamical system. \S\ref{Sect:2.6} outlines the numerical procedure for reconstructing $H(z)$, including the first–order reduction, the computation of $E(\tau)$, the time–redshift relation, the definition of normalized variables, and the analysis of critical points. \S\ref{Sect:2.7} presents the data and methods used for parameter estimation. The effective dynamics for $\mu=0$ is investigated in \S\ref{effective-dyn}.

\S\ref{Sect:4.1} examines the full system, covering the phase space, the slow–fast structure, critical points, stability, global dynamics, and observational implications. Section~\ref{Sect:4.2} studies the regularized system, describing its assumptions, equilibria, stability, geometric flow, numerical implementation, and application to the specific model.

In \S\ref{Sect:5.2}, a normalized form of the evolution equation for $\mathcal{R}$ is introduced, with physical interpretation and structural analysis. In \S\ref{Sect:5.3}, the critical points of the normalized system are classified and their cosmological meaning discussed. In \S\ref{Sect:5.4}, the crossing of $u=0$ is addressed through desingularization, and alternative global strategies to compactify phase space and represent the slow manifold are presented. In \S\ref{Sect:5.7} we analyze how the temporal variation of $G$ modifies the growth of perturbations: \S\ref{Sect:5.7.1} defines the observable $f\sigma_8$ and its calculation from the system solutions; \S\ref{Sect:5.7.2} proposes a fractional closure relation $f\simeq\Omega_m^{\gamma_{\rm frac}}$ that generalizes the standard law; \S\ref{Sect:5.7.3} describes the procedure to determine the fractional growth index $\gamma_{\rm frac}$ through numerical fits and robustness criteria; and \S\ref{Sect:5.7.4} presents practical parametrizations of the dimensionless age and of $H(t)$ that facilitate comparison between models and data.

\S\ref{Sect:4.8} summarizes the main results: the geometry of the flow, the existence and stability of cosmological attractors, the slow–fast structure, and their implications for cosmic evolution, together with limitations and directions for future research.

\section{Basic Setups}
\label{Sec.2}
We consider a four-dimensional flat, isotropic, and homogeneous universe described the FLRW spacetime metric in agreement with observations described by the usual metric \$with signature $(-,+,+,+)$):

\begin{equation}
ds^2= -dt^2 +a(t)^2 \left(d r^2 +r^2(d\theta^2 +\sin^2 \theta d\phi^2)\right) \label{tag1}
\end{equation}
where $a(t)$ is the scale factor. We assume that a scalar, $\varphi$, dominates the Universe. Matter and gravity are described by the following fractional Einstein-Hilbert action (in units $\hbar=c=1$ and with a lapse $N=1$):
\begin{align}
L=\frac{1}{8\pi G}\Big(-3a\dot{a}^2 - a^3\Lambda(G) + \tfrac{1}{2}\nu a^3\frac{\dot{G}^2}{G^2}
+4\pi G\,a^3\dot{\varphi}^{\,2}-8\pi G\,a^3V(\varphi)\Big)-a^3\rho,
\label{tag2}
\end{align}
where the last term represents the Lagrangian of matter given by $L_m=-a^3 \rho$, $V(\varphi)$ is the scalar potential, and $\mu$ is a real parameter. The resulting field equations are deduced from the fractional Euler-Lagrange equation \cite{71,72}:
\begin{equation}
\frac{\partial L}{\partial q}-\frac{d}{d\tau}\left(\frac{\partial L}{\partial{\dot{q}}}\right)= \frac{1-\alpha}{t-\tau }\frac{\partial L}{\partial{\dot{q}}},  \label{tag3}
\end{equation}
after varying the action for $\{G,\varphi, a\}$. To designate the temporary independent variables, we use the rule $t-\tau \to t, \tau\to t$ \cite{27}.

Varying with respect to $q=G$, we obtain 
\begin{align}
& 3\frac{\dot a^{\,2}}{a^2}+\Lambda(G)+\nu\frac{\dot G^{\,2}}{G^2}+G\Lambda'(G)
-8\pi G\dot\varphi^{\,2}+8\pi G V(\varphi) \nonumber\\
& 
 -\nu\!\left(\left(3\frac{\dot a}{a}+\frac{1-\alpha}{t}\right)\frac{\dot G}{G}
+\frac{\ddot G}{G}-4\frac{\dot G^{\,2}}{G^2}\right)=0.
\label{tag4}
\end{align}
Varying with respect to $q=\varphi$, we obtain 
\begin{equation}
\ddot{\varphi}+\left(3\frac{\dot a}{a}+\frac{1-\alpha}{t}\right)\dot{\varphi}
+V'(\varphi)=0.
\label{tag5}
\end{equation}
Varying with respect to $q=a$,  we obtain
\begin{align}
&\frac{\dot a^{\,2}}{a^2}+2\frac{1-\alpha}{t}\frac{\dot a}{a}
-\Lambda(G)+\tfrac{1}{2}\nu\frac{\dot G^{\,2}}{G^2}
+4\pi G\dot\varphi^{\,2}-8\pi G V(\varphi)-8\pi G\rho
 +2\left(\frac{\ddot a}{a}-3\frac{\dot a}{a}\frac{\dot G}{G}\right)=0.
\label{tag6}
\end{align}

Using the equation of state $p=\gamma \rho$, $\gamma$ being the equation of state (EoS) parameter, then it is easy to check that within FAC based on FALVA, the density varies as $\rho =\rho_0 a^{-3(1+\gamma)} t^{-(\alpha-1)(1+\gamma)}$ \cite{16}. Usually, a decreasing density is favored by observations, and this implies an increase in the gravitational constant.  After summing equations \eqref{tag4} and \eqref{tag6} we obtain:
\begin{align}
& 4\frac{\dot a^{\,2}}{a^2}+2\frac{1-\alpha}{t}\frac{\dot a}{a}
-4\pi G\dot\varphi^{\,2}-8\pi G\rho
 +\tfrac{11}{2}\nu\frac{\dot G^{\,2}}{G^2}
+G\Lambda'(G)
 +2\left(\frac{\ddot a}{a}-3\frac{\dot a}{a}\frac{\dot G}{G}\right)
\nonumber\\
& -\nu\!\left(\left(3\frac{\dot a}{a}+\frac{1-\alpha}{t}\right)\frac{\dot G}{G}
+\frac{\ddot G}{G}\right)=0.
\label{tag7}
\end{align}

Motivated by various scalar field and quintessence cosmological models \cite{73,74,75}, we suggest the generalized ansatz
\begin{equation}
H=H_0 +\xi \varphi +\varepsilon/t, \label{anzat}
\end{equation}
$H_0$ is the Hubble parameter at the present epoch, and $\xi$ and $\varepsilon$ are real constants. We assume the quartic form of the potential \begin{equation}
V(\varphi)=V_0+\tfrac{1}{2}m^2\varphi^2+\tfrac{f}{4}\varphi^4, \quad m^2\propto H_0^2, \quad
f \in\mathbb{R}. 
\end{equation} ($V_0$ is a real parameter) which is representative of the most standard kind of quintessence \cite{76}. Here $m^2\propto H_0^2$ (massive scalar field) as predicted from the inflationary paradigm and from modified gravity models \cite{77,78}, and $f$ is a real parameter. It is notable that, based on the recent analysis done in \cite{79} supported by the SN Ia observational data, a very low inferior limit to the DM mass is obtained, which is about $m \geq 1.56\times 10^{-33}$eV. Although slightly smaller than the constraints $m\approx 10^{-22}$eV obtained in \cite{80,81,82}, it has been proved that any DM mass larger than this value is also compatible with SNIa astronomical observations. Notably, a very small mass of DM may disagree with the cosmological formation of the Universe and various astrophysical structures. Equation \eqref{tag5} is written now as:
\begin{equation}
\ddot{\varphi}+\left(3\xi\varphi+\frac{1-\alpha+3\varepsilon}{t}+3H_0\right)\dot{\varphi}+m^2\varphi+f\varphi^3=0,
\label{tag8}
\end{equation}
which for $\varepsilon=(\alpha-1)/3$, It is a first-order solution for large field values $\xi\varphi\gg m \approx H_0$ \cite{83}, is given by \cite{84}:
\begin{equation}
\varphi(t)=\frac{A\sin(\omega t+\delta)}{1-\frac{kA}{3\omega}\cos(\omega t+\delta)},
\label{tag9}
\end{equation}
where we consider the specific parameters satisfying $0\leq A<3\omega/k$, $k=3\xi=3\sqrt{f}$, $\omega=m$ and $\delta$ is an arbitrary constant. Equation \eqref{tag8} belongs to the class of Levinson-Smith type differential equations \cite{85}. We let $m=\zeta H_0$ ($\zeta$ is a real parameter) which appears in quintessence models \cite{86,87,88}, in extended supergravities with unstable de Sitter vacua \cite{89}, in $N = 2$ gauged supergravities with stable dS
vacua \cite{90,91,92}. The cosmological constant is of the same order of magnitude as the Hubble time; therefore, a scalar field conformally coupled to gravity would acquire $m\propto H_0$. The Hubble parameter varies as:
\begin{align}
H(t)=\sqrt{f}\,A\frac{\sin(\zeta H_0 t+\delta)}{1-\frac{\sqrt{f}A}{\zeta H_0}\cos(\zeta H_0 t+\delta)}+\frac{\alpha-1}{3t}+H_0.
\label{tag10}
\end{align}

\begin{Remark}
\label{Remark1}
If, for instance, $m^2<0$ (Higgs potential), then for $\varepsilon=(\alpha-1)/3$ and $\xi\varphi\gg |m| \approx H_0$,
the solution of equation \eqref{tag8} is given by:
\begin{equation*}
\varphi(t)=\frac{\sqrt{|m^2|}\left(K_1 e^{2 \sqrt{|m^2|} t}-1\right)}{f\left(K_2 e^{ \sqrt{|m^2|} t}+1+K_1 e^{2 \sqrt{|m^2|} t}\right)},
\end{equation*}
where $K_1$ and $K_2$ are constants of integration. Observe that, for a considerable time, the scalar field tends asymptotically to a constant value which is $\sqrt{|m^2|}/f$. Consequently, the Hubble parameter tends asymptotically to $H=H_0 + \sqrt{|m^2|}/f= \left(1+\zeta/f\right)H_0$, and the universe expands exponentially with time.
\end{Remark}
\begin{Remark}
\label{Remark2} 
In general, the explicit periodic solution of the differential equation  \eqref{tag8} for $\varepsilon=(\alpha-1)/3$ is given by \cite{84}:
\begin{equation*}
\varphi_{e}=\frac{A \sin \left(W t +\delta\right)}{1-\frac{k A}{3 W}\cos\left(W t +\delta\right)},
\end{equation*}
where $\varphi_{e}(t)= \varphi(t)+H_0\xi$ and $W= \sqrt{m^2-3 H_0^2}$. Therefore, the constraint $m^2>3 H_0^2$ must
hold in that case.
\end{Remark}

Letting $\Lambda=\Lambda_0 G^{\beta}$ where $\Lambda_0$ is a real parameter and $\beta$ is a real constant, we find from equation \eqref{tag6}, and in particular for large time (mainly for $\delta=0$):
\begin{align}
G(t)\approx G_N\!\left(
\frac{4\pi k^2A^2}{9\Lambda_0}
\frac{\left(1-\frac{3\omega}{kA}\cos\omega t\right)^2}
{\left(1-\frac{kA}{3\omega}\cos\omega t\right)^4}
+\frac{8\pi\rho_0}{\Lambda_0}a^{-3(1+\gamma)}t^{-(\alpha-1)(1+\gamma)}
\right)^{\!\frac{1}{\beta-1}},
\label{tag11}
\end{align}
where $G_N$ is Newton's gravitational constant. Observe that when $\gamma=-1$, the effective gravitational constant is governed by the oscillatory part. When $(\alpha-1)(\gamma+1)<0$, i.e. $\alpha>1, \gamma<-1$ or
$0<\alpha<1, \gamma>-1$, then at the origin of time, the effective gravitational constant is given by:
\begin{equation}
G(t=0) \approx G_N \left(\frac{4\pi k^2 A^2}{9}\frac{\left(1-\frac{3\omega}{k A}\right)^2}{\left(1-\frac{k A}{3\omega})\right)^4}\right)^{\frac{1}{\beta-1}}, \label{tag12}
\end{equation}
and therefore the constraint:
\begin{equation}
\frac{4 \pi k^2 A^2}{9} \left(1-\frac{3\omega}{k A}\right)^2= \left(1-\frac{k A}{3\omega}\right)^4, \label{tag13}
\end{equation}
must hold to that $G(t=0)\approx G_N$.
Besides, at a singular time
\begin{equation}
t_s= \frac{2\pi n \pm \cos^{-1}\left(\frac{k A}{3\omega}\right)}{\omega}, \quad n\in \mathbb{Z}\label{tag14}
\end{equation}
the gravitational constant tends to infinity. The scale factor varies as:
\begin{small}
\begin{equation}
a(t)=A_0 e^{H_0 t}(\zeta H_0 t)^{\alpha-1}
\left(1-\frac{\sqrt{f}A}{\zeta H_0}\cos(\zeta H_0 t)\right)^3.
\label{tag15}
\end{equation}
\end{small}
The local strength of gravity is expected to be large during the early epoch; therefore, in our approach, the Universe is cyclic, and at the beginning of each cycle, the gravitational constant tends to a large value, whereas the scale factor tends to zero. Observe that equation \eqref{tag10} restricts $\Lambda_0 >0, \beta>0$ (positive cosmological constant) or $\Lambda_0<0, \beta<0$ (negative cosmological constant). Recent studies based on the full CMB likelihood, including the lensing likelihood obtained by the Planck-2018 data release, the Pantheon data, and the BAO measurements, as well as the~$ H_0$~measurement by the power spectrum of the redshifted HT $21-cm$ brightness temperature maps from the post-reionization epoch as a cosmological probe, suggest the existence of a negative cosmological constant \cite{93,94}. From a theoretical perspective, several phenomenological theories (such as those inspired by string theory) suggest a negative $\Lambda$ to explain observations from the James Webb Space Telescope, which have revealed an unexpectedly large abundance of extremely massive galaxies at redshifts $z\approx 5$ \cite{95,96}. Additional theories based on supergravity and higher-dimensional arguments support the idea of a negative cosmological constant \cite{97,98,99,100,101,102,103,104,105}.

\begin{Remark}
\label{Remark3}
Throughout this study, various parameters have been introduced. Let us summarize their physical interpretations to make
it clear:
\begin{enumerate}
\item $\alpha$ is the fractional constant parameter representing the fractional order of the mathematical operations. $\alpha=1$, the standard cosmology is recovered.
  
\item $\nu$ is a dimensionless interaction parameter without any observational constraint, since it is significantly different from zero mainly at very high energy limits \cite{55}. Its numerical estimate depends on the model under consideration.

\item $\xi$ is a parameter much less than unity. It is motivated by various cosmological arguments that suggest that the Hubble parameter varies as a power law of the scalar field 
  \cite{106,107,108,109}.
  
\item $\varepsilon$ is a real parameter motivated by cosmological arguments, which assumes that the Hubble parameter is made up of two terms: the $H_0$ constant term (as in an inflationary epoch of time) and a time-dependent one, which is expected during the radiation- and matter-dominated cosmological epochs of time \cite{110}. In our approach, the Hubble parameter comprises three parameters due to the presence of the scalar field. Notably, several cosmological arguments point toward a decreasing rate of expansion of the Universe with cosmic time \cite{111}. It is notable that, during the inflation-oscillation-dominated \\ phase, $H=2/(3t)$ mainly when the inflaton oscillates in the quadratic potential after inflation \cite{112}. In this study, we have selected $\varepsilon=(\alpha-1)/3$ for two main reasons: first, it considerably simplifies the analytical solution, and second, it eliminates the time-dependent friction term in the differential equation, so we will deal only with constant friction.
\item $\beta$  is a real constant introduced in the phenomenological law $\Lambda=\Lambda_0 G^{\beta}$. For $\beta=1$, the cosmological constant is proportional to the gravitational constant. This case has been observed in brane cosmology \cite{113}. The law $\Lambda=\Lambda_0 G^{\beta}$ suggests a connection between the cosmological and gravitational constant in FRGIC. In the basic formalism \cite{55}, we have $\beta=-2$, yet in the present fractional model, we will show that $\beta>0$.
\item Finally, the parameter $\zeta$ is a real parameter as mentioned previously. Its numerical estimate is somewhat tricky. However, some phenomenological arguments argue that $\zeta\approx 1$ in a radiation-dominated era \cite{114}. In the minimal supersymmetric standard model plasma model, the parameter $\zeta$ has been constrained to $10^{-3}<\zeta <10^{-2}$.  This parameter plays a leading role in inflationary cosmology dominated by the Higgs field, mainly when the number of e-folds during inflation is not too large \cite{112}. In extended supergravities \cite{86,87,88,89,90,91,92}, we have $1<\zeta <2$. Some arguments suggest large values of the parameter $\zeta$, mainly when the gravitational field interacts with the inflaton during the coherent oscillations in the kinematic regime \cite{116}. Hence, measuring this parameter requires additional astronomical observations and astrophysical constraints.
\end{enumerate}
\end{Remark}

The Hubble parameter as a function of the redshift $z$, with $1+z = a_0/a$ and $a_0$ the present value of the scale factor, is written as \cite{120}
\begin{equation}
H(z) = -(1+z)^{-1}\frac{dz}{dt}, 
\qquad 
E(z)= H_0\,H(z).
\label{eq:Hubble-z}
\end{equation}

From the scale factor, we define the deceleration, jerk, and snap parameters:
\begin{equation}
q(t) := -\frac{\ddot{a}\,a}{\dot{a}^2}, 
\qquad 
j(t) := \frac{\dddot{a}\,a^2}{\dot{a}^3}, 
\qquad 
s(t) := \frac{a^{(4)}\,a^{3}}{\dot a^{4}}.
\end{equation}

Using $E(z)$, the deceleration parameter becomes
\begin{equation}\label{eq:qz}
q(z) = -1 + (1+z)\,\frac{1}{E(z)}\,\frac{dE}{dz},
\end{equation}
where the derivative $dE/dz$ is computed numerically using finite differences.  
The jerk parameter is
\begin{equation}\label{eq:jz}
j(z) =(1+z)\frac{dq(z)}{dz} + q(z)\bigl(1+2q(z)\bigr),
\end{equation}
and the snap parameter is
\begin{equation}\label{eq:sz}
s(z) = j(z)\,\bigl(2 + 3q(z)\bigr) + (1+z)\,\frac{dj}{dz},
\end{equation}
where the derivatives $dq/dz$ and $dj/dz$ are also evaluated numerically.

From equations \eqref{tag8} and \eqref{anzat}, with $\varepsilon=(\alpha-1)/3$ and assuming $f\neq\xi^2$, we introduce the rescaled quantities
\begin{equation}
\xi=\Upsilon H_0,\qquad 
m=\zeta H_0,\qquad 
\varrho=-f H_0^2,
\end{equation}
and rescale time using $H_0 t=\tau$, so that
\begin{equation}
\dot{f}=H_0\frac{df}{d\tau}.
\end{equation}

The dimensionless Hubble parameter becomes
\begin{equation}
E(\tau):=\frac{H}{H_0}
=1+\Upsilon\varphi(\tau)+\frac{\alpha-1}{3\tau},
\label{eq:E-tau-def}
\end{equation}
and satisfies
\begin{equation}
\frac{d\ln a}{d\tau}=E(\tau).
\end{equation}

At the present time $t=t_0$ we define $\tau_0=H_0 t_0$.  Since $E(\tau_0)=1$, we obtain the initial condition
\begin{equation}
\varphi(\tau_0)=\frac{1-\alpha}{3\tau_0\Upsilon}.
\label{eq:IC1}
\end{equation}

From \eqref{anzat}, the present deceleration parameter is
\begin{equation}
q_0=-1-\Upsilon\frac{d\varphi}{d\tau}(\tau_0)+\frac{\alpha-1}{3\tau_0^2},
\end{equation}
which gives
\begin{equation}
\frac{d\varphi}{d\tau}(\tau_0)
=\frac{1}{\Upsilon}\left(-1-q_0+\frac{\alpha-1}{3\tau_0^2}\right).
\label{eq:IC2}
\end{equation}

From the original system \eqref{tag8} we obtain the second–order equation
\begin{equation}
\frac{d^2\varphi}{d\tau^2}
+3(\Upsilon\varphi+1)\frac{d\varphi}{d\tau}
+\varphi\bigl(\zeta^2-\varrho\varphi^2\bigr)=0,
\label{eq:main}
\end{equation}
with initial conditions \eqref{eq:IC1}–\eqref{eq:IC2}.  
This equation describes a nonlinear damped oscillator with frequency $\zeta$, nonlinearity $\varrho$, and damping coefficient $\Upsilon$.

Taking $a_0=1$ for convenience, the redshift satisfies
\begin{equation}
\frac{1}{1+z}=\frac{a}{a_0}.
\end{equation}
The relation between $\tau$ and $z$ is
\begin{align}
-\frac{d\ln(1+z)}{d\tau}&=E(\tau),\\
z(\tau_0)&=0.
\end{align}
Using \eqref{eq:IC1} one checks that $z'(\tau_0)=-1$.  
A quadrature formula for $z(\tau)$ is
\begin{equation}
z(\tau)
=-\int_{\tau_0}^{\tau} 
E(\kappa_2)\exp\!\Big(-\int_{\kappa_2}^{\tau} E(\kappa_1)\,d\kappa_1\Big)\,d\kappa_2.
\label{eq:quadrature}
\end{equation}
We analyse the nonlinear equation \eqref{eq:main} with initial conditions \eqref{eq:IC1}–\eqref{eq:IC2}, assuming that $\Upsilon$, $\zeta$, and $\varrho$ are small parameters. For $0<\Upsilon\ll1$, the initial conditions become singular as $\Upsilon\to0$, which motivates introducing suitable rescaled variables.

We define
\begin{equation}
\Phi=\Upsilon\varphi,\qquad 
\xi=\Upsilon H_0,\qquad 
m=\zeta H_0,\qquad 
\varrho=-f H_0^2,\qquad 
\mu=\frac{\varrho}{\Upsilon^2},\qquad 
\tau=H_0 t,
\end{equation}
together with
\begin{equation}
\alpha=1+3\tau_{\rm rel},\qquad 
\tau_{\rm rel}=\delta_0\tau_0^2.
\end{equation}

The inverse relations are
\begin{equation}
\Upsilon=\frac{\xi}{H_0},\qquad
\varrho=\mu\frac{\xi^2}{H_0^2},\qquad
f=-\frac{\mu\xi^2}{H_0^4},\qquad
\Phi=\frac{\xi}{H_0}\varphi,\qquad
t=\frac{\tau}{H_0}.
\label{eq:originals}
\end{equation}

Under these transformations, the field equation becomes
\begin{equation}
\frac{d^2\Phi}{d\tau^2}
+3(1+\Phi)\frac{d\Phi}{d\tau}
+\zeta^2\Phi-\mu\Phi^3=0,
\label{eq:phi-eq}
\end{equation}
We define 
\begin{equation}
\lambda=\frac{1}{\delta_0\tau_0^2}.
\label{lambda_u}
\end{equation}

The parameter $\lambda$ enters the initial conditions as
\begin{equation}
\Phi(\tau_0)=-\frac{1}{\lambda\tau_0},
\qquad
\frac{d\Phi}{d\tau}(\tau_0)
=-1-q_0+\frac{1}{\lambda\tau_0^{2}},
\label{eq:phi-ic}
\end{equation}

Since both the initial displacement $\Phi(\tau_0)$ and the initial slope $d\Phi/d\tau(\tau_0)$ scale as $1/\lambda$, the quantity
\begin{equation}
\lambda := \frac{1}{\tau_{\rm rel}}
\end{equation}
acts as a \emph{relaxation rate} for the rescaled field $\Phi$.  Equivalently, $\tau_{\rm rel}$ is the characteristic time (in the dimensionless variable $\tau$) over which the initial perturbation of $\Phi$ relaxes toward the attractor. A larger value of $\lambda$ (smaller $\tau_{\rm rel}$) corresponds to a smaller initial deviation and a faster return to equilibrium, while a smaller $\lambda$ (larger $\tau_{\rm rel}$) implies a larger initial departure and a slower relaxation.

For the numerical study, the system is expressed in terms of the variable $N$ (e‑folds) when convenient, and we apply the regularization and hybrid asymptotic–numerical scheme described in Section~\ref{Sect:2.6} to control stiffness and ensure accurate reconstruction of $H(z)$.

\section{Numerical procedure to obtain $H(z)$}
\label{Sect:2.6}

\noindent
In this section we describe a step–by–step numerical procedure to compute the Hubble function $H(z)$.  
To do this, we must solve the second–order differential equation for $\Phi(\tau)$ in \eqref{eq:phi-eq} with the initial conditions \eqref{eq:phi-ic}.

\subsection{Step A — Reduction to a first–order system}
\label{Sect:2.6.1}

The second–order equation \eqref{eq:phi-eq}, together with the initial conditions \eqref{eq:phi-ic}, can be written as an autonomous system by introducing $(x,y)=\left(\Phi, \frac{d\Phi}{d\tau}\right)$: 
\begin{align}
\frac{dx}{d\tau} &= y, \label{eq:stepA-x2}\\
\frac{dy}{d\tau} &= -3(x+1)y - \zeta^2 x + \mu x^3, \label{eq:stepA-y2}
\end{align}
with initial conditions
\begin{equation}
x(\tau_0) = -\tau_0^{-1}\lambda^{-1}, 
\qquad
y(\tau_0) = -1 - q_0 + \tau_0^{-2}\lambda^{-1}.
\label{eq:stepA-ic2}
\end{equation}

If we treat $x_0$ and $y_0$ as free initial parameters, together with $\lambda$, we obtain the derived quantities
\begin{equation}\label{defofalpha}
\alpha = \frac{\lambda+3}{\lambda}, 
\qquad
\tau_0 = -\frac{1}{\lambda x_0}, 
\qquad
q_0 = \lambda x_0^2 - y_0 - 1.
\end{equation}

Here, $\alpha$ is the effective order of the fractional derivative, $\tau_0$ is the age parameter of the universe (with $t_0=\tau_0/H_0$), and $q_0$ is the present deceleration parameter.

\subsection{Step B — Computing the effective rate $E(\tau)$}
\label{Sect:2.6.2}

The function $E(\tau)$ can be obtained directly from its definition. However, when terms of order $\tau^{-1}$ appear or the system becomes stiff, it is more stable to integrate the equivalent equation
\begin{equation}
E'(\tau)
= y - \lambda\,(1 - E + x)^2,
\qquad
\lambda := \frac{1}{\tau_{\rm rel}}.
\label{eq:Etau-deriv2}
\end{equation}

\subsubsection{Stability analysis of the reduced system}

The autonomous system \eqref{eq:stepA-x2}--\eqref{eq:stepA-y2} has a critical point at
\begin{equation}
(x,y)=(0,0),
\end{equation}
which is important for understanding the local behaviour of the solutions of \eqref{eq:phi-eq}.  
The linearization at the origin is
\begin{equation}
\begin{pmatrix} x' \\ y' \end{pmatrix}
=
\begin{pmatrix}
0 & 1 \\
-\zeta^2 & -3
\end{pmatrix}
\begin{pmatrix} x \\ y \end{pmatrix},
\end{equation}
with characteristic polynomial
\begin{equation}
\lambda^2 + 3\lambda + \zeta^2 = 0.
\end{equation}
The eigenvalues are
\begin{equation}
\lambda_{1,2}=\frac{-3\pm\sqrt{9-4\zeta^2}}{2}.
\end{equation}

\begin{itemize}
\item If $\zeta\neq 0$, then $\zeta^2>0$ and both eigenvalues have negative real part:
\begin{equation}
\Re(\lambda_{1,2})=-\frac{3}{2}<0.
\end{equation}
Thus, the origin is an \emph{asymptotically stable} critical point.  
More precisely:
\begin{itemize}
\item If $0<\zeta^2<9/4$, the eigenvalues are real and negative $\Rightarrow$ \emph{stable node}.
\item If $\zeta^2>9/4$, the eigenvalues are complex with negative real part $\Rightarrow$ \emph{stable focus}.
\end{itemize}

\item If $\zeta=0$, the polynomial becomes $\lambda(\lambda+3)=0$, so the origin is \emph{non‑hyperbolic}.  
In this case, stability depends on the nonlinear terms, especially on the sign of $\mu$ in the cubic term $\mu x^3$.
\end{itemize}

\subsubsection{Coupling with the equation for $E(\tau)$.}

The full system includes the extra equation \eqref{eq:Etau-deriv2}.  
The equilibrium associated with the origin in $(x,y)$ is obtained by imposing $E'=0$, which gives
\begin{equation}
(1-E)^2=0 \quad\Rightarrow\quad E=1.
\end{equation}
Thus the critical point of the extended system is
\begin{equation}
(x,y,E)=(0,0,1).
\end{equation}

Near this point, the stability of $(x,y)$ determines the overall behaviour:  
if $(x,y)\to(0,0)$ asymptotically (the case $\zeta\neq 0$), then $y\to 0$ and the quadratic term in \eqref{eq:Etau-deriv2} forces
\begin{equation}
E(\tau)\longrightarrow 1.
\end{equation}
Therefore, the equilibrium $(0,0,1)$ is \emph{asymptotically stable} whenever $\zeta\neq 0$.

If $\zeta=0$, stability depends on the nonlinear terms and must be studied using alternative procedures.

\subsection{Step C — Direct numerical integration in $z$ and reformulation of the system}
\label{Sect:2.6.3}

To construct the expansion function $E(z)$, there are two possible approaches:

\textbf{Option 1: Quadrature in $\tau$.}
\begin{itemize}
\item Integrate the quadrature \eqref{eq:quadrature} to obtain $z(\tau)$.
\item Interpolate $E(\tau)$ on the grid $z(\tau)$.
\end{itemize}

\textbf{Option 2: Direct integration in $z$ (recommended).}

The system \eqref{eq:stepA-x2}–\eqref{eq:Etau-deriv2} can be rewritten as
\begin{subequations}
\label{eq:system-x-y-E}
\begin{align}
\frac{dx}{dz} &= -\frac{y}{(1+z)E},\\
\frac{dy}{dz} &= -\frac{-3(x+1)y - \zeta^{2}x + \mu x^{3}}{(1+z)E},\\
\frac{dE}{dz} &= -\frac{y - \lambda(1+x-E)^{2}}{(1+z)E},
\end{align}
\end{subequations}
with initial conditions at $z=0$:
\begin{equation}\label{con-eq:system-x-y-E}
x(0)=-\frac{1}{\tau_0\lambda},\qquad
y(0)=-1-q_0+\frac{1}{\tau_0^2\lambda},\qquad
E(0)=1.
\end{equation}

\textbf{Change of variable $z\to N$}

To remove the prefactor $(1+z)$ we introduce
\begin{equation}
N=-\ln(1+z),\qquad \frac{dX}{dN}=-(1+z)\frac{dX}{dz}.
\end{equation}
The system becomes
\begin{subequations}
\label{eq:system-x-y-E-N}
\begin{align}
\frac{dx}{dN} &= \frac{y}{E},\\
\frac{dy}{dN} &= \frac{-3(x+1)y - \zeta^{2}x + \mu x^{3}}{E},\\
\frac{dE}{dN} &= \frac{y - \lambda(1+x-E)^{2}}{E}.
\end{align}
\end{subequations}

\subsubsection{Local series expansion around \texorpdfstring{$E=1$}{E=1}}

To analyse the local behaviour of \eqref{eq:system-x-y-E-N}, we construct Taylor expansions around a point where $E(0)=1$. Since the right-hand side is analytic for $E\neq 0$, the solution admits a convergent series in a neighborhood of $N=0$.

We write
\begin{align}
x(N) &= x_{0} + x_{1}N + \frac{x_{2}}{2}N^{2} + \mathcal{O}(N^{3}),\\
y(N) &= y_{0} + y_{1}N + \frac{y_{2}}{2}N^{2} + \mathcal{O}(N^{3}),\\
E(N) &= 1 + E_{1}N + \frac{E_{2}}{2}N^{2} + \mathcal{O}(N^{3}),
\end{align}
where $x_{0}=x(0)$ and $y_{0}=y(0)$ are arbitrary initial data.

\textbf{First-order coefficients}

Evaluating \eqref{eq:system-x-y-E-N} at $N=0$ gives
\begin{align}
x_{1} &= y_{0},\\
y_{1} &= -3(x_{0}+1)y_{0} - \zeta^{2}x_{0} + \mu x_{0}^{3},\\
E_{1} &= y_{0} - \lambda x_{0}^{2}.
\end{align}

\textbf{Second-order coefficients}

From
\begin{equation}
x'' = \frac{y'E - yE'}{E^{2}},
\end{equation}
and using $E(0)=1$, we obtain
\begin{equation}
x_{2} = y_{1} - y_{0}E_{1}
      = -3(x_{0}+1)y_{0} - \zeta^{2}x_{0} + \mu x_{0}^{3}
        - y_{0}^{2} + \lambda x_{0}^{2}y_{0}.
\end{equation}

For $y$, writing $F(x,y)=-3(x+1)y - \zeta^{2}x + \mu x^{3}$ so that $y'=F/E$, differentiation yields
\begin{equation}
y'' = F_{x}x' + F_{y}y' - F E'.
\end{equation}
Evaluating at $N=0$ gives
\begin{equation}
y_{2}
= F_{x}(x_{0},y_{0})\,x_{1}
  + F_{y}(x_{0},y_{0})\,y_{1}
  - F(x_{0},y_{0})\,E_{1},
\end{equation}
with
\begin{align}
F(x_{0},y_{0}) &= y_{1},\\
F_{x}(x_{0},y_{0}) &= -3y_{0} - \zeta^{2} + 3\mu x_{0}^{2},\\
F_{y}(x_{0},y_{0}) &= -3(x_{0}+1).
\end{align}

For $E$, writing
\begin{equation}
G(x,y,E)=\frac{y - \lambda(1+x-E)^{2}}{E},
\end{equation}
we have
\begin{equation}
E'' = G_{x}x' + G_{y}y' + G_{E}E'.
\end{equation}
At $(x_{0},y_{0},1)$,
\begin{align}
G_{x} &= -2\lambda x_{0},\\
G_{y} &= 1,\\
G_{E} &= 2\lambda x_{0} - y_{0} + \lambda x_{0}^{2},
\end{align}
so
\begin{equation}
E_{2}
= -2\lambda x_{0}x_{1} + y_{1}
  + \bigl(2\lambda x_{0} - y_{0} + \lambda x_{0}^{2}\bigr)E_{1}.
\end{equation}

\textbf{Resulting expansion}

Collecting the above, the local series expansion around $E(0)=1$ is
\begin{align}
x(N) &= x_{0}
       + y_{0}N
       + \frac{1}{2}\Bigl[
          -3(x_{0}+1)y_{0}
          - \zeta^{2}x_{0}
          + \mu x_{0}^{3}
          - y_{0}^{2}
          + \lambda x_{0}^{2}y_{0}
       \Bigr]N^{2}
       + \mathcal{O}(N^{3}),\\
y(N) &= y_{0}
       + y_{1}N
       + \frac{1}{2}y_{2}N^{2}
       + \mathcal{O}(N^{3}),\\
E(N) &= 1
       + E_{1}N
       + \frac{1}{2}E_{2}N^{2}
       + \mathcal{O}(N^{3}),
\end{align}
where $y_{1}$, $y_{2}$ and $E_{1}$ are those already defined above. All coefficients are explicit polynomials in the initial data $(x_{0},y_{0})$ and the parameters $(\zeta,\mu,\lambda)$. Higher-order terms follow recursively from successive differentiation of the system, although the expressions grow rapidly in complexity. This expansion is useful for local stability analysis, approximate invariant manifolds, and high-accuracy numerical initialization.

\subsubsection{Puiseux-Type Expansion Near the Singular Point \texorpdfstring{$(0,0,0)$}{(0,0,0)}}

The system \eqref{eq:system-x-y-E-N} becomes singular at $(x,y,E)=(0,0,0)$ because each equation contains the factor $1/E$.  
Along trajectories satisfying
\begin{equation}
x(N)\to 0,\qquad y(N)\to 0,\qquad E(N)\to 0,
\end{equation}
the numerator of the $E$–equation tends to $-\lambda$ and $1+x-E\to 1$.  
The dominant balance is therefore
\begin{equation}
\frac{dE}{dN}\sim -\frac{\lambda}{E},
\qquad
\frac{d}{dN}(E^{2})\sim -2\lambda,
\end{equation}
which integrates near some $N_{0}$ to
\begin{equation}
E(N)\sim a_{1}\,\Delta^{1/2},\qquad 
\Delta:=N-N_{0},\qquad
a_{1}=\pm\sqrt{-2\lambda}.
\end{equation}
Thus, the singularity is a square-root branch point, and the appropriate local representation is a Puiseux expansion
\begin{equation}
E(N)=\sum_{k=1}^{\infty}a_{k}\,\Delta^{k/2},
\label{eq:Puiseux-E}
\end{equation}
with analogous expansions for $x$ and $y$:
\begin{align}
x(N)&=\sum_{k=0}^{\infty}b_{k}\,\Delta^{k/2},\\
y(N)&=\sum_{k=0}^{\infty}c_{k}\,\Delta^{k/2}.
\end{align}

Substituting these series into \eqref{eq:system-x-y-E-N} and matching coefficients of $\Delta^{k/2}$ yields a recursive algebraic system for the triples $(a_{k},b_{k},c_{k})$.  

Below, we list all coefficients up to the third order.

\textbf{Order $\Delta^{-1/2}$.}
From $x'=y/E$,
\begin{equation}
x'=\tfrac12 b_{1}\Delta^{-1/2}+\cdots,
\qquad
\frac{y}{E}=\frac{c_{0}}{a_{1}}\Delta^{-1/2}+\cdots,
\end{equation}
so
\begin{equation}
b_{1}=\frac{2c_{0}}{a_{1}}.
\label{eq:b1}
\end{equation}

From $y'=F/E$ with $F=-3(x+1)y-\zeta^{2}x+\mu x^{3}$,
\begin{equation}
y'=\tfrac12 c_{1}\Delta^{-1/2}+\cdots,
\qquad
\frac{F}{E}=\frac{F_{0}}{a_{1}}\Delta^{-1/2}+\cdots,
\end{equation}
where
\begin{equation}
F_{0}=-3(b_{0}+1)c_{0}-\zeta^{2}b_{0}+\mu b_{0}^{3}.
\end{equation}
Thus
\begin{equation}
c_{1}=\frac{2F_{0}}{a_{1}}.
\label{eq:c1}
\end{equation}

\textbf{Order $\Delta^{0}$ in the $E$–equation.}
Expanding the numerator of $E'$ gives
\begin{equation}
y-\lambda(1+x-E)^{2}
=N_{0}+N_{1}\Delta^{1/2}+\cdots,
\end{equation}
with
\begin{equation}
N_{0}=c_{0}-\lambda(1+b_{0})^{2},\qquad
N_{1}=c_{1}-2\lambda(1+b_{0})(b_{1}-a_{1}).
\end{equation}
The leading-order condition reproduces $a_{1}^{2}=-2\lambda$. The next order requires
\begin{equation}
c_{1}=2\lambda(1+b_{0})(b_{1}-a_{1}),
\label{eq:N1}
\end{equation}
which is a consistency condition relating the leading amplitudes.

\textbf{Order $\Delta^{0}$ in $x'$ and $y'$.}
From $x'=y/E$,
\begin{equation}
b_{2}=\frac{c_{1}}{a_{1}}.
\label{eq:b2}
\end{equation}

From $y'=F/E$,
\begin{equation}
c_{2}=\frac{F_{1}}{a_{1}},
\end{equation}
where
\begin{equation}
F_{1}
=-3\bigl[(b_{0}+1)c_{1}+b_{1}c_{0}\bigr]
 -\zeta^{2}b_{1}
 +3\mu b_{0}^{2}b_{1}.
\end{equation}

\textbf{Order $\Delta^{1/2}$ in the $E$–equation.}
This determines $a_{2}$:
\begin{equation}
a_{2}=\frac{N_{2}}{3a_{1}},
\end{equation}
where $N_{2}$ is the coefficient of $\Delta$ in the expansion of $y-\lambda(1+x-E)^{2}$.  
Its explicit form is polynomial in $(b_{0},c_{0})$ and the parameters $(\zeta,\mu,\lambda)$.

\textbf{Summary up to third order}

Up to order $\Delta^{3/2}$ in $E$ and order $\Delta$ in $x$ and $y$, the Puiseux expansions are
\begin{align}
E(N)&=a_{1}\Delta^{1/2}+a_{2}\Delta^{3/2}+\mathcal{O}(\Delta^{5/2}),\\
x(N)&=b_{0}+b_{1}\Delta^{1/2}+b_{2}\Delta+\mathcal{O}(\Delta^{3/2}),\\
y(N)&=c_{0}+c_{1}\Delta^{1/2}+c_{2}\Delta+\mathcal{O}(\Delta^{3/2}),
\end{align}
with
\begin{align}
a_{1}&=\pm\sqrt{-2\lambda},\\
b_{1}&=\frac{2c_{0}}{a_{1}},\\
c_{1}&=\frac{2\bigl[-3(b_{0}+1)c_{0}-\zeta^{2}b_{0}+\mu b_{0}^{3}\bigr]}{a_{1}},\\
b_{2}&=\frac{c_{1}}{a_{1}},\\
c_{2}&=\frac{-3[(b_{0}+1)c_{1}+b_{1}c_{0}]
              -\zeta^{2}b_{1}
              +3\mu b_{0}^{2}b_{1}}{a_{1}},\\
a_{2}&=\frac{N_{2}}{3a_{1}},
\end{align}
and the consistency condition
\begin{equation}
c_{1}=2\lambda(1+b_{0})(b_{1}-a_{1})
\end{equation}
must hold.

All coefficients are explicit polynomials in $(b_{0},c_{0})$ and the parameters $(\zeta,\mu,\lambda)$.  
The singularity at $(0,0,0)$ is therefore a genuine branch point, and the correct local representation is the Puiseux expansion \eqref{eq:Puiseux-E}.

\subsection{Step D — Reconstruction of $H(z)$}
\label{Sect:2.6.4}

The Hubble function is obtained directly from
\begin{equation}
H(z)=H_0\,E(z).
\end{equation}

If the integration was done in $\tau$, we first interpolate $E(\tau)$ on the grid in $z$ and then evaluate the expression above.

\section{Dynamical cosmological model and Bayesian inference}
\label{Sect:2.7}
In this paper, we test whether the fractional model can reproduce current cosmological observations. To achieve this, we compute the best-fit parameters at the $1\sigma\,(68.3\%)$ confidence level (CL) using the affine-invariant Markov chain Monte Carlo (MCMC) method \cite{Goodman_Ensemble_2010}, implemented in the pure-Python code \textit{emcee} \cite{Foreman-Mackey:2012any}. In this procedure, we consider 40 chains or ``walkers" with a total of $7500$ steps. Due to a strong degeneracy between certain parameters, as we can see later, we only use the mean acceptance fraction of the chains as a " convergence test, which must have a value between $0.2$ and $0.5$ \cite{Foreman-Mackey:2012any} and can be modified by the stretch move provided by the \textit{emcee} module. The first $1500$ steps are discarded as ``burn-in" steps, and the statistical analysis is performed on a flattened chain. Therefore, for this MCMC analysis, we consider the following Gaussian distribution as the likelihood function:
\begin{equation}\label{likelihood}
\mathcal{L}_{I}\left(\boldsymbol{X},\boldsymbol{\theta}\right)\propto\exp{\left(-\frac{\chi_{\text{I}}^{2}\left(\boldsymbol{X},\boldsymbol{\theta}\right)}{2}\right)},
\end{equation}
where $\chi_{\text{I}}^{2}$ is the merit function, with $\boldsymbol{\theta}$ representing the parameter space of the model and $\boldsymbol{X}$ the observational data, and I stands for each data set consider for the constraint, including the joint analysis in which $\chi^{2}_{\text{joint}}=\sum_{I}\chi_{\text{I}}^{2}$. 

Since the best-fit parameters minimize the merit function, we can use the evaluation of the best-fit parameters in the merit function, $\chi_{\text{min}}^{2}$, as an indicator of the goodness of the fit: the smaller the value of $\chi_{\text{min}}^{2}$ is, the better the fit. Nevertheless, in principle, the value of $\chi_{\text{min}}^{2}$ obtained for the best-fit parameters can be reduced by adding free parameters to the model under study, potentially resulting in overfitting. Hence, we compute the Bayesian Information Criterion (BIC) \cite{Schwarz:1978tpv} to compare the goodness of the fit statistically. This criterion adds a penalization to the value of $\chi_{\text{min}}^{2}$, which depends on the total number of free parameters of the model ($\theta_{N}$), according to the expression
\begin{equation}
    BIC=\theta_{N}\ln{(n)}+\chi_{\text{min}}^{2},
\end{equation}
where $n$ is the total number of data points in the corresponding data sample. Thus, when two different models are compared, the one most consistent with the observations is the one with the smallest BIC. In general, a difference of $2-6$ in BIC is evidence against the model with the higher BIC, a difference of $6-10$ is strong evidence, and a difference of $>10$ is very strong evidence.

\subsection{\label{subsec:Data}Observational data}
In this section, we provide a brief overview of the merit function for each data set included in the cosmological constraints.

\subsubsection{\label{subsec:SNeIa}Type Ia supernovae}
Supernovae (SNe) are highly energetic explosions of certain stars and play an important role in astrophysics and cosmology, as they have been used as cosmic distance indicators. In particular, Type Ia Supernovae (SNe Ia) are considered standard candles to measure the geometry and the late-time dynamics of the Universe \cite{Liu:2023qmw}. For SNe Ia data, we consider the Pantheon+ sample \cite{Brout:2022vxf}, which consists of 1701 data points in the redshift range $0.001\leq z\leq 2.26$, whose respective merit function can be conveniently constructed in matrix notation (denoted by bold symbols) as
\begin{equation}\label{meritSNe}
\chi_{\text{SNe}}^{2}=\mathbf{\Delta D}(z,\boldsymbol{\theta},M)^{\dagger}\mathbf{C}^{-1}\mathbf{\Delta D}(z,\boldsymbol{\theta},M),
\end{equation}
where $\left[\mathbf{\Delta D}(z,\boldsymbol{\theta},M)\right]_{i}= m_{B,i}-M-\mu_{th}(z_{i},\boldsymbol{\theta})$ and $\mathbf{C}=\mathbf{C}_{\text{stat}}+\mathbf{C}_{\text{sys}}$, with $\mathbf{C}$ being the total uncertainty covariance matrix. The matrices $\mathbf{C}_{\text{stat}}$ and $\mathbf{C}_{\text{sys}}$ account for the statistical and systematic uncertainties, respectively. The quantity $\mu_{i}=m_{B, i}-M$ corresponds to the observational distance modulus of the Pantheon+ sample, which is obtained by a modified version of Trip's formula \cite{Tripp:1997wt} and the BBC (BEAMS with Bias Corrections) approach \cite{Kessler:2016uwi}. In contrast, $m_{B, i}$ is the corrected apparent B-band magnitude of a fiducial SNe Ia at redshift $z_{i}$, and $M$ is the fiducial magnitude of a SNe Ia, which must be jointly estimated with the free parameters of the model under study. On the other hand, the theoretical  distance modulus for a spatially flat FLRW spacetime is given by
\begin{equation}\label{theoreticaldistance}
\mu_{th}(z_{i},\boldsymbol{\theta})=5\log_{10}{\left[\frac{d_{L}(z_{i},\boldsymbol{\theta})}{\text{Mpc}}\right]}+25,
\end{equation}
with $d_{L}(z_{i},\boldsymbol{\theta})$ the  luminosity distance given by
\begin{equation}\label{luminosity}
d_{L}(z_{i},\boldsymbol{\theta})=c(1+z_{i})\int_{0}^{z_{i}}{\frac{dz'}{H_{th}(z',\boldsymbol{\theta})}},
\end{equation}
where $c$ is the speed of light given in units of $\text{km/s}$.

In principle, there is a degeneracy between $M$ and $H_{0}$. Hence, to constrain $H_{0}$ using SNe Ia data alone, it is necessary to include the SH0ES (Supernovae and $H_{0}$ for the Equation of State of the dark energy program) Cepheid host distance anchors, with a merit function of the form
\begin{equation}\label{Cepheidmerit}
\chi^{2}_{\text{Cepheid}}=\mathbf{\Delta D}_{\text{Cepheid}}\left(M\right)^{\dagger}\textbf{C}^{-1}\mathbf{\Delta D}_{\text{Cepheid}}\left(M\right),
\end{equation}
where 
$\left[\mathbf{\Delta D}_{\text{Cepheid}}\left(M\right)\right]_{i}=\mu_{i}\left(M\right)-\mu_{i}^{\text{Cepheid}}$, where $\mu_{i}^{\text{Cepheid}}$ is the Cepheid calibrated host-galaxy distance obtained by SH0ES \cite{Riess:2021jrx}. Thus, we use the Cepheid distances as the ``theory model'' to calibrate $M$, considering that the difference $\mu_{i}\left(M\right)-\mu_{i}^{\text{Cepheid}}$ is sensitive to $M$ and largely insensitive to other parameters of the cosmological model. Therefore, taking into account that the total uncertainty covariance matrix for Cepheid is contained in the total uncertainty covariance matrix $\mathbf{C}$, we define the merit function for the SNe Ia data as
\begin{equation}\label{SNemeritfull}
\chi_{\text{SNe}}^{2}=\mathbf{\Delta D'}(z,\boldsymbol{\theta},M)^{\dagger}\mathbf{C}^{-1}\mathbf{\Delta D'}(z,\boldsymbol{\theta},M),
\end{equation}
where
\begin{equation}\label{SNeresidual}
\Delta\mathbf{D'}_{i}=\left\{\begin{array}{ll}
m_{B,i}-M-\mu_{i}^{\text{Cepheid}} & i\in\text{Cepheid host} \\
\\ m_{B,i}-M-\mu_{th}(z_{i},\boldsymbol{\theta}) & \text{otherwise}
\end{array}
\right..
\end{equation}

\subsubsection{\label{subsec:CC}Cosmic Chronometers}
Even though SNe Ia data provide consistent evidence about the existence of a transition epoch in cosmic history where the expansion rate of the Universe changes, it is important to highlight that this conclusion is obtained in a model-dependent way \cite{Moresco:2016mzx}. The study of the expansion rate of the Universe in a model-independent way can be carried out through measurements of the Hubble parameter compiled in the Cosmic Chronometers (CC) data sample.
In this paper, we consider the data set from Ref. \cite{Capozziello:2017nbu}, which comprises 31 Hubble parameter data points spanning the redshift range $0.0708\leq z\leq 1.965$. In this case, the merit function can be directly constructed as follows:
\begin{equation}\label{meritCC}
    \chi_{CC}^{2}=\sum_{i=1}^{31}{\left[\frac{H_{i}-H_{th}(z_{i},\boldsymbol{\theta)}}{\sigma_{H,i}}\right]^{2}},
\end{equation}
where $H_{i}$ represents the observational Hubble parameter data at redshift $z_{i}$, with an associated uncertainty $\sigma_{H,i}$, all provided by the CC sample, and $H_{\text{th}}$ denotes the theoretical Hubble parameter value at the same redshift. It is important to mention that these Hubble parameter data points are derived using the differential age method, a model-independent approach \cite{Jimenez:2001gg}.

\subsubsection{\label{subsec:BAO}Baryon acoustic oscillation}
Measurements of baryon acoustic oscillations (BAO) provide a powerful tool for studying the expansion history of the Universe. The method is based on a characteristic scale that was imprinted on matter clustering by pressure waves propagating through the coupled photon-baryon fluid of the pre-recombination era \cite{Eisenstein:1998tu,Blake:2003rh,Seo:2003pu}. In other words, BAO exploits the enhancement of clustering at the scale of the pre-recombination sound horizon \cite{Weinberg:2013agg}, given by
\begin{equation}\label{sh}
    r_{d}=\int_{z_{d}}^{\infty}{\frac{C_{s}(z)}{H(z)}}dz,
\end{equation}
where $C_{s}(z)$ is the speed of sound in the photon-baryon fluid and $z_{d}\approx 1060$ \cite{Planck:2018vyg} is the redshift at which acoustic waves stall because photons no longer ``drag'' the baryons. Assuming a standard cosmological model in the pre-recombination epoch, in which the universe is composed of baryons, CDM, photons, and other relativistic species (which does not necessarily imply $\Lambda$CDM cosmology) \cite{Brieden:2022heh}, the sound horizon above can be written as
\begin{equation}\label{shstandard}
    r_{d}=147.05\,Mpc\left(\frac{\Omega_{B,0}h^{2}}{0.02236}\right)^{-0.13}\left[\frac{(\Omega_{B,0}+\Omega_{DM,0})h^{2}}{0.1432}\right]^{-0.23}\left(\frac{N_{eff}}{3.04}\right)^{-0.1},
\end{equation}
where $H_{0}=100\frac{km/s}{Mpc}h$, with $h$ being the reduced Hubble parameter, and $N_{eff}=3.044$ represents the effective number of relativistic degrees of freedom for three neutrinos at $z>z_{d}$. The above equation is scaled to the best-fit values obtained from the Planck collaboration \cite{Planck:2018vyg}, and $\Omega_{B,0}$ and $\Omega_{DM,0}$ represent the density parameters of baryons and DM, respectively. Nevertheless, since the model under study does not explicitly depend on $\Omega_{B,0}$ and $\Omega_{DM,0}$, we consider $r_{d}$ as a free parameter which must be jointly estimated with the free parameters of the model under study.

Concerning the BAO measurements, the BAO scale in the transverse direction for a spatially flat FLRW spacetime is given by $d_{L}(z_{i},\boldsymbol{\theta})=(1+z_{i})d_{M}(z_{i},\boldsymbol{\theta})$, or equivalently, according to Eq. \eqref{luminosity},
\begin{equation}\label{BAODM}
    d_{M}(z_{i},\boldsymbol{\theta})=c\int_{0}^{z_{i}}{\frac{dz'}{H_{th}(z',\boldsymbol{\theta})}dz'},
\end{equation}
which allows for constraining the transverse comoving distance. On the other hand, the BAO measurement in the line-of-sight direction is given by
\begin{equation}\label{BAODH}
    d_{H}(z_{i},\boldsymbol{\theta})=\frac{c}{H_{th}(z_{i},\boldsymbol{\theta})},
\end{equation}
which allows us to constrain the expansion rate of the universe, $H(z)$.

In this paper, we consider the DESI DR2 sample (see Table IV of Ref. \cite{DESI:2025zgx}), whose respective inferred BAO distances are given relative to the sound horizon, i.e., $d_{M,i}/r_{d}$ and $d_{H,i}/r_{d}$, except for one data point, which corresponds to the isotropic BAO distance, given by
\begin{equation}\label{BAODV}
    d_{V}(z_{i},\boldsymbol{\theta})=\left[z_{i}d_{M}^{2}(z_{i},\boldsymbol{\theta})d_{H}(z_{i},\boldsymbol{\theta})\right]^{1/3}.
\end{equation}
Therefore, the merit function for the BAO measurements is constructed as
\begin{equation}\label{meritBAO}
    \chi_{\text{BAO}}^{2}=\sum_{i=1}^{13} 
{\left[\frac{\Delta d_{i}(z_{i},\boldsymbol{\theta})}{\sigma_{D,i}}\right]^{2}},
\end{equation}
where $\sigma_{D,i}$ is the associated uncertainty and
\begin{equation}\label{BAOdata}
\Delta d_{i}(z_{i},\boldsymbol{\theta})=\left\{\begin{array}{lll}
\left[d_{M,i}-d_{M}(z_{i},\boldsymbol{\theta})\right]/r_{d}, & i\in A \\
\\ \left[d_{H,i}-d_{H}(z_{i},\boldsymbol{\theta})\right]/r_{d}, & i\in A \\
\\ \left[d_{V,i}-d_{V}(z_{i},\boldsymbol{\theta})\right]/r_{d}, & i\in B
\end{array}
\right.,
\end{equation}
with $A=\{\text{LRG1},\text{LRG2},\text{LRG3}+\text{ELG1},\text{ELG2},\text{QSO},\allowbreak\text{Lya}\}$ and $B=\{\text{BGS}\}$.

\subsubsection{\label{subsec:lensing}Gravitational Lensing}
Multiple images are produced when a background object (the source) is lensed by the gravitational field of a massive body (the lens). Therefore, the light rays emitted from the source will take different paths through spacetime at different image positions and arrive at the observer at different times. In this sense, the time delay of two different images $k$ and $l$ depends on the mass distribution along the line-of-sight of the lensing object, which can be calculated as follows:
\begin{equation}\label{lensing}
    \Delta t_{kl}=\frac{d_{\Delta t}}{c}\left[\frac{\left(\phi_{k}-\beta\right)^{2}}{2}-\psi(\phi_{k})-\frac{\left(\phi_{l}-\beta\right)^{2}}{2}+\psi(\phi_{l})\right],
\end{equation}
where $\phi_{k}$ and $\phi_{l}$ are the angular positions of the images, $\beta$ is the angular positions of the source, $\psi(\phi_{k})$ and $\psi(\phi_{l})$ are the lens potentials at the image positions, and $d_{\Delta t}$ is the ``time-delay distance'', which is theoretically given by the expression \cite{Treu:2016ljm}
\begin{equation}\label{time-delay}
    d_{\Delta t}^{th}(\mathbf{z},\boldsymbol{\theta})=\left(1+z_{l}\right)\frac{d_{A,l}(z_{l},\boldsymbol{\theta})d_{A,s}(z_{s},\boldsymbol{\theta})}{d_{A,ls}(z_{ls},\boldsymbol{\theta})},
\end{equation}
where the subscripts $l$, $s$, and $ls$ stand for the lens, the source, and between the lens and the source, respectively; $\mathbf{z}=(z_{l},z_{s},z_{ls})$ and $d_{A,j}$ is the angular diameter distance, which can be written in terms of the luminosity distance \eqref{luminosity} as $d_{L}(z_{j},\boldsymbol{\theta})=d_{A,j}(1+z_{j})^{2}$, or
\begin{equation}\label{angulardistance}
    d_{A,j}(z_{j},\boldsymbol{\theta})=\frac{c}{(1+z_{j})}\int_{0}^{z_{j}}{\frac{dz'}{H_{th}(z',\boldsymbol{\theta})}}.
\end{equation}

In this paper, we consider the Gravitational Lensing (GL) compilation provided by the H0LiCOW collaboration \cite{Wong:2019kwg}, which consists of six lensed quasars: B1608+656 \cite{Jee:2019hah}, SDSS 1206+4332 \cite{Birrer:2018vtm}, WFI2033-4723 \cite{Rusu:2019xrq}, RXJ1131-1231, HE 0435-1223, and PG 1115-080 \cite{Chen:2019ejq}; whose respective merit function can be written as
\begin{equation}\label{H0LiCOWmerit}
    \chi^{2}_{\text{GL}}=\sum_{i=1}^{10}{\left[\frac{d_{\Delta t,i}-d_{\Delta t}^{th}(\mathbf{z}_{i},\boldsymbol{\theta})}{\sigma_{d_{\Delta t},i}}\right]^{2}},
\end{equation}
where $d_{\Delta t, i}$ is the observational time-delay distance of the lensed quasar at redshift $\mathbf{z}_{i}=(z_{l, i};z_{s, i};z_{ls, i})$ with an associated uncertainty $\sigma_{d_{\Delta t}, i}$ (for more details, see Ref. \cite{Wong:2019kwg}). It is important to note that, for $z\to 0$, the angular diameter distance \eqref{angulardistance} tends to $d_{A}\to cz/H_{0}$ and, therefore, the gravitational lensing data of the H0LiCOW collaboration is sensitive to $H_{0}$, with a weak dependency on other cosmological parameters.

\subsubsection{\label{subsec:BHS}Black Hole Shadows}
The Black Hole Shadows (BHS) data are of interest for studying our local universe since their dynamics are quite simple and can be seen as standard rulers if the angular size redshift $\alpha$, the relation between the size of the shadow and the mass of the supermassive black hole that produces it, is established \cite{Escamilla-Rivera:2022mkc}. In this paper, we are interested in two measurements: the first one was made on the M87* supermassive black hole by The Event Horizon Telescope Collaboration \cite{EventHorizonTelescope:2019dse} (the first detection of a BHS), and the second one corresponds to the detection of Sagittarius A* (Sgr A*) \cite{EventHorizonTelescope:2022wkp}.

Light rays curve around its event horizon in a black hole (BH), creating a ring with a black spot at its center, the so-called shadow of the BH. In this sense, the angular radius of the BHS for a Schwarzschild (SH) BH at redshift $z_{i}$ can be written as
\begin{equation}\label{angularradius}
    \alpha_{SH}\left(z_{i},\boldsymbol{\theta}\right)=\frac{3\sqrt{3}m}{d_{A}(z_{i},\boldsymbol{\theta})},
\end{equation}
where $d_{A}(z_{i},\boldsymbol{\theta})$ is given by Eq. \eqref{angulardistance} (note that the sub-index $j$ is not necessary in this case) and $m=GM_{BH}/c^{2}$ is the mass parameter of the BH, with $M_{BH}$ the mass of the BH in solar masses units and $G$ the gravitational constant. It is common to write Eq. \eqref{angularradius} in terms of the shadow radius $\alpha_{SH}(z_{i},\boldsymbol{\theta})=R_{SH}/d_{A}(z_{i},\boldsymbol{\theta})$, where $R_{SH}=3\sqrt{3}GM_{BH}/c^{2}$ (the speed of light is given in units of $\text{m/s}$ in this case). Therefore, the merit function for the BHS data can be constructed as
\begin{equation}\label{BHSmerit}
    \chi^{2}_{BHS}=\sum_{i=1}^{2}{\left[\frac{\alpha_{i}-\alpha_{SH}(z_{i},\boldsymbol{\theta})}{\sigma_{\alpha,i}}\right]^{2}},
\end{equation}
where $\alpha_{i}$ is the observational angular radius of the BHS at redshift $z_{i}$ with an associated uncertainty $\sigma_{\alpha,i}$. It is important to note that for $z\to 0$ the angular radius \eqref{angularradius} tends to $\alpha_{SH}\to R_{SH}H_{0}/cz$ and, therefore, similar to the GL data, the BHS data is sensitive to $H_{0}$, with a weak dependency on other cosmological parameters. On the other hand, Eq. \eqref{angularradius} is divided by a factor of $1.496\times 10^{11}$ to obtain $\alpha_{SH}$ in units of $\mu as$.

\subsection{\label{subsec:priors}Theoretical Hubble parameter and priors}
Regardless of the dataset considered in the constraints, we need to define the theoretical Hubble parameter for the cosmological model under study. In our case, we integrate the system \eqref{eq:system-x-y-E} with the initial conditions given by Eq. \eqref{con-eq:system-x-y-E}, which ensures the model is correctly normalized at present. The function $E(z)$ gives the evolution of the expansion rate as a function of redshift, through
\begin{equation}
H(z) = H_0\,E(z),
\end{equation}
where $H_0=100\frac{km/s}{Mpc}h$ is the current Hubble parameter. We derive the solutions using the \textit{numbalsoda} package, a Python wrapper for the LSODA method in ODEPACK, which is used to solve ordinary differential equation initial value problems (\url{https://github.com/Nicholaswogan/numbalsoda}). Also, we use the definition given by Eq. \eqref{defofalpha} to constrain the parameter $\alpha$ instead of $\lambda$. Therefore, the respective free parameters of the model are $\boldsymbol{\theta}=\{h,r_{d},\tau_{0},q_{0},\alpha,\zeta,\mu,M\}$, for which we consider the following flat (F) priors in our MCMC analysis:
\begin{itemize}
    \item $h\in F(0.55,0.85)$,
    \item $r_{d}\in F(120,170)\,Mpc$,
    \item $\tau_{0}\in F(0.5,4)$,
    \item $q_{0}\in F(-1,0)$,
    \item $\alpha\in F(1,4)$,
    \item $\zeta\in F(0,4)$,
    \item $\mu\in F(-3,10)$,
    \item $M\in F(-20,-18)$,
\end{itemize}

It is important to highlight that the parameter $\zeta$ is related to the $\phi ^{2}$ term of the scalar field potential, while $\mu$ is related to the $\phi ^{4}$ term. So, for the constraint, we consider two cases: (i) the model in which the $\phi^{2}$ term dominates at the late-universe, which correspond to a case where $\mu =0$, and (ii) the model in which the $\phi^{4}$ term influence in the late-time evolution of the universe, i.e., the case where $\mu \neq 0$. Furthermore, we also constrain the $\Lambda$CDM model for comparison, whose respective theoretical Hubble parameter at late-times is given by
\begin{equation}\label{HLCDM}
    H(z)=H_{0}\sqrt{\Omega_{m,0}(1+z)^{3}+1-\Omega_{m,0}},
\end{equation}
where we have applied the constraint $\Omega_{\Lambda,0}=1-\Omega_{m,0}$ and, consequently, its respective free parameters are $\boldsymbol{\theta}=(h,r_{d},\Omega_{m,0},M)$, for which we use the same priors as in the model under study.

\subsection{\label{sec:results}Results and Discussion}
\begin{figure}[H]
    \centering
    \includegraphics[width=\columnwidth]{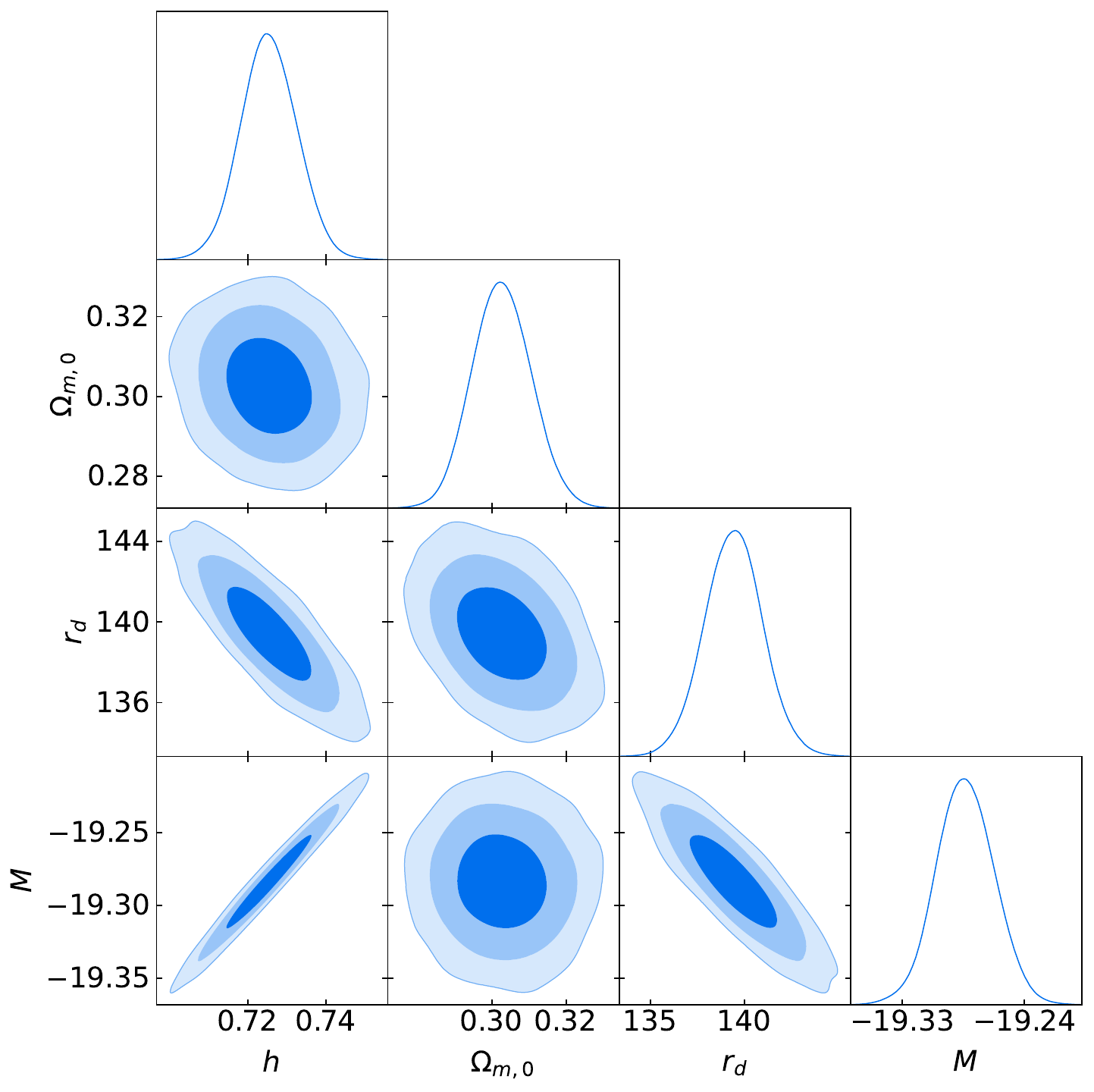}
    \caption{The posterior 1D distributions and joint marginalized regions for the free parameter space of the $\Lambda$CDM model, obtained via the MCMC analysis described in Section \ref{Sect:2.7}. The admissible joint regions correspond to the $1\sigma$, $2\sigma$, and $3\sigma$ CL, respectively. The best-fit values for each free parameter are shown in Table \ref{tab:best-fits}.}
    \label{fig:TriangleLCDM}
\end{figure}

\begin{figure}[H]
    \centering
    \includegraphics[width=\columnwidth]{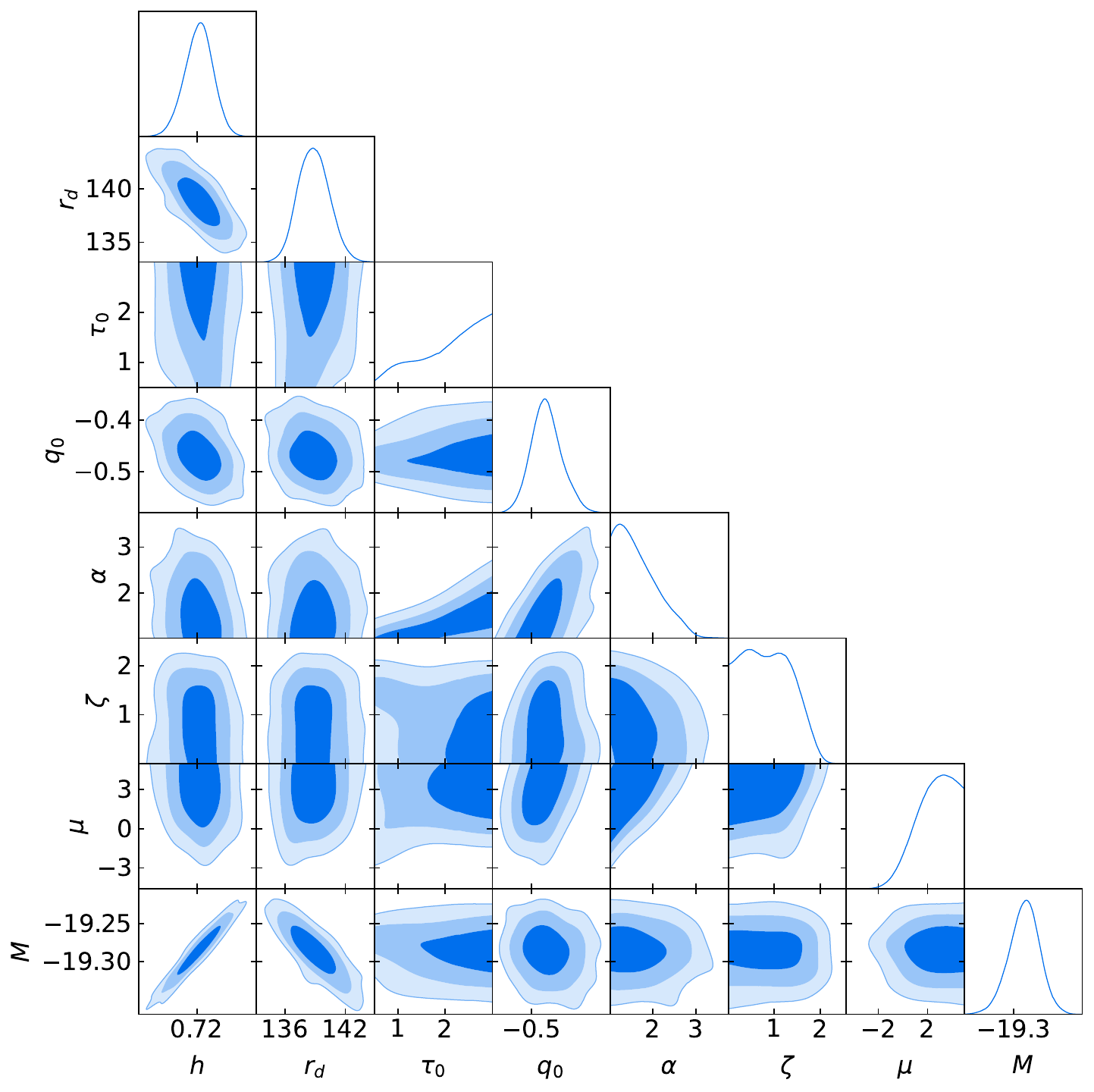}
    \caption{The posterior 1D distributions and joint marginalized regions for the free parameter space of the Fractional model with $\mu\neq 0$, obtained via the MCMC analysis described in Section \ref{Sect:2.7}. The admissible joint regions correspond to the $1\sigma$, $2\sigma$, and $3\sigma$ CL, respectively. The best-fit values for each free parameter are shown in Table \ref{tab:best-fits}.}
    \label{fig:TriangleFractional}
\end{figure}

\begin{figure}[H]
    \centering
    \includegraphics[width=\columnwidth]{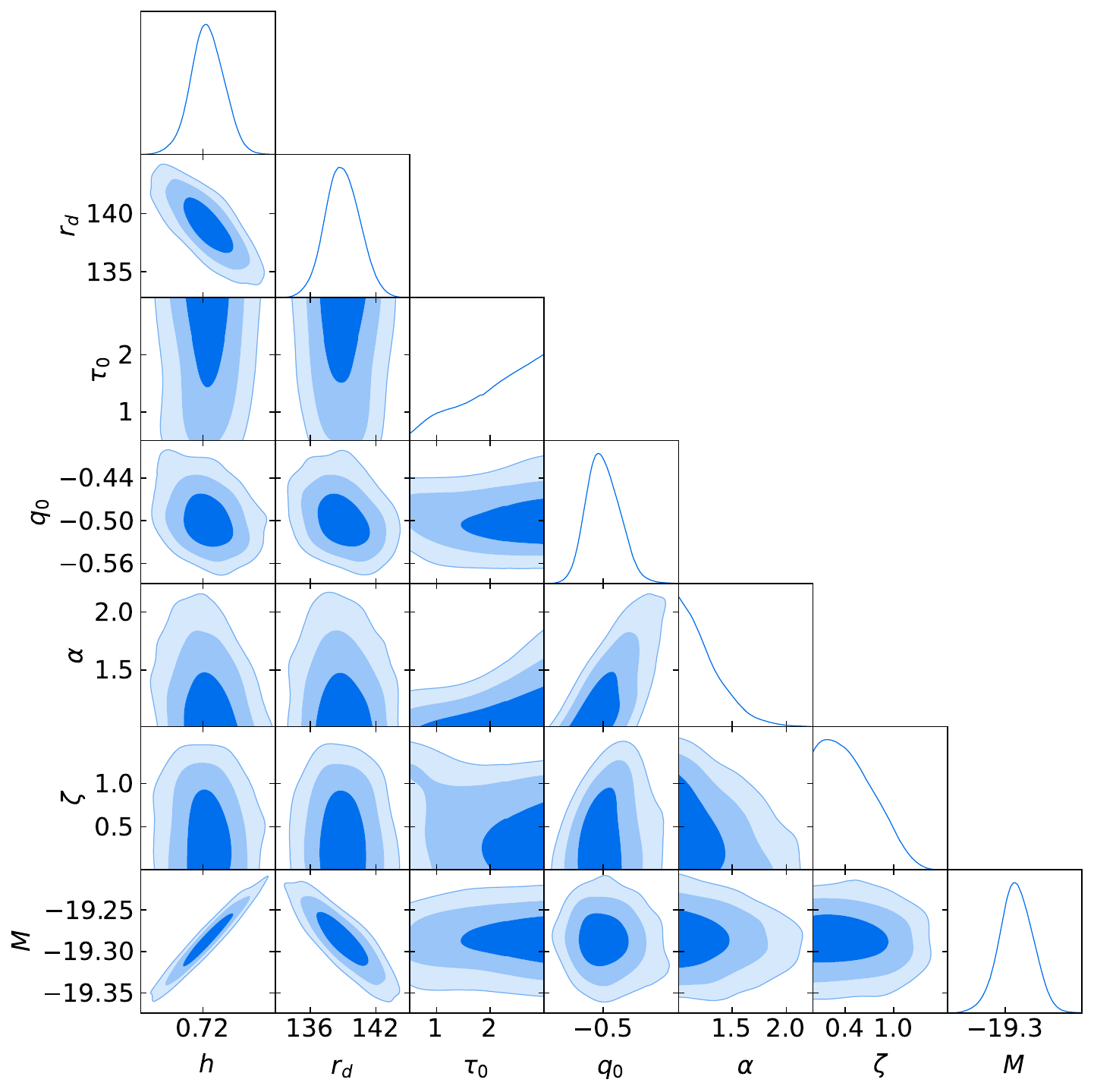}
    \caption{The posterior 1D distributions and joint marginalized regions for the free parameter space of the Fractional model with $\mu=0$, obtained via the MCMC analysis described in Section \ref{Sect:2.7}. The admissible joint regions correspond to the $1\sigma$, $2\sigma$, and $3\sigma$ CL, respectively. The best-fit values for each free parameter are shown in Table \ref{tab:best-fits}.}
    \label{fig:TriangleFractionalmu}
\end{figure}

In Table \ref{tab:best-fits}, we present the best-fit values and selected inferred parameters, both at the $1\sigma$ CL, alongside the goodness-of-fit criteria for the $\Lambda$CDM and Fractional models with $\mu\neq 0$ and $\mu=0$, respectively. In Figs. \ref{fig:TriangleLCDM}, \ref{fig:TriangleFractional}, and \ref{fig:TriangleFractionalmu}, we depict the 1D posterior distributions and joint marginalized regions at the $1\sigma$, $2\sigma$ ($95.5\%$), and $3\sigma$ ($99.7\%$) CL for the free parameter spaces of the $\Lambda$CDM and Fractional models with $\mu\neq 0$ and $\mu=0$, respectively. These results were obtained using the MCMC analysis described in Section \ref{Sect:2.7}.

\begin{table}[H]
    \centering
    \renewcommand{\arraystretch}{1.25}
    \begin{tabularx}{\columnwidth}{YYYY}
        \hline\hline
        \multirow{2}{*}{Parameter} & \multirow{2}{*}{$\Lambda$CDM} & Fractional model & Fractional model \\
         & & ($\mu\neq 0$) & ($\mu=0$) \\
         \hline
         $h$  & $0.7253_{-0.0071}^{+0.0074}$ & $0.7211_{-0.0080}^{+0.0072}$ & $0.7219_{-0.0072}^{+0.0076}$ \\
         $\Omega_{m,0}$ & $0.3026_{-0.0078}^{+0.0081}$ & $\cdots$ & $\cdots$ \\
         $r_{d}\,(Mpc)$ & $139.4\pm 1.5$ & $138.8_{-1.5}^{+1.6}$ & $138.9_{-1.5}^{+1.7}$ \\
         $\tau_{0}$ & $\cdots$ & $2.95_{-1.17}^{+0.78}$ & $2.94_{-1.18}^{+0.77}$ \\
         $q_{0}$ & $\cdots$ & $-0.469_{-0.030}^{+0.032}$ & $-0.502_{-0.023}^{+0.027}$ \\
         $\alpha$ & $\cdots$ & $1.57_{-0.38}^{+0.60}$ & $1.20_{-0.14}^{+0.25}$ \\
         $\zeta$ & $\cdots$ & $0.85\pm 0.57$ & $0.43_{-0.29}^{+0.39}$ \\
         $\mu$ & $\cdots$ & $4.0_{-2.5}^{+2.8}$ & $\cdots$ \\
         $M$ & $-19.284_{-0.021}^{+0.022}$ & $-19.284_{-0.022}^{+0.021}$ & $-19.286_{-0.021}^{+0.022}$ \\
         \hline
         $t_{0}\;(Gyr)$ & $\cdots$ & $39.9_{-15.7}^{+10.8}$ & $39.8_{-16.0}^{+10.5}$ \\
         $m\;\mathrm{(km\,s^{-1}\,Mpc^{-1})}$ & $\cdots$ & $61.8_{-40.9}^{+41.2}$ & $30.8_{-20.9}^{+28.0}$ \\
         $\Gamma\;\mathrm{(km\,s^{-1}\,Mpc^{-1})}$ & $\cdots$ & $\cdots$ & $108.3\pm 1.1$ \\
         $\tau_{\rm rel}\;(Gyr)$ & $\cdots$ & $\cdots$ & $9.037_{-0.094}^{+0.091}$ \\
         $\Delta\;(\mathrm{(km\,s^{-1}\,Mpc^{-1})^{2}})$ & $\cdots$ & $\cdots$ & $43090_{-10146}^{+3424}$ \\
         \hline
         $\chi_{\text{min}}^{2}$ & $1693.3$ & $1699.3$ & $1692.8$ \\
         BIC & $1723.2$ & $1759.05$ & $1745.13$ \\
         \hline\hline
    \end{tabularx}
    \caption{The best-fit values and selected inferred parameters, both at the $1\sigma$ CL, alongside the goodness-of-fit criteria for the $\Lambda$CDM and Fractional models with $\mu\neq 0$ and $\mu=0$, respectively. These values were obtained through the MCMC analysis described in Section \ref{Sect:2.7}.}
    \label{tab:best-fits}
\end{table}

From Table \ref{tab:best-fits}, and focusing on the goodness-of-fit criteria, it can be seen that the Fractional model with $\mu=0$ exhibits a lower $\chi_{\text{min}}^{2}$ value than the $\Lambda$CDM model, while the Fractional model with $\mu\neq 0$ exhibits a much larger value in comparison to the same model. This indicates that the $\mu=0$ case is more competitive in describing the cosmological data from SNe Ia, CC, BAO, GL, and BHS. Nevertheless, $\Lambda$CDM is the most favored by the data, since the difference in BIC is greater than $10$, providing very strong evidence against the Fractional model. This latter result is expected because the Fractional model has four more parameters than the $\Lambda$CDM model (three in the case of $\mu=0$). Between the Fractional models, there is strong evidence against the Fractional model with $\mu\neq 0$. While this can be explained by the fact that the latter has one extra free parameter in comparison to the former, a possible explanation that makes the Fractional model with $\mu=0$ more competitive is that it is not possible to obtain a well-defined best-fit parameter for $\mu$ in the general case, as can be seen from Fig. \ref{fig:TriangleFractional}. In fact, from that figure, we can see that we have only obtained best-fit parameters for $h$, $r_{d}$, and $q_{0}$; upper bounds for $\alpha$ and $\zeta$; and a lower bound for $\mu$. For $\tau_{0}$, it is not possible to obtain a best fit at $2\sigma$ and $3\sigma$, but rather a lower bound at $1\sigma$. These results are not a consequence of the priors used but rather arise from the strong degeneracy between the model parameters, especially between $\alpha$ and $\tau$, as can be seen from Eqs. \eqref{defofalpha} and \eqref{con-eq:system-x-y-E}. Similar constraints are obtained for the Fractional model with $\mu=0$, as shown in Fig. \ref{fig:TriangleFractionalmu}.

Now, focusing on some of the best-fit parameters, we see no appreciable differences in the values of the reduced Hubble parameter $h$ among the models. The same occurs for the scale of the pre-recombination sound horizon $r_{d}$, $\tau_{0}$, and $M$. The largest difference concerns the parameters $\alpha$ and $\zeta$ of the Fractional models. Interestingly, we see that the Fractional model with $\mu=0$ has a lower deceleration parameter than the general case, implying that it represents a universe that expands more rapidly than the one described by the Fractional model with $\mu\neq 0$. Finally, the age of the universe is inferred through the expression $\tau_{0}=H_{0}t_{0}$, yielding a similar value for both Fractional cases of $t_{0}\approx 40\,\text{Gyr}$. While this value is greater than the age of the universe in the $\Lambda$CDM model, since it is not possible to obtain a best fit at $2\sigma$ and $3\sigma$, and we only have a lower bound at $1\sigma$ due to the strong degeneracy of this parameter with $\alpha$, we cannot reach a definitive conclusion about this value. Note that, considering the lower bound of the uncertainty presented, the minimum value for the age of the universe in the Fractional cases is $t_{0}\approx 24\;\text{Gyr}$.

\begin{figure}[H]
    \centering
    \includegraphics[width=\columnwidth]{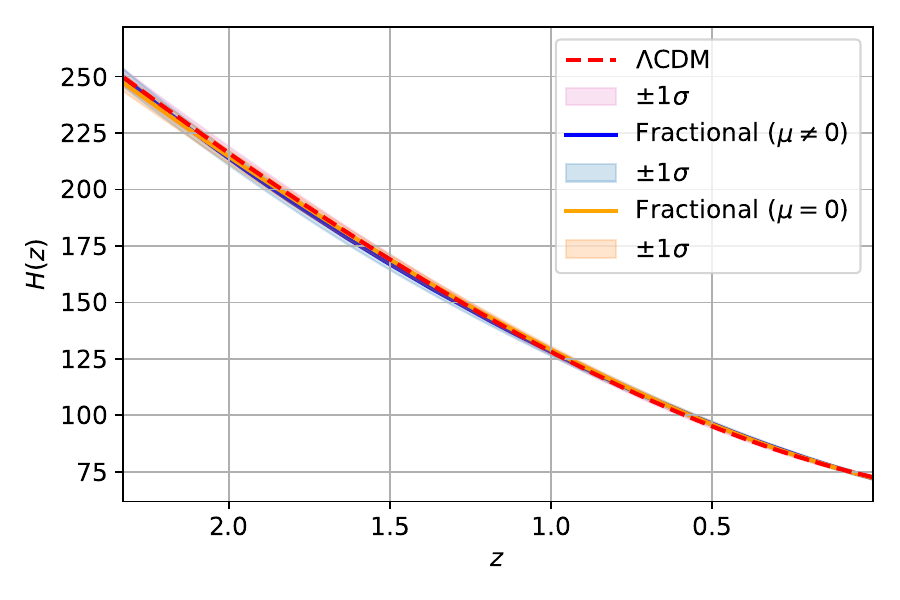}
    \caption{The Hubble parameter ($H$) as a function of redshift ($z$) for the $\Lambda$CDM (red dashed line) and Fractional models with $\mu\neq 0$ (blue solid line) and $\mu=0$ (orange solid line). The shaded regions represent the confidence intervals of the Hubble parameter at the $1\sigma$ CL. The figure was generated using the chains from the MCMC analysis described in Section \ref{Sect:2.7}.}
    \label{fig:Hubble}
\end{figure}

\begin{figure}[H]
    \centering
    \includegraphics[width=\columnwidth]{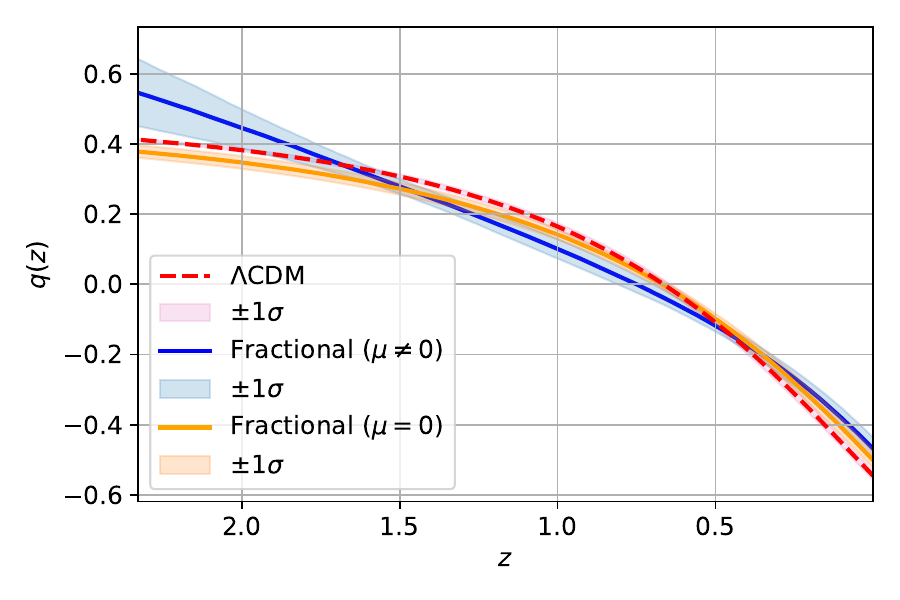}
    \caption{The deceleration parameter ($q$) as a function of redshift ($z$) for the $\Lambda$CDM (red dashed line) and Fractional models with $\mu\neq 0$ (blue solid line) and $\mu=0$ (orange solid line). The shaded regions represent the confidence intervals of the deceleration parameter at the $1\sigma$ CL. The figure was generated using the chains from the MCMC analysis described in Section \ref{Sect:2.7}.}
    \label{fig:deceleration}
\end{figure}

\begin{figure}[H]
    \centering
    \includegraphics[width=\columnwidth]{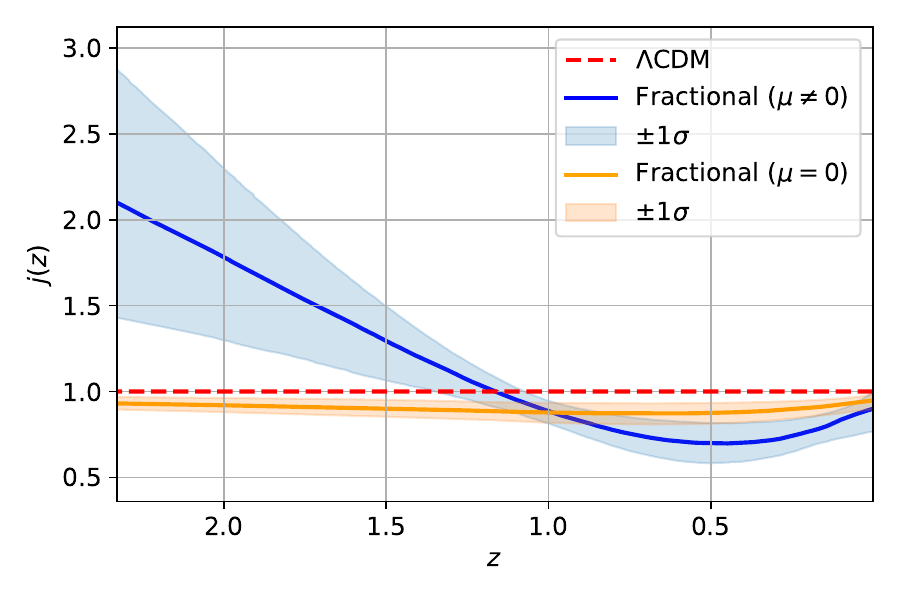}
    \caption{The jerk parameter ($j$) as a function of redshift ($z$) for the $\Lambda$CDM (red dashed line) and Fractional models with $\mu\neq 0$ (blue solid line) and $\mu=0$ (orange solid line). The shaded regions represent the confidence intervals of the jerk parameter at the $1\sigma$ CL. Note that, for the $\Lambda$CDM model, the jerk parameter is strictly constant with a value of $j=1$. The figure was generated using the chains from the MCMC analysis described in Section \ref{Sect:2.7}.}
    \label{fig:jerk}
\end{figure}

\begin{figure}[H]
    \centering
    \includegraphics[width=\columnwidth]{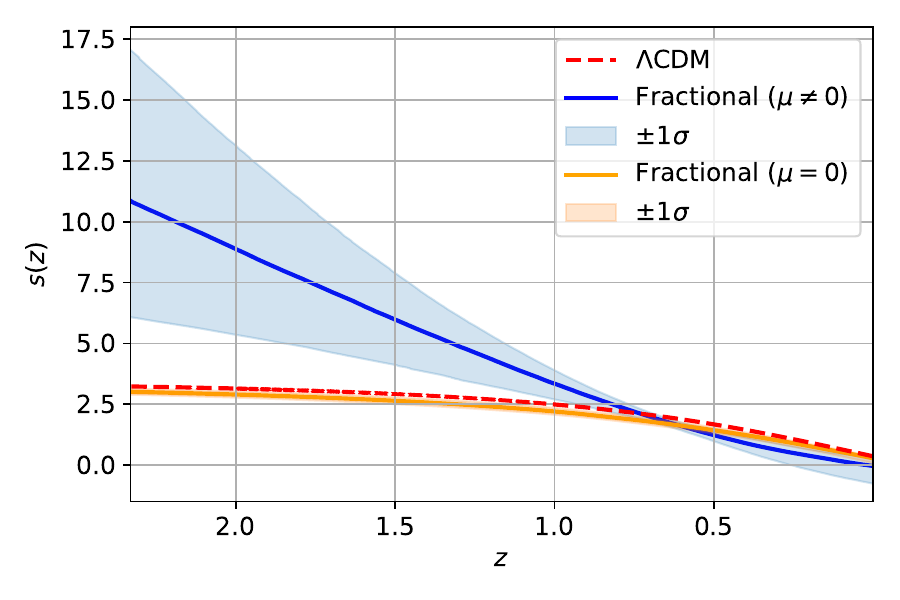}
    \caption{The snap parameter ($s$) as a function of redshift ($z$) for the $\Lambda$CDM (red dashed line) and Fractional models with $\mu\neq 0$ (blue solid line) and $\mu=0$ (orange solid line). The shaded regions represent the confidence intervals of the snap parameter at the $1\sigma$ CL. The figure was generated using the chains from the MCMC analysis described in Section \ref{Sect:2.7}.}
    \label{fig:snap}
\end{figure}

On the other hand, in Figs. \ref{fig:Hubble}, \ref{fig:deceleration}, \ref{fig:jerk}, and \ref{fig:snap}, we depict the Hubble parameter ($H$) and the cosmographic parameters: deceleration ($q$), jerk ($j$), and snap ($s$), respectively, as functions of redshift ($z$) for the $\Lambda$CDM (red dashed line) and Fractional models with $\mu\neq 0$ (blue solid line) and $\mu=0$ (orange solid line). The shaded regions represent the confidence intervals at the $1\sigma$ CL. These figures were generated using the chains from the MCMC analysis described in Section \ref{Sect:2.7}.

From Fig. \ref{fig:Hubble}, it can be seen that the Fractional model with $\mu=0$ exhibits a Hubble parameter behavior more similar to the $\Lambda$CDM model, while the general case ($\mu\neq 0$) shows greater deviations in comparison to the standard scenario. These departures of the Fractional model from the $\Lambda$CDM case are more visible in Fig. \ref{fig:deceleration}. As can be seen, while the deceleration parameter for the Fractional case with $\mu=0$ again evolves similarly to that in the $\Lambda$CDM model, the former represents a less accelerated universe than the $\Lambda$CDM model for $z\lessapprox 0.5$ and a more accelerated universe for $z\gtrapprox 0.5$. On the other hand, the Fractional model with $\mu\neq 0$ represents a less accelerated universe between $z\gtrapprox 1.5$ and $z\lessapprox 0.5$, while representing a more accelerated universe in the range $1.5\gtrapprox z\gtrapprox 0.5$, in comparison to the $\Lambda$CDM model. From the jerk parameter presented in Fig. \ref{fig:jerk}, we can conclude that, while the behavior of the Fractional model with $\mu=0$ is similar to the $\Lambda$CDM model, it is clearly not identical since it exhibits a departure from the value $j=1$, which is the constant value associated with the $\Lambda$CDM model. Here, the difference between the Fractional model in the general case ($\mu\neq 0$) and the $\Lambda$CDM model is even more evident. This departure of the Fractional model from $j=1$ indicates that we are in the presence of a form of dark energy other than the cosmological constant. All these conclusions are, again, reinforced by the snap parameter presented in Fig. \ref{fig:snap}. It is important to highlight that the confidence regions for the Fractional model with $\mu\neq 0$ are much larger than those for the Fractional model with $\mu =0$ and the $\Lambda$CDM model, and they grow wider for each successive cosmographic parameter. The reason is that the Fractional model with $\mu\neq 0$ imposes tighter constraints than the other models and carries greater uncertainty. Since each cosmographic parameter corresponds to a derivative of the previous one, this propagates a cumulative numerical error in the computation of each parameter.

\section{Effective dynamics for $\mu=0$}
\label{effective-dyn}
Setting $\mu=0$ implies $\varrho=f=0$, and the potential reduces to the simple quadratic form
\begin{equation}
V(\varphi)=V_0+\tfrac12 m^2\varphi^2.
\label{eq:V}
\end{equation}

Introducing
\begin{equation}
\varepsilon=\frac{\alpha-1}{3},
\label{eq:eps}
\end{equation}
the Hubble ansatz becomes
\begin{equation}
H(t)=H_0+\xi\varphi(t)+\frac{\alpha-1}{3t}.
\label{eq:H-ansatz}
\end{equation}
Substituting \eqref{eq:H-ansatz} into the scalar-field equation yields
\begin{equation}
\ddot{\varphi}
+\left(3\xi\varphi+3H_0\right)\dot{\varphi}
+m^2\varphi=0.
\label{eq:phi-mu0}
\end{equation}

When $|\xi\varphi|\ll H_0$, the friction term is approximately constant and
\eqref{eq:phi-mu0} reduces to the linear damped oscillator
\begin{equation}
\ddot{\varphi}+2\Gamma\dot{\varphi}+m^2\varphi=0,
\qquad \Gamma=\tfrac32 H_0.
\label{eq:damped}
\end{equation}
The characteristic equation is
\begin{equation}
\lambda^{2}+2\Gamma\lambda+m^{2}=0,
\end{equation}
and we define the discriminant as
\begin{equation}
\Delta := 4(\Gamma^{2}-m^{2}),
\end{equation}
so that the eigenvalues take the form
\begin{equation}
\lambda_{\pm}
= -\Gamma \pm \tfrac12\sqrt{\Delta}.
\end{equation}

The sign of $\Delta$ determines the dynamical regime:
\begin{itemize}
\item \textbf{Overdamped case} ($\Delta>0$):  
\begin{equation}
\lambda_{\pm}
= -\Gamma \pm \sqrt{\Gamma^{2}-m^{2}},
\end{equation}
two distinct real negative eigenvalues. The solution is a linear combination of two decaying exponentials,
\begin{equation}
\varphi(t)=C_{1}e^{\lambda_{+}t}+C_{2}e^{\lambda_{-}t}.
\end{equation}

\item \textbf{Critically damped case} ($\Delta=0$):  
\begin{equation}
\lambda_{+}=\lambda_{-}=-\Gamma,
\end{equation}
a repeated real eigenvalue, giving
\begin{equation}
\varphi(t)=(C_{1}+C_{2}t)e^{-\Gamma t}.
\end{equation}

\item \textbf{Underdamped case} ($\Delta<0$):  
\begin{equation}
\lambda_{\pm}
= -\Gamma \pm i\sqrt{m^{2}-\Gamma^{2}},
\end{equation}
complex conjugate eigenvalues with negative real part. The solution exhibits exponentially damped oscillations,
\begin{equation}
\varphi(t)\sim e^{-\Gamma t}
\sin\!\left(\sqrt{m^{2}-\Gamma^{2}}\,t+\delta\right).
\end{equation}
\end{itemize}
For  $m\gg\Gamma$, the oscillatory approximation becomes
\begin{equation}
\varphi(t)\simeq A\,e^{-\Gamma t}\sin(mt+\delta).
\label{eq:phi-sol}
\end{equation}
Substituting \eqref{eq:phi-sol} into \eqref{eq:H-ansatz} yields
\begin{equation}
H(t)\simeq H_0+\xi A e^{-\tfrac32 H_0 t}\sin(mt+\delta)
+\frac{\alpha-1}{3t}.
\label{eq:H-sol}
\end{equation}

With $f=0$, the effective gravitational constant evolves as
\begin{equation}
G(t)\approx 
G_N\!\left(
\frac{8\pi\rho_0}{\Lambda_0}\,
a^{-3(1+\gamma)}\,
t^{-(\alpha-1)(1+\gamma)}
\right)^{\frac{1}{\beta-1}}.
\label{eq:G}
\end{equation}
Integrating $H=\dot a/a$ gives
\begin{equation}
a(t)=e^{H_0 t}(\zeta H_0 t)^{\alpha-1}.
\label{eq:a}
\end{equation}

Finally, since $\Gamma>0$ and $m^{2}>0$, the real parts of the eigenvalues always satisfy
\begin{equation}
\Re(\lambda_{\pm})<0,
\end{equation}
so the equilibrium point $\varphi=\dot{\varphi}=0$ is a stable attractor.

\subsection{Numerical regime from the best-fit parameters}

Using the best-fit values in Table~\ref{tab:best-fits},
\begin{equation}
m \simeq 30.8^{+28.0}_{-20.9}\;\mathrm{km\,s^{-1}\,Mpc^{-1}},
\qquad
\Gamma \simeq 108.3 \pm 1.1\;\mathrm{km\,s^{-1}\,Mpc^{-1}},
\end{equation}
the discriminant of the damped oscillator is obtained directly from the fit as
\begin{equation}
\Delta = (4.309^{+0.342}_{-1.015})\times 10^{4}\;
\mathrm{(km\,s^{-1}\,Mpc^{-1})^{2}}.
\end{equation}
Since $\Delta>0$, the system lies in the \emph{overdamped} regime.

The corresponding eigenvalues,
\begin{equation}
\lambda_{\pm}
= -\Gamma \pm \tfrac12\sqrt{\Delta},
\end{equation}
evaluate, for the central values, to
\begin{equation}
\lambda_{+}\approx -4.4,
\qquad
\lambda_{-}\approx -212.2,
\end{equation}
both real and negative, with a clear separation of scales. The slow mode $\lambda_{+}$ controls the late-time behaviour and defines the effective damping timescale,
\begin{equation}
\tau_{d}=\frac{1}{|\lambda_{+}|}
\approx \frac{1}{4.4}\;\mathrm{Mpc\,s\,km^{-1}}
\approx 7.01\times 10^{18}\;\mathrm{s}
\approx 9.0\;\mathrm{Gyr}
\simeq \tau_{\rm rel}
=9.037^{+0.091}_{-0.094}\;\mathrm{Gyr},
\end{equation}
in excellent agreement with the fitted value reported in Table~\ref{tab:best-fits}. This timescale is of the order of the Hubble time, indicating that the scalar-field amplitude decays monotonically but only very gradually on cosmological timescales, while the background expansion remains governed by $H_{0}$.

\subsection{Rescaled first-order formulation}

Writing \eqref{eq:phi-mu0} as a first-order system,
\begin{equation}
\dot{\varphi}=v,\qquad
\dot{v}=-(3\xi\varphi+3H_0)v-m^2\varphi,
\label{sys_first_order_rescaled}
\end{equation}
and rescaling time via $\tau=H_0 t$ with $u=d\varphi/d\tau$, we obtain
\begin{equation}
\frac{d\varphi}{d\tau}=u,\qquad
\frac{du}{d\tau}
=-3\Big(1+\frac{\xi}{H_0}\varphi\Big)u
-\Big(\frac{m}{H_0}\Big)^{\!2}\varphi.
\label{sys_tau_rescaled}
\end{equation}

Introducing
\begin{equation}
\Upsilon=\frac{\xi}{H_0},\qquad
\Phi=\Upsilon\varphi,\qquad
W=\Upsilon u,
\end{equation}
the system becomes
\begin{equation}
\frac{d\Phi}{d\tau}=W,\qquad
\frac{dW}{d\tau}=-3(1+\Phi)W-\Lambda\Phi,
\label{sys_Phi_W}
\end{equation}
where the natural scale of the system is
\begin{equation}
\Lambda=\Upsilon\left(\frac{m}{H_0}\right)^2
=\Upsilon\zeta^2
=\frac{\xi}{H_0}\,\zeta^2.
\label{Lambda_def}
\end{equation}

The initial conditions become
\begin{equation}
\Phi(\tau_0)=\frac{1-\alpha}{3\tau_0},\qquad
W(\tau_0)=-1-q_0+\frac{\alpha-1}{3\tau_0^2},
\label{ICs_Phi_W}
\end{equation}
which are finite and independent of $\Upsilon$.

We consider the system \eqref{sys_Phi_W} and analyse the stability of the equilibrium point $(\Phi,W)=(0,0)$.

Linearizing \eqref{sys_Phi_W} around the origin gives
\begin{equation}
\frac{d\Phi}{d\tau}=W,\qquad
\frac{dW}{d\tau}=-3W-\Lambda\Phi.
\end{equation}

The Jacobian matrix at $(0,0)$ is
\begin{equation}
J(0,0)=
\begin{pmatrix}
0 & 1\\[4pt]
-\Lambda & -3
\end{pmatrix}.
\label{Jacobian}
\end{equation}

The characteristic equation is
\begin{equation}
\lambda^2+3\lambda+\Lambda=0,
\label{char_eq}
\end{equation}
with eigenvalues
\begin{equation}
\lambda_{1,2}
=\frac{-3\pm\sqrt{9-4\Lambda}}{2}.
\label{eigenvalues}
\end{equation}

\textbf{Stability classification.}
\begin{itemize}
    \item \textbf{If $\Lambda>0$}: both eigenvalues have negative real part.
    \begin{itemize}
        \item If $0<\Lambda<\tfrac{9}{4}$, then $9-4\Lambda>0$ and the eigenvalues are real and negative:  
        \textbf{stable node}.
        \item If $\Lambda>\tfrac{9}{4}$, then $9-4\Lambda<0$ and the eigenvalues are complex with real part $-3/2<0$:  
        \textbf{stable focus} (damped oscillations).
    \end{itemize}

    \item \textbf{If $\Lambda=0$}: the eigenvalues are $0$ and $-3$.  
    The equilibrium is \textbf{non-hyperbolic}, and nonlinear terms determine the behaviour.

    \item \textbf{If $\Lambda<0$}: the product of eigenvalues is negative, so one eigenvalue is positive and the other negative:  
    \textbf{saddle point} (unstable).
\end{itemize}
Summarizing, for the cosmologically relevant case $\Lambda>0$, the origin is always \emph{asymptotically stable}:  
a \textbf{stable node} if $0<\Lambda<9/4$ and a \textbf{stable focus} if $\Lambda>9/4$.

From the best-fit values in Table~\ref{tab:best-fits},
\begin{equation}
m \simeq 30.8\;\mathrm{km\,s^{-1}\,Mpc^{-1}},
\qquad
\Gamma \simeq 108.3\;\mathrm{km\,s^{-1}\,Mpc^{-1}},
\end{equation}
we obtain
\begin{equation}
H_{0}=\frac{2}{3}\Gamma \simeq 72.2\;\mathrm{km\,s^{-1}\,Mpc^{-1}},
\qquad
\frac{m}{H_{0}} \simeq 0.43,
\qquad
\left(\frac{m}{H_{0}}\right)^{2}\simeq 0.18.
\end{equation}
In the rescaled system \eqref{sys_Phi_W}, the parameter
\begin{equation}
\Lambda=\Upsilon\left(\frac{m}{H_{0}}\right)^{2},
\qquad
\Upsilon=\frac{\xi}{H_{0}}<1,
\end{equation}
is therefore bounded by
\begin{equation}
\Lambda \lesssim 0.18,
\end{equation}
with the maximum attained when $\Upsilon=1$. Hence $\Lambda$ is small for any $\Upsilon$ of order unity.

The dynamics near the equilibrium point $(\Phi,W)=(0,0)$ are governed by the linearised system with eigenvalues \begin{equation}
\lambda_{1,2}
=\frac{-3\pm\sqrt{9-4\Lambda}}{2}.
\end{equation}

For $\Lambda\ll 1$, we expand the square root as
\begin{equation}
\sqrt{9-4\Lambda}
=3\sqrt{1-\tfrac{4\Lambda}{9}}
\simeq 3\left(1-\tfrac{2\Lambda}{9}\right)
=3-\tfrac{2\Lambda}{3},
\end{equation}
using the first-order approximation $\sqrt{1-x}\simeq 1-\tfrac{x}{2}$ for $|x|\ll 1$. Substituting this into the expression for $\lambda_{1,2}$ gives
\begin{equation}
\lambda_{1}
=\frac{-3+\left(3-\tfrac{2\Lambda}{3}\right)}{2}
=-\frac{\Lambda}{3},
\qquad
\lambda_{2}
=\frac{-3-\left(3-\tfrac{2\Lambda}{3}\right)}{2}
=-3+\frac{\Lambda}{3}.
\end{equation}

Since $\Lambda\ll 1$, the correction $\Lambda/3$ in $\lambda_{2}$ is negligible, and we obtain the approximations
\begin{equation}
\lambda_{1}\approx -\frac{\Lambda}{3},
\qquad
\lambda_{2}\approx -3,
\end{equation}
showing a clear separation between a slow and a fast decay mode. The origin is therefore a \emph{stable node} in the rescaled phase space: the fast mode ($\lambda_{2}\approx -3$) drives a rapid decay of $W$, while the slow mode ($\lambda_{1}\approx -\Lambda/3$) reflects the long damping timescale $\tau_{d}$ obtained in the physical variables.

These results confirm that the numerical values $(m,\Gamma,\Delta,\tau_{d})$ are fully consistent with the qualitative structure of the rescaled system: the scalar field decays monotonically in the overdamped regime, while the variables $(\Phi,W)$ approach the equilibrium along two distinct decay directions.

\section{Qualitative and numerical analysis}
\label{Sect:4.1}

In this section, we analyze the global dynamics of the fractional model using the autonomous system that governs its evolution.  
To do so, we introduce dimensionless variables that regularize the cosmological equations and reveal a \emph{slow--fast} structure controlled by the variable $u$, which separates the dynamics into slow and fast modes.  
With this structure, we identify the system's critical points and study their stability.  
We then examine the scalar subsystem in the $(v_1,v_2)$ phase plane, thereby describing the slow dynamics that organize the global evolution.  
Together, these analyses provide a clear picture of the different cosmological regimes that emerge in the model.

\subsection{Phase–space analysis}
\label{Sect:4.1.1}

To study the model's dynamics, it is useful to work with dimensionless variables.  
We define
\begin{equation}
(u, v_1, v_2, v_3, v_4)
=\left(\frac{H_0}{H},\, \Upsilon \varphi,\, \Upsilon\frac{\dot{\varphi}}{H_0},\, G^\beta,\, \frac{\dot{G}}{H_0 G}\right),
\label{vars-dimensionless}
\end{equation}
so that each quantity is expressed in natural units of the system.  
We also introduce the parameter
\begin{equation}
\Omega_{0}= \frac{\Lambda_{0}}{3 H_0^2},
\end{equation}
and use the number of e–foldings
\begin{equation}
N := \ln a = -\ln(1+z), \qquad 
\frac{d}{dN} = -(1+z)\frac{d}{dz}=u\,\frac{d}{d\tau},
\end{equation}
which allows us to rewrite the time evolution in a simpler form.

With these definitions, the dynamical system takes the autonomous form
\begin{subequations}
\label{Aut}
\begin{align}
\frac{du}{dN} &=-u\!\left[v_2 u^2 -\frac{1}{\delta_0\tau_0^2}\bigl(1-(1+v_1)u\bigr)^2\right],\label{Aut:u}\\
\frac{dv_1}{dN} &= u\, v_2,\label{Aut:v1}\\
\frac{dv_2}{dN} &= u\!\left[-3(v_1+1)v_2 - \zeta^2 v_1 + \mu v_1^3\right],\label{Aut:v2}\\
\frac{dv_3}{dN} &= \beta u\, v_3 v_4,\label{Aut:v3}\\
\frac{dv_4}{dN} &= \frac{3}{\mu u}
+\frac{1}{2}u v_4(-6v_1+v_4-6)
-\frac{3(\beta-1)u v_3\Omega_0}{\mu}.\label{Aut:v4}
\end{align}
\end{subequations}

Using the ansatz \eqref{anzat}, we obtain the relation
\begin{equation}
\frac{H}{H_0}=1+\Upsilon \varphi+\frac{(\alpha - 1)}{3\tau} 
\implies
\frac{1}{u} = 1 + v_1 + \frac{\alpha - 1}{3\tau},
\label{eq:manifold-hyp}
\end{equation}
which links the expansion of the universe with the scalar field and the fractional parameter.

\noindent
Moreover, if $u\geq 0$ and we use the time variable $\tau = H_0 t$, the system can be studied through the equivalent equations
\begin{subequations}
\label{eq:slow-system}
\begin{align}
\frac{dv_1}{d\tau} &= v_2,\\
\frac{dv_2}{d\tau} &= \mu v_1^3 - \zeta^2 v_1 - 3(v_1+1)v_2,\\
\frac{dv_3}{d\tau} &= \beta v_3 v_4,\\
\frac{dv_4}{d\tau} &= \frac{3 (v_1+1)^2 - 3(\beta -1)\Omega_0 v_3}{\mu}
+\frac{1}{2}v_4(v_4 - 6v_1 - 6).
\end{align}
\end{subequations}

\subsection{Slow–fast structure}
\label{Sect:4.1.2}

We group the slow variables as
\begin{equation}
X := (u,v_1,v_2,v_3)^{\mathsf T},
\end{equation}
leaving $v_4$ as the fast variable.  
With this notation, the system \eqref{Aut} can be written as
\begin{equation}
\label{eq:block-form-fastslow}
\begin{aligned}
\frac{dX}{dN} &= u\,F(X,v_4),\\
\frac{dv_4}{dN} &= \frac{3}{\mu u} + R(X,v_4,u),
\end{aligned}
\end{equation}
which makes the scale separation explicit.  
Near $u=0$ we obtain
\begin{equation}
\frac{dX}{dN} = \mathcal{O}(u), \qquad
\frac{dv_4}{dN} = \frac{3}{\mu u} + \mathcal{O}(1),
\end{equation}
so $v_4$ evolves rapidly while $X$ changes slowly.  
Thus, the system has a clear \emph{slow–fast} structure: $u$, $v_1$, $v_2$, and $v_3$ are slow variables, and $v_4$ is the fast one.

\subsection{Critical points of the full system}
\label{Sect:4.1.3}

To understand the system's global structure, we first identify its critical points.  
These are obtained by imposing
\begin{equation}
\frac{dX}{dN}=0, 
\qquad 
\frac{dv_4}{dN}=0.
\end{equation}
From the autonomous system \eqref{Aut}, these conditions give
\begin{equation}
v_2^\ast=0,\qquad
-\zeta^2 v_1^\ast + \mu (v_1^\ast)^3=0,\qquad
\beta v_3^\ast v_4^\ast=0,
\end{equation}
together with the scalar equation that sets $\dot v_4=0$.  
Solving this algebraic system produces two families of critical points: (i) points with $u^\ast=0$, dominated by the fast dynamics, and (ii) points with $u^\ast>0$, where expansion and the scalar field balance each other.

For clarity, all equilibrium points of the reduced system are summarized in Table~\ref{tab:mathematica_data}, where we list the existence conditions and the corresponding values of $(v_1^\ast,v_2^\ast,v_3^\ast,v_4^\ast)$.  
Structurally, points A–C belong to the branch with $v_4^\ast=0$, where the gravitational dynamics freeze, and the equilibrium value of $G$ is set only by $v_3^\ast$.  
These points exist whenever $\beta>1$ and represent static configurations of the coupling.

In contrast, points D–I belong to the branch with $v_3^\ast=0$, where the gravitational evolution is controlled by $v_4^\ast$.  
Their existence requires $\mu>2/3$, and the sign of $v_4^\ast$ distinguishes between growing and decaying coupling regimes.  
The scalar shift $\pm\zeta/\sqrt{\mu}$ in points F–I modifies the amplitude of $v_4^\ast$, directly affecting the strength of the fast dynamics.

Overall, the table shows that the phase–space structure is strongly shaped by the parameters $(\mu,\zeta,\beta,\Omega_0)$, and that each critical point corresponds to a different cosmological regime determined by the interaction between the scalar field and the varying gravitational coupling.

\begin{table}[htbp]
\centering
\begin{tabular}{|c|c|c|c|c|c|}
\hline
Label & Conditions & $v_1^\ast$ & $v_2^\ast$ & $v_3^\ast$ & $v_4^\ast$ \\
\hline
A & $\beta>1$ & $0$ & $0$ & $\dfrac{1}{(\beta-1)\Omega_0}$ & $0$ \\ \hline
B & $\beta>1$ & $\dfrac{\zeta}{\sqrt{\mu}}$ & $0$ & $\dfrac{(1+\frac{\zeta}{\sqrt{\mu}})^2}{(\beta-1)\Omega_0}$ & $0$ \\ \hline
C & $\beta>1$ & $-\dfrac{\zeta}{\sqrt{\mu}}$ & $0$ & $\dfrac{(1-\frac{\zeta}{\sqrt{\mu}})^2}{(\beta-1)\Omega_0}$ & $0$ \\ \hline
D & $\mu>\tfrac{2}{3}$ & $0$ & $0$ & $0$ & $3\left[1-\sqrt{1-\tfrac{2}{3\mu}}\right]$ \\ \hline
E & $\mu>\tfrac{2}{3}$ & $0$ & $0$ & $0$ & $3\left[1+\sqrt{1-\tfrac{2}{3\mu}}\right]$ \\ \hline
F & $\mu>\tfrac{2}{3}$ & $\dfrac{\zeta}{\sqrt{\mu}}$ & $0$ & $0$ & $3\!\left(1+\tfrac{\zeta}{\sqrt{\mu}}\right)\!\left[1-\sqrt{1-\tfrac{2}{3\mu}}\right]$ \\ \hline
G & $\mu>\tfrac{2}{3}$ & $\dfrac{\zeta}{\sqrt{\mu}}$ & $0$ & $0$ & $3\!\left(1+\tfrac{\zeta}{\sqrt{\mu}}\right)\!\left[1+\sqrt{1-\tfrac{2}{3\mu}}\right]$ \\ \hline
H & $\mu>\tfrac{2}{3}$ & $-\dfrac{\zeta}{\sqrt{\mu}}$ & $0$ & $0$ & $3\!\left(1-\tfrac{\zeta}{\sqrt{\mu}}\right)\!\left[1-\sqrt{1-\tfrac{2}{3\mu}}\right]$ \\ \hline
I & $\mu>\tfrac{2}{3}$ & $-\dfrac{\zeta}{\sqrt{\mu}}$ & $0$ & $0$ & $3\!\left(1-\tfrac{\zeta}{\sqrt{\mu}}\right)\!\left[1+\sqrt{1-\tfrac{2}{3\mu}}\right]$ \\ \hline
\end{tabular}
\caption{Equilibrium points of the reduced system \eqref{eq:slow-system}.}
\label{tab:mathematica_data}
\end{table}
 
\subsection{Stability analysis}
\label{Sect:4.1.4}

Once the critical points have been identified, we study their linear stability.  
The Jacobian of the system naturally splits into two blocks: a slow block associated with $(v_1,v_2)$ and a fast block associated with $(v_3,v_4)$.  
This decomposition directly reflects the \emph{slow–fast} structure of the system.

\textbf{Slow modes.}
The eigenvalues of the slow block are
\begin{equation}
\sigma_{1,2}
= \frac12\left[-3(v_1^\ast+1)\pm\sqrt{9-4\zeta^2+12\mu (v_1^\ast)^2}\right].
\end{equation}

\textbf{Fast modes.}
The eigenvalues of the fast block are
\begin{equation}
\sigma_{3,4}
= \frac12\left(
\operatorname{tr}J_{\text{fast}}
\pm
\sqrt{(\operatorname{tr}J_{\text{fast}})^2 - 4\det J_{\text{fast}}}
\right),
\end{equation}
where
\begin{equation}
\operatorname{tr}J_{\text{fast}}=\beta v_4^\ast - 3(v_1^\ast+1),\qquad
\det J_{\text{fast}}
=\beta v_4^\ast(-3(v_1^\ast+1)+v_4^\ast)
+\frac{3\beta(\beta-1)\Omega_0}{\mu}v_3^\ast.
\end{equation}

Table~\ref{tab:eigenvalues} lists the Jacobian eigenvalues at each critical point of the reduced system, allowing the linear nature of each equilibrium to be identified immediately.

\begin{table}[htbp]
\centering
\resizebox{\textwidth}{!}{
\begin{tabular}{|c|c|c|c|c|}
\hline
\textbf{Label} & $\sigma_1$ & $\sigma_2$ & $\sigma_3$ & $\sigma_4$ \\
\hline
A & $\frac12[-3-\sqrt{9-4\zeta^2}]$ & $\frac12[-3+\sqrt{9-4\zeta^2}]$ & $0$ & $-3$ \\ \hline
B & $\frac12[-3(1+\frac{\zeta}{\sqrt{\mu}})-\sqrt{9-4\zeta^2+12\zeta^2}]$ &
$\frac12[-3(1+\frac{\zeta}{\sqrt{\mu}})+\sqrt{9-4\zeta^2+12\zeta^2}]$ &
$0$ & $-3(1+\frac{\zeta}{\sqrt{\mu}})$ \\ \hline
C & $\frac12[-3(1-\frac{\zeta}{\sqrt{\mu}})-\sqrt{9-4\zeta^2+12\zeta^2}]$ &
$\frac12[-3(1-\frac{\zeta}{\sqrt{\mu}})+\sqrt{9-4\zeta^2+12\zeta^2}]$ &
$0$ & $-3(1-\frac{\zeta}{\sqrt{\mu}})$ \\ \hline
D & $\frac12[-3-\sqrt{9-4\zeta^2}]$ & $\frac12[-3+\sqrt{9-4\zeta^2}]$ &
$\beta v_4^\ast - 3$ & $v_4^\ast$ \\ \hline
E & $\frac12[-3-\sqrt{9-4\zeta^2}]$ & $\frac12[-3+\sqrt{9-4\zeta^2}]$ &
$\beta v_4^\ast - 3$ & $v_4^\ast$ \\ \hline
F & $\frac12[-3(1+\frac{\zeta}{\sqrt{\mu}})-\sqrt{9-4\zeta^2+12\zeta^2}]$ &
$\frac12[-3(1+\frac{\zeta}{\sqrt{\mu}})+\sqrt{9-4\zeta^2+12\zeta^2}]$ &
$\beta v_4^\ast - 3(1+\frac{\zeta}{\sqrt{\mu}})$ & $v_4^\ast$ \\ \hline
G & $\frac12[-3(1+\frac{\zeta}{\sqrt{\mu}})-\sqrt{9-4\zeta^2+12\zeta^2}]$ &
$\frac12[-3(1+\frac{\zeta}{\sqrt{\mu}})+\sqrt{9-4\zeta^2+12\zeta^2}]$ &
$\beta v_4^\ast - 3(1+\frac{\zeta}{\sqrt{\mu}})$ & $v_4^\ast$ \\ \hline
H & $\frac12[-3(1-\frac{\zeta}{\sqrt{\mu}})-\sqrt{9-4\zeta^2+12\zeta^2}]$ &
$\frac12[-3(1-\frac{\zeta}{\sqrt{\mu}})+\sqrt{9-4\zeta^2+12\zeta^2}]$ &
$\beta v_4^\ast - 3(1-\frac{\zeta}{\sqrt{\mu}})$ & $v_4^\ast$ \\ \hline
I & $\frac12[-3(1-\frac{\zeta}{\sqrt{\mu}})-\sqrt{9-4\zeta^2+12\zeta^2}]$ &
$\frac12[-3(1-\frac{\zeta}{\sqrt{\mu}})+\sqrt{9-4\zeta^2+12\zeta^2}]$ &
$\beta v_4^\ast - 3(1-\frac{\zeta}{\sqrt{\mu}})$ & $v_4^\ast$ \\ \hline
\end{tabular}}
\caption{\label{tab:eigenvalues} Jacobian eigenvalues evaluated at the critical points of the reduced system.}
\end{table}

The slow modes, $\sigma_1$ and $\sigma_2$, depend only on $v_1^\ast$ and reproduce the typical structure of the scalar subsystem: for $v_1^\ast=0$ (point A) the eigenvalues are independent of $\mu$, while for $v_1^\ast=\pm\zeta/\sqrt{\mu}$ (points B, C, F–I) the stability depends on $\zeta$ but not on $\mu$.

The fast modes, $\sigma_3$ and $\sigma_4$, clearly distinguish the two branches of the system.  
In Branch I (A–C), one fast eigenvalue is zero, and the other is strictly negative, reflecting a neutral mode associated with the static gravitational coupling ($v_4^\ast=0$).  
In Branch II (D–I), both eigenvalues depend explicitly on $v_4^\ast$, introducing exponential growth or decay depending on its sign.  
In particular, positive values of $v_4^\ast$ (points E, G, I) lead to rapid instability, while negative values (points D, H) can partially stabilize the dynamics.

Overall, the table shows that the stability of each critical point is controlled by the interaction between the scalar displacement $v_1^\ast$ and the gravitational dynamics encoded in $v_4^\ast$, confirming the \emph{slow–fast} structure of the system.

\subsection{Interpretation of the critical points}
\label{Sect:4.1.5}

Once the critical points and their linear stability are characterized, it is useful to interpret their physical meaning.  
Table~\ref{tab:G_interpretation} summarizes the behavior of the gravitational coupling $G(t)$ in each case, as well as the type of equilibrium.

\begin{table}[htbp]
\centering
\begin{tabular}{|c|c|c|c|p{3.5cm}|}
\hline
Label & $G(t)=G_0 e^{H_0 v_4^\ast (t-t_0)}$ & Limit of $G=v_3^{1/\beta}$ & Type & Interpretation \\ \hline
A & $G(t)=G_0$ & $(\beta-1)^{-1/\beta}\Omega_0^{-1/\beta}$ & Static & Constant gravitational coupling. \\ \hline
B & $G(t)=G_0$ & $\left[\frac{(1+\frac{\zeta}{\sqrt{\mu}})^2}{(\beta-1)\Omega_0}\right]^{1/\beta}$ & Static & Shifted coupling equilibrium. \\ \hline
C & $G(t)=G_0$ & $\left[\frac{(1-\frac{\zeta}{\sqrt{\mu}})^2}{(\beta-1)\Omega_0}\right]^{1/\beta}$ & Static & Symmetric counterpart of B. \\ \hline
D & $G(t)=G_0 e^{H_0 v_4^\ast (t-t_0)}$ & $0$ & Dynamic & Critical decay from $G=0$. \\ \hline
E & $G(t)=G_0 e^{H_0 v_4^\ast (t-t_0)}$ & $0$ & Dynamic & Exponential growth from $G=0$. \\ \hline
F & $G(t)=G_0 e^{H_0 v_4^\ast (t-t_0)}$ & $0$ & Dynamic & Non‑trivial evolution from $G=0$. \\ \hline
G & $G(t)=G_0 e^{H_0 v_4^\ast (t-t_0)}$ & $0$ & Dynamic & Enhanced growth of the coupling. \\ \hline
H & $G(t)=G_0 e^{H_0 v_4^\ast (t-t_0)}$ & $0$ & Dynamic & Decay of the coupling. \\ \hline
I & $G(t)=G_0 e^{H_0 v_4^\ast (t-t_0)}$ & $0$ & Dynamic & Partial recovery of the coupling. \\ \hline
\end{tabular}
\caption{Physical interpretation of the equilibrium points of the reduced system.}
\label{tab:G_interpretation}
\end{table}

The table shows that points A–C correspond to configurations in which the gravitational coupling is exactly stabilized ($v_4^\ast=0$).  
In these cases, $G(t)$ remains constant, and the equilibrium value depends only on $v_3^\ast$, reflecting a static regime in which gravitational dynamics does not contribute to the evolution of the system.  
Points B and C represent shifted scalar equilibria that produce different values of $G$ while remaining stationary.

In contrast, points D–I describe dynamic regimes in which $G(t)$ evolves exponentially because $v_4^\ast\neq 0$.  
The sign of $v_4^\ast$ determines whether the coupling grows or decays: points E and G, with $v_4^\ast>0$, correspond to exponential growth, while points D and H, with $v_4^\ast<0$, describe gravitational decay.  
Points F–I also include the scalar shift $\pm\zeta/\sqrt{\mu}$, which modulates the strength of the gravitational evolution and leads to richer behaviors, such as partial recovery of the coupling (point I).

Overall, Table~\ref{tab:G_interpretation} shows that the dynamics of the gravitational coupling are tightly linked to the slow–fast structure of the system: static points belong to the slow branch, while dynamic points reflect the dominant action of the fast mode associated with $v_4$.  
This distinction allows each critical point to be interpreted as a well‑defined cosmological regime in phase space.

In the next section, we analyze how these local behaviors interact with the evolution of the gravitational variables to produce the model's global regimes.
 
\subsection{Global dynamics}
\label{Sect:4.1.6}

We now combine the previous results to describe the system's global behaviour.  
Table~\ref{tab:qualitative_stability} summarizes the qualitative stability of each critical point, bringing together the slow and fast modes and showing the different dynamical regimes that appear in phase space.

From this classification, three characteristic behaviours emerge:

\begin{itemize}
\item \textbf{$u\ll 1$}: the fast variable $v_4$ dominates, producing sharp transitions controlled by the gravitational branch.
\item \textbf{$u\sim\mathcal{O}(1)$}: the system behaves like a regular autonomous system, with slow and fast modes interacting on similar scales.
\item \textbf{$u\to 0$}: a critical manifold appears, and the slow scalar sector governs the global evolution.
\end{itemize}

\begin{table}[h!]
\centering
\begin{tabularx}{\textwidth}{|c|c|X|}
\hline
\textbf{Point} & \textbf{Branch} & \textbf{Qualitative stability} \\
\hline
A & I ($v_4^\ast=0$) &
Weak saddle: one neutral and one stable fast mode; slow stability set by $v_1^\ast=0$. \\
\hline
B & I ($v_4^\ast=0$) &
Weak saddle: one neutral and one stable fast mode; slow modes are stable if $\zeta>0$. \\
\hline
C & I ($v_4^\ast=0$) &
Weak saddle: one neutral and one stable fast mode; slow modes are stable if $\zeta>0$. \\
\hline
D & II ($v_3^\ast=0$) &
Unstable: $v_4^\ast>0$ gives a positive trace in the fast block. \\
\hline
E & II ($v_3^\ast=0$) &
Strongly unstable: $v_4^\ast$ large and positive. \\
\hline
F & II ($v_3^\ast=0$) &
Saddle: one unstable and one stable fast mode; slow modes are stable if $\zeta>0$. \\
\hline
G & II ($v_3^\ast=0$) &
Unstable: $v_4^\ast>0$ dominates and produces exponential growth. \\
\hline
H & II ($v_3^\ast=0$) &
Saddle: $v_4^\ast<0$ stabilizes one fast mode, but the other may be unstable. \\
\hline
I & II ($v_3^\ast=0$) &
Saddle: a combination of one stable and one unstable fast mode. \\
\hline
\end{tabularx}
\caption{Qualitative stability classification of the critical points of the reduced system.}
\label{tab:qualitative_stability}
\end{table}

Table~\ref{tab:qualitative_stability} helps interpret these regimes.  
Points A–C, in Branch I ($v_4^\ast=0$), always have one neutral and one stable fast mode, making them \emph{weak saddles}. Their stability is mainly controlled by the scalar sector: when $v_1^\ast=0$ (point A), the slow dynamics is marginally stable, while for $v_1^\ast=\pm\zeta/\sqrt{\mu}$ (points B and C), the stability improves if $\zeta>0$. These points describe situations in which the gravitational coupling is frozen, and the slow modes dominate the evolution.

In contrast, points D–I belong to Branch II ($v_3^\ast=0$), where gravitational dynamics is active.  
Points D and H, with $v_4^\ast<0$, have one stabilized fast mode and one potentially unstable mode, producing saddles with clear attracting and repelling directions.  
Points E and G, with $v_4^\ast>0$, are strongly unstable due to exponential growth driven by the fast mode.  
Points F and I mix one stable and one unstable fast mode, influenced by the scalar displacement, giving richer saddle structures in phase space.

Overall, these results show that the slow scalar sector shapes the system's global structure, while the fast variable $v_4$ introduces amplification or damping mechanisms that determine local stability. The interaction between both sectors is essential to understanding the full evolution of the model.

\subsection{Physical interpretation}
\label{Sect:4.1.7}

The critical points of the reduced system represent the possible asymptotic states of the cosmological dynamics when the trajectory approaches the slow manifold $u=(1+v_1)^{-1}$.  
The stability of each point determines whether the model can remain in that state or is pushed toward another regime.  
Below, we interpret each branch in terms of the points A–I.

\textbf{Branch I ($v_4^\ast=0$): points A, B and C.}
Here, the gravitational coupling becomes exactly constant, since $v_4^\ast=0$ implies
\begin{equation}
G(t)=G_0, \qquad G^\ast=(v_3^\ast)^{1/\beta}.
\end{equation}
Point A corresponds to the trivial scalar equilibrium $v_1^\ast=0$, while points B and C represent positive and negative scalar shifts.  
In all cases, the fast dynamics switches off, and the system enters a neutral gravitational mode.  
Stability comes only from the scalar sector, so these points act as partial attractors whenever $v_1^\ast>-1$, which holds for A, B, and C.

\textbf{Branch II ($v_3^\ast=0$): points D, E, F, G, H, and I.}
In this branch, the gravitational coupling vanishes, and the evolution is driven by the rate $v_4^\ast$.  
Points D and E correspond to the trivial scalar equilibrium, while F, G, H, and I include positive or negative scalar shifts.

The sign of $v_4^\ast$ determines the gravitational behaviour:
\begin{itemize}
    \item points E, G and I: $v_4^\ast>0$ leads to exponential growth of $G(t)$ and fast instability;
    \item points D and H: $v_4^\ast<0$ produces decay of the coupling and saddles with one stable and one unstable fast mode;
    \item point F: one stable and one unstable fast mode, both influenced by the scalar displacement.
\end{itemize}

\textbf{Slow–fast structure.}
Global stability is controlled by two complementary mechanisms:
\begin{itemize}
    \item cosmological damping in the slow modes $(v_1,v_2)$, which sets partial stability for points A, D, and E;
    \item gravitational feedback in the fast modes $(v_3,v_4)$, which clearly separates the static branch (A–C) from the dynamic branch (D–I).
\end{itemize}

In summary, only the points in Branch I (A, B, C) can act as partial attractors, while the points in Branch II (D–I) are typically unstable or saddles due to the sign of $v_4^\ast$.  
The global dynamics of the model are therefore organized by the interaction between the slow manifold and the fast gravitational mode.

\section{Analysis of the Regularized Dynamical System}
\label{Sect:4.2}

Let
\begin{equation}
N := \ln a = -\ln(1+z),
\qquad 
\frac{d}{dN} = -(1+z)\frac{d}{dz},
\end{equation}
which transforms the regularized system \eqref{eq:system-x-y-E-N} into
\begin{equation}
\begin{aligned}
\frac{dx}{dN} &= u\,y,\\
\frac{dy}{dN} &= u\,F^{[y]}(x,y),\\
\frac{du}{dN} &= -\,u\,F^{[u]}(x,y,u),
\end{aligned}
\label{eq:sistemaN}
\end{equation}
where
\begin{align}
   F^{[y]}(x,y)&:=-3(x+1)y-\zeta^2 x+\mu x^3,\label{eq:Fxy}\\
   F^{[u]}(x,y,u)&:=y u^2 -\frac{\bigl(u(1+x)-1\bigr)^2}{\delta_0\tau_0^2}. \label{eq:Gxyu}
\end{align}
With initial conditions at $N=0$:
\begin{equation}
{\,x(0)=-\tau_0\delta_0,\qquad y(0)=-1-q_0+\delta_0,\qquad u(0)=1\,}. 
\label{eq:stepC-ic}
\end{equation}

This system is autonomous and smooth, and it exhibits a geometric structure that facilitates its qualitative analysis. In particular, the equations for $(x,y,u)$ are completely decoupled from the equations for $(v_3,v_4)$, so the dynamics of the scalar sector can be studied independently.

\subsection{General Structure of the System}
\label{Sect:4.2.1}

Each equation in \eqref{eq:sistemaN} contains a common factor $u$. From this it follows that:
\begin{itemize}
    \item The hyperplane $u=0$ is \emph{invariant}.
    \item The sign of $u$ is preserved along trajectories.
    \item For $u>0$ one may write
    \begin{equation}
    \frac{d\ln u}{dN} = -F^{[u]}(x,y,u),
    \end{equation}
    which is useful for asymptotic analysis and numerical integration.
\end{itemize}

The system is nonlinear and contains non-hyperbolic critical points, so its analysis requires special care.

\subsection{Equilibrium Points}
\label{Sect:4.2.2}

Critical points satisfy simultaneously
\begin{equation}
u\,y = 0, \qquad
u\,F^{[y]}(x,y) = 0, \qquad
u\,F^{[u]}(x,y,u) = 0.
\end{equation}

\subsubsection{(a) Critical Manifold $u=0$}

If $u=0$, all three equations vanish independently of $x$ and $y$. Therefore,
\begin{equation}
\mathcal{M}_0 = \{(x,y,u)\in\mathbb{R}^3 : u=0\}
\end{equation}
is a \emph{two-dimensional equilibrium manifold}. This degeneracy is typical of regularized systems.

\subsubsection{(b) Equilibria with $u\neq 0$}

If $u\neq 0$, one must satisfy
\begin{equation}
y=0, \qquad F(x,0)=0, \qquad G(x,0,u)=0.
\end{equation}

From \eqref{eq:Fxy},
\begin{equation}
F(x,0) = -\zeta^2 x + \mu x^3 = x(\mu x^2 - \zeta^2),
\end{equation}
whose roots are
\begin{equation}
x_\ast \in \left\{0,\; \pm \frac{\zeta}{\sqrt{\mu}}\right\}.
\end{equation}

Evaluating \eqref{eq:Gxyu} at $y=0$:
\begin{equation}
G(x,0,u)
=
-\frac{1}{\delta_0\tau_0^2}\bigl(1-(1+x)u\bigr)^2,
\end{equation}
so that
\begin{equation}
G(x,0,u)=0
\quad\Longleftrightarrow\quad
u = \frac{1}{1+x}.
\end{equation}

Thus, the non-degenerate equilibrium points are
\begin{equation}
(x_\ast,0,u_\ast),\qquad 
u_\ast=\frac{1}{1+x_\ast},\qquad
x_\ast \in \left\{0,\; \pm \frac{\zeta}{\sqrt{\mu}}\right\}.
\end{equation}

\subsection{Linear Stability}
\label{Sect:4.2.3}

Let $\mathbf{f}(x,y,u)$ denote the vector field in \eqref{eq:sistemaN}. The general Jacobian is
\begin{equation}
J=
\begin{pmatrix}
0 & u & y\\
u F^{[y]}_x & u F^{[y]}_y & F^{[y]}\\
- u F^{[u]}_x & - u F^{[u]}_y & -G -u F^{[u]}_u
\end{pmatrix}.
\end{equation}

\subsubsection{(a) Stability on the Manifold $u=0$}

At $(x_0,y_0,0)$ one obtains
\begin{equation}
J(x_0,y_0,0)=
\begin{pmatrix}
0 & 0 & y_0\\
0 & 0 & F^{[y]}(x_0,y_0)\\
0 & 0 & -F^{[u]}(x_0,y_0,0)
\end{pmatrix}.
\end{equation}

Since
\begin{equation}
F^{[u]}(x_0,y_0,0) = -\frac{1}{\delta_0\tau_0^2},
\end{equation}
there is a positive transversal eigenvalue:
\begin{equation}
\lambda_u = \frac{1}{\delta_0\tau_0^2} > 0.
\end{equation}

Thus, the manifold $u=0$ is \emph{transversely unstable}.

\subsubsection{(b) Stability at an Equilibrium with $u\neq 0$}

At $(x_\ast,0,u_\ast)$, with $u_\ast=1/(1+x_\ast)$ and $F^{[y]}(x_\ast,0)=0$, the Jacobian reduces to
\begin{equation}
J(x_\ast,0,u_\ast)=
\begin{pmatrix}
0 & u_\ast & 0\\
u_\ast F^{[y]}_x(x_\ast,0) & u_\ast F^{[y]}_y(x_\ast,0) & 0\\
0 & 0 & 0
\end{pmatrix}.
\end{equation}

The relevant block is
\begin{equation}
A=
\begin{pmatrix}
0 & 1\\
F^{[y]}_x(x_\ast,0) & F^{[y]}_y(x_\ast,0)
\end{pmatrix},
\end{equation}
whose eigenvalues determine stability in the $(x,y)$ subspace, multiplied by $u_\ast$ in the full system.

The third eigenvalue is always $0$, associated with the $u$ direction.

\subsection{Geometric Structure of the Flow}
\label{Sect:4.2.4}

The structure of \eqref{eq:sistemaN} allows a clear geometric description:
\begin{itemize}
    \item The hyperplane $u=0$ is an \emph{unstable critical manifold}: any orbit approaching it transversely is repelled in the $u$ direction.
    \item The regions $u>0$ and $u<0$ are invariant; the sign of $u$ is fixed by initial conditions.
    \item The evolution of $u$ is governed by
    \begin{equation}
    \frac{d\ln u}{dN} = -F^{[u]}(x,y,u),
    \end{equation}
    which allows us to interpret $G$ as an effective scalar field that regulates expansion or contraction in the $u$ direction.
    \item Equilibria with $u\neq 0$ lie on the surface
    \begin{equation}
    u = \frac{1}{1+x}. 
    \label{eq:superficie-critica}
    \end{equation}
    This surface acts as a \emph{slow manifold} and intersects the cubic curve $F(x,0)=0$.
\end{itemize}

This structure reflects the slow–fast nature of the system: the dynamics in $(x,y)$ is smooth and controlled by $F$, while the evolution in $u$ is dominated by $G$, introducing a non-hyperbolic central direction.

\subsection{Classification of Equilibria with $u\neq 0$}
\label{Sect:4.2.5}

The non-degenerate critical points of \eqref{eq:sistemaN} are
\begin{equation}
(x_\ast,0,u_\ast),\qquad 
x_\ast \in \left\{0,\; \pm \frac{\zeta}{\sqrt{\mu}}\right\},\qquad
u_\ast=\frac{1}{1+x_\ast}.
\end{equation}

To study their stability, consider the block
\begin{equation}
A=
\begin{pmatrix}
0 & 1\\
F^{[y]}_x(x_\ast,0) & F^{[y]}_y(x_\ast,0)
\end{pmatrix},
\qquad
\lambda_{1,2} = \frac{F^{[y]}_y(x_\ast,0)\pm\sqrt{F^{[y]}_y(x_\ast,0)^2+4F^{[y]}_x(x_\ast,0)}}{2}.
\end{equation}

Recall
\begin{equation}
F^{[y]}_x(x,0) = -\zeta^2 + 3\mu x^2,
\qquad
F^{[y]}_y(x,0) = -3(x+1).
\end{equation}

We analyze each point:

\subsubsection{Critical Point $x_\ast = 0$}

Here,
\begin{equation}
F^{[y]}_x(0,0) = -\zeta^2,
\qquad
F^{[y]}_y(0,0) = -3.
\end{equation}

The discriminant is
\begin{equation}
\Delta_0 = 9 - 4\zeta^2.
\end{equation}

\begin{itemize}
    \item If $\zeta^2 < \tfrac{9}{4}$, then $\Delta_0>0$ and the eigenvalues are real:
    \begin{equation}
    \lambda_{1,2} = \frac{-3\pm\sqrt{9-4\zeta^2}}{2}.
    \end{equation}
    Both are negative: $(0,0,u_\ast)$ is a \emph{stable node}.
    \item If $\zeta^2 > \tfrac{9}{4}$, then $\Delta_0<0$ and the eigenvalues are complex with negative real part:
    \begin{equation}
    \lambda_{1,2} = -\frac{3}{2} \pm i\sqrt{\zeta^2 - \tfrac{9}{4}},
    \end{equation}
    so $(0,0,u_\ast)$ is a \emph{stable focus}.
\end{itemize}

In both cases, the third eigenvalue is $0$, so the point is \emph{non-hyperbolic}.

\subsubsection{Critical Point $x_\ast = +\dfrac{\zeta}{\sqrt{\mu}}$}

Here,
\begin{equation}
F^{[y]}_x\!\left(\frac{\zeta}{\sqrt{\mu}},0\right)
= 2\zeta^2 > 0,
\qquad
F^{[y]}_y\!\left(\frac{\zeta}{\sqrt{\mu}},0\right)
= -3\left(1+\frac{\zeta}{\sqrt{\mu}}\right) < 0.
\end{equation}

The discriminant is positive:
\begin{equation}
\Delta_+ = {F^{[y]}_y}^2 + 4F^{[y]}_x > 0,
\end{equation}
and the eigenvalues have opposite signs:
\begin{equation}
\lambda_1 < 0 < \lambda_2.
\end{equation}

Thus the point
\begin{equation}
\left(\frac{\zeta}{\sqrt{\mu}},\,0,\,\frac{1}{1+\frac{\zeta}{\sqrt{\mu}}}\right)
\end{equation}
is a \emph{saddle} in the $(x,y)$ subspace, with an additional zero eigenvalue.

\subsubsection{3. Critical Point $x_\ast = -\dfrac{\zeta}{\sqrt{\mu}}$}

Here,
\begin{equation}
F^{[y]}_x\!\left(-\frac{\zeta}{\sqrt{\mu}},0\right)
= 2\zeta^2 > 0,
\qquad
F^{[y]}_y\!\left(-\frac{\zeta}{\sqrt{\mu}},0\right)
= -3\left(1-\frac{\zeta}{\sqrt{\mu}}\right).
\end{equation}

The sign of $F^{[y]}_y$ depends on $\zeta/\sqrt{\mu}$:

\begin{itemize}
    \item If $\dfrac{\zeta}{\sqrt{\mu}} < 1$, then $F^{[y]}_y<0$ and the point is a \emph{saddle}.
    \item If $\dfrac{\zeta}{\sqrt{\mu}} = 1$, then $F^{[y]}_y=0$ and the point is \emph{non-hyperbolic} with symmetric eigenvalues:
    \begin{equation}
    \lambda_{1,2} = \pm \sqrt{2}\,\zeta.
    \end{equation}
    \item If $\dfrac{\zeta}{\sqrt{\mu}} > 1$, then $F^{[y]}_y>0$ and the point is again a \emph{saddle}.
\end{itemize}

In all cases, the third eigenvalue is $0$.

\subsubsection{Stability summary}
\begin{itemize}
    \item The point $(0,0,u_\ast)$ is stable in $(x,y)$ (a node or a focus depending on $\zeta$), but it is non-hyperbolic due to the zero eigenvalue in the $u$ direction.
    \item The point $\left(\frac{\zeta}{\sqrt{\mu}},0,u_\ast\right)$ is always a saddle.
    \item The point $\left(-\frac{\zeta}{\sqrt{\mu}},0,u_\ast\right)$ is typically a saddle, except in the limiting case $\zeta=\sqrt{\mu}$, where it becomes non-hyperbolic.
\end{itemize}
These results fully characterize the local structure of the flow around the non-degenerate critical points of the regularized system for $u>0$.

The regularized system exhibits a rich structure: an unstable critical manifold, non-hyperbolic equilibrium points, and dynamics controlled by the function $G$. The formulation in terms of $N=\ln a$ enables a clear qualitative analysis and a robust numerical implementation. The stability of equilibria with $u\neq 0$ depends on the derivatives of $F$ along the line $y=0$, while the $u$ direction introduces a central mode whose interpretation requires a more refined analysis, especially when connected with the fast dynamics of the original system.

This observation suggests that the flow geometry is not fully captured by the local linear analysis and instead reflects a deeper slow–fast structure. In particular, the relationship between $u$ and $x$ that emerges in the asymptotic analysis indicates the existence of a critical surface that organizes the global dynamics. To understand this phenomenon, it is useful to reinterpret the geometric identity \eqref{eq:manifold-hyp} as an approximation of the invariant manifolds associated with the non-degenerate critical points, providing a natural link between the regularized system and the effective dynamics on the fast time scale $\tau$.

\subsubsection{Final remarks and connection with slow–fast dynamics}

The relation \eqref{eq:manifold-hyp} should not be understood as an exact identity of the system, but rather as a \emph{structural approximation} capturing the dominant shape of the invariant manifolds associated with the non-degenerate critical points. Its usefulness lies in the fact that:
\begin{itemize}
    \item It provides an explicit description of how trajectories move away from or approach the critical surface \eqref{eq:superficie-critica}.
    \item It allows the dynamics to be interpreted in terms of an auxiliary variable $\tau$ that regularizes the fast direction of the system.
    \item It naturally connects the regularized formulation with the slow–fast analysis developed on the time scale $\tau$.
\end{itemize}

Indeed, the slow–fast structure of the original system implies that the combination
\begin{equation}
\Psi(x,u) := \frac{1}{u} - (1+x)
\end{equation}
evolves according to a degenerate Riccati equation on the fast scale, leading to an asymptotic behaviour of the form
\begin{equation}
\Psi(\tau) \sim \frac{C}{\tau},
\qquad \tau\to\infty,
\end{equation}
for a constant $C$ determined by the model parameters. The hypothesis
\begin{equation}
\Psi(\tau) = \frac{\alpha-1}{3\tau}
\end{equation}
is therefore consistent with the qualitative structure of the dynamics and reproduces the algebraic decay characteristic of slow manifolds in systems with scale separation.

Moreover, the local analysis near a critical point $(x_\ast,0,u_\ast)$ shows that
\begin{equation}
\Psi(x,u) \approx -X - \frac{U}{u_\ast^2},
\end{equation}
so that the condition $\Psi(\tau)=\Phi(\tau)$ selects a curve (or surface) in phase space that approximates the stable or unstable manifold depending on the sign of $\tau$. This description is consistent with invariant manifold theory: the function $\Phi(\tau)$ acts as a parameter that controls the distance to the critical surface and determines the direction (stable or unstable) of the trajectory.

Taken together, the relation \eqref{eq:manifold-hyp} provides a useful geometric representation of the regularized dynamics. It allows one to visualize how trajectories organize around the critical surface \eqref{eq:superficie-critica}, how they approach it in the asymptotic regime $\tau\to\infty$, and how they separate rapidly when $\tau\to 0$. Furthermore, it provides a conceptual bridge between local analysis (linearization and invariant manifolds) and global analysis (slow–fast structure and reduced systems), demonstrating that both approaches capture complementary aspects of the same underlying geometry.

In the absence of a complete derivation, the relation \eqref{eq:manifold-hyp} should be interpreted as a \emph{geometric ansatz} consistent with the structure of the system, effectively capturing the transition between the stable and unstable manifolds in terms of the critical surface \eqref{eq:superficie-critica} and the scale characterized by $\tau$.

\subsubsection{Formal validation of the ansatzes and construction of the stable manifold}

We now justify the relation
\begin{equation}
\frac{1}{u}-(1+x)=\Xi(\tau),
\label{eq:Phi-ansatz}
\end{equation}
using invariant manifold theory and show how it naturally arises from the slow–fast structure of the regularized system.

Let $(x_\ast,0,u_\ast)$ be a nondegenerate equilibrium satisfying
\begin{equation}
\frac{1}{u_\ast}=1+x_\ast.
\label{eq:equilibrio-superficie}
\end{equation}
Introduce shifted variables
\begin{equation}
X=x-x_\ast,\qquad Y=y,\qquad U=u-u_\ast,
\end{equation}
so that the equilibrium is moved to the origin. In these coordinates the system becomes
\begin{equation}
\frac{d}{dN}
\begin{pmatrix}X\\Y\\U\end{pmatrix}
=
\begin{pmatrix}
F_1(X,Y,U)\\F_2(X,Y,U)\\F_3(X,Y,U)
\end{pmatrix},
\qquad F_i(0,0,0)=0,
\end{equation}
with Jacobian
\begin{equation}
Df(0,0,0)=
\begin{pmatrix}
a_{11} & a_{12} & a_{13}\\
a_{21} & a_{22} & a_{23}\\
a_{31} & a_{32} & a_{33}
\end{pmatrix}.
\end{equation}

If the $(X,Y)$–subsystem has at least one eigenvalue with negative real part, invariant manifold theory guarantees the existence of a local stable manifold
\begin{equation}
U=h_s(X,Y),\qquad h_s(0,0)=0,\quad Dh_s(0,0)=0,
\label{eq:stable-manifold-graph}
\end{equation}
and similarly, an unstable manifold.

The deviation from the critical surface is measured by
\begin{equation}
\Psi(x,u)=\frac{1}{u}-(1+x).
\end{equation}
Expanding in $(X,U)$ gives
\begin{equation}
\Psi(x,u)\approx -X-\frac{U}{u_\ast^2},
\label{eq:Psi-linearizada}
\end{equation}
using \eqref{eq:equilibrio-superficie}.

On $\mathcal{W}^s$ we have $U=h_s(X,Y)$, and invariance requires
\begin{equation}
\frac{dU}{dN}
=\partial_X h_s\,\frac{dX}{dN}
+\partial_Y h_s\,\frac{dY}{dN}
=F_3(X,Y,h_s(X,Y)).
\end{equation}
Thus
\begin{align}
F_3(X,Y,h_s)
&=\partial_X h_s\,F_1(X,Y,h_s)
+\partial_Y h_s\,F_2(X,Y,h_s).
\label{eq:invariance-equation}
\end{align}
A first–order approximation
\begin{equation}
h_s(X,Y)=\alpha_X X+\alpha_Y Y+\mathcal{O}(\|(X,Y)\|^2)
\end{equation}
substituted into \eqref{eq:invariance-equation} determines $(\alpha_X,\alpha_Y)$ and the local orientation of the stable manifold.

We obtain the connection with the geometric ansatz by using \eqref{eq:Psi-linearizada} and \eqref{eq:stable-manifold-graph}, on $\mathcal{W}^s$,
\begin{equation}
\Psi(x,u)\approx -X-\frac{h_s(X,Y)}{u_\ast^2}.
\end{equation}
The ansatz \eqref{eq:Phi-ansatz} assumes that this quantity evolves as
\begin{equation}
\Psi(x,u)=\Xi(\tau),\qquad 
\Xi(\tau)=\frac{\alpha-1}{3\tau},
\end{equation}
which decays to zero as $\tau\to\infty$ and diverges as $\tau\to 0$, matching the behavior of stable and unstable directions.

\subsubsection{Slow–fast interpretation.}

The invariance condition
\begin{equation}
\frac{d}{dN}\Bigl(\tfrac{1}{u}-(1+x)\Bigr)=0
\tag{\ref{eq:invariance-Psi}}
\end{equation}
combined with $\frac{d}{dN}=u\,\frac{d}{d\tau}$ and $\Psi=\Xi(\tau)$ yields an effective equation
\begin{equation}
u\,\frac{d\Phi}{d\tau}=\mathcal{G}(\Phi,\text{parameters}),
\label{eq:Psi-riccati-especifica}
\end{equation}
whose first–order solution is consistent with
\begin{equation}
\Xi(\tau)=\frac{\alpha-1}{3\tau}.
\end{equation}

Thus the relation
\begin{equation}
\frac{1}{u}-(1+x)=\frac{\alpha-1}{3\tau}
\end{equation}
is a geometrically motivated approximation of the invariant manifold. The variable $u$ encodes the scale separation, while $\Xi(\tau)$ captures how trajectories approach or depart from the critical surface within the slow–fast structure.

\subsection{Summary for the specific system}
\label{Sect:4.2.7}

For the system \eqref{eq:sistemaN}--\eqref{eq:Gxyu}:

\begin{itemize}
\item The critical surface $\frac{1}{u}=1+x$ contains all nondegenerate equilibria and is characterized by the exact invariance condition $\frac{d\Psi}{dN}=0$.
\item For $u>0$, the deviation $\Psi=\frac{1}{u}-(1+x)$ decreases strictly along trajectories, so the surface acts as a transversal attractor.
\item On the fast scale $\tau$, $\Psi$ satisfies a Riccati-type equation whose solution yields
\begin{equation}
\frac{1}{u}-(1+x)=\frac{\alpha-1}{3\tau},
\end{equation}
providing an explicit expression for $\alpha$ in terms of the model parameters.
\end{itemize}

Hence, the geometric framework naturally specializes to the regularized system, clarifying the roles of the critical surface, the variable $\Psi$, and the slow–fast ansatz in cosmological dynamics.

In this paper, a systematic study of a cosmological model of modified fractional gravity formulated within FALVA is presented, from which modifications to the Friedmann equations are derived.
An adimensional dynamical system is constructed to describe the evolution of the scalar field and the Hubble parameter, identifying the quantity $\mathcal{R}\equiv \frac{d\ln G}{d\ln a}$
as the diagnostic of the model.
 
To connect this formulation with the phase–space analysis, we use the dimensionless variables \eqref{vars-dimensionless} and the manifold relation \eqref{eq:manifold-hyp}. In these variables, one has
\begin{equation}
\frac{\dot G}{G}=H_0 v_4,\qquad H=\frac{H_0}{u},
\end{equation}
so the physical condition $\dot G/G\gg H$ can be compactly rewritten as
\begin{equation}
u\,v_4 \gg 1.
\end{equation}
The dimensionless product $u v_4$ therefore quantifies the relevance of the variation of $G$ relative to dilution by expansion.

Starting from the autonomous system in the e–folding variable $N$ given in equations \eqref{Aut}, the discussion of continuity and of the dimensionless quantity
\begin{equation}
\mathcal{R}(N):=u(N)\,v_4(N)=\frac{\dot G/G}{H}
\end{equation}
is reformulated with $N$ as the independent variable while keeping the system's standard notation.

\subsection{Derivation of the evolution equation for $\mathcal{R}(N)$}
\label{Sect:5.1.1}

Applying the product rule and using the relation $\dfrac{d}{dN}=u\,\dfrac{d}{d\tau}$, we obtain
\begin{equation}\label{R:prod}
\frac{d\mathcal{R}}{dN}
= v_4\frac{du}{dN} + u\frac{dv_4}{dN}.
\end{equation}
Substituting $\dfrac{du}{dN}$ and $\dfrac{dv_4}{dN}$ from \eqref{Aut}, we arrive at
\begin{align}\label{R:explicit}
\frac{d\mathcal{R}}{dN}
&= -v_4\,u\!\left[v_2 u^2 -\lambda\bigl(1-(1+v_1)u\bigr)^2\right] \nonumber\\
&\quad + u\left[\frac{3}{\mu u}
+ \frac{1}{2}u v_4 (v_4 - 6 v_1 - 6)
- \frac{3(\beta -1)u v_3 \Omega_{0}}{\mu}\right].
\end{align}

Using $v_4=\mathcal{R}/u$ and simplifying, we obtain a closed equation for $\mathcal{R}$:
\begin{align}\label{R:closed}
\frac{d\mathcal{R}}{dN}
&= -\mathcal{R}\!\left[v_2 u^2
- \lambda\bigl(1-(1+v_1)u\bigr)^2\right]  + \frac{3}{\mu}
+ \frac{1}{2}\mathcal{R}^2
- \frac{1}{2}(6v_1+6)\,\mathcal{R}
- \frac{3(\beta-1)\Omega_0\,u^2 v_3}{\mu}.
\end{align}

This equation depends only on $\mathcal{R}$, on $u$, and on the slow variables $v_1,v_2,v_3$, and it allows one to study directly the influence of the variation of $G$ on energy continuity. The terms in \eqref{R:closed} have the following interpretation:
\begin{itemize}
\item $3/\mu$: dominant source when $u\ll1$, inherited from the singular term $3/(\mu u)$ in \eqref{Aut:v4}.
\item $\tfrac{1}{2}\mathcal{R}^2$: positive feedback.
\item $-3(v_1+1)\mathcal{R}$: damping depending on the scalar field.
\item $-\mathcal{R}v_2 u^2$ and $-3(\beta-1)\Omega_0 u^2 v_3/\mu$: couplings that modulate the dynamics when $u$ is not small.
\end{itemize}

\section{Alternative evolution equation: normalized form}
\label{Sect:5.2}

For qualitative analysis it is useful to rearrange \eqref{R:closed} isolating the dominant source $3/\mu$:
\begin{equation}\label{R:normalized}
\frac{d\mathcal{R}}{dN}
= \frac{3}{\mu}
+ \frac{1}{2}\mathcal{R}^2
- 3(v_1+1)\mathcal{R}
- \mathcal{R}\,v_2 u^2
+ \lambda\mathcal{R}\bigl(1-(1+v_1)u\bigr)^2
- \frac{3(\beta-1)\Omega_0\,u^2 v_3}{\mu}.
\end{equation}

This form makes explicit the different mechanisms governing the evolution of $\mathcal{R}$:
\begin{itemize}
\item \textbf{Growth sources:} the constant term $3/\mu$ and the nonlinear feedback $\tfrac{1}{2}\mathcal{R}^2$.
\item \textbf{Damping terms:} the linear term $-3(v_1+1)\mathcal{R}$ and the coupling $-\mathcal{R}v_2 u^2$.
\item \textbf{Additional corrections:} the terms proportional to $u^2 v_3$ and to $(1-(1+v_1)u)^2$, relevant when $u$ departs from zero.
\end{itemize}

\subsection{Physical interpretation and practical use}
\label{Sect:5.2.1}

\begin{itemize}
  \item When $u\ll 1$, the source $3/\mu$ dominates the evolution of $\mathcal{R}$, reproducing the fast behaviour associated with the singular term $3/(\mu u)$ in \eqref{Aut:v4}. This explains why $v_4$ becomes the system's fast variable near $u=0$.

  \item The quadratic term $\tfrac{1}{2}\mathcal{R}^2$ introduces positive feedback: if $\mathcal{R}$ grows, its growth accelerates, potentially generating sharp peaks.

  \item The linear terms and the couplings with $u^2 v_2$ and $u^2 v_3$ modulate the dynamics when $u$ is not extremely small, either regulating or amplifying the evolution depending on the phase–space trajectory.

  \item The operational bound $|\mathcal{R}|<\varepsilon$ during BBN is naturally expressed in \eqref{R:closed}, allowing one to evaluate how the parameters $(\mu,\beta,\Omega_0,\delta_0,\tau_0,\zeta)$ and the initial conditions affect the cosmological viability of the model.
\end{itemize}

\subsection{Qualitative structure of the system and formal analysis}
\label{Sect:5.2.2}

The autonomous system
\begin{subequations}
\label{Aut-R}
\begin{align}
u' &= -u\!\left[v_2 u^2 -\lambda\bigl(1-(1+v_1)u\bigr)^2\right], \label{Aut:u-R} \\
v_1' &= u\, v_2, \label{Aut:v1-R} \\
v_2' &= u\!\left[-3(v_1+1)v_2 - \zeta^2 v_1 + \mu v_1^3\right], \label{Aut:v2-R}\\
v_3' &= \beta v_3\,\mathcal{R}, \label{Aut:v3-R} \\
\mathcal{R}' &= \frac{3}{\mu}
+ \frac{1}{2}\mathcal{R}^2
- 3(v_1+1)\mathcal{R}
- \mathcal{R}\,v_2 u^2
+ \lambda\mathcal{R}\bigl(1-(1+v_1)u\bigr)^2
- \frac{3(\beta-1)\Omega_0\,u^2 v_3}{\mu} \label{R:normalized-R}
\end{align}
\end{subequations}
describes the joint evolution of the effective geometry $(u,v_1,v_2)$, the auxiliary variable $v_3$, and the relative variation of $G$, encoded in $\mathcal{R}$. Its structure exhibits properties that allow a clear and systematic understanding of its global behaviour.

\textbf{Closed dynamics of $\mathcal{R}$ and stability.}  
The closed equation for $\mathcal{R}$ shows that, when $u\ll1$, the dominant instability arises from the source term $3/(\mu u)$, inherited from \eqref{tag4}. In the fractional framework, this term combines with the antifriction $(1-\alpha)/t$, which can amplify peaks in $\mathcal{R}$ if $\alpha>1$ or attenuate them if $\alpha<1$.

The slow manifold acts as a geometric attractor: once the solution enters it, the subsequent evolution is dominated by the dynamics of $\mathcal{R}$. Therefore, the stability of the critical points on the slow manifold determines the late–time cosmological behaviour.

\textbf{Slow–fast structure.}  
The background equations $(u,v_1,v_2)$ contain a common factor of $u$. This implies that:
\begin{itemize}
    \item when $u\approx 0$, the evolution of $(u,v_1,v_2)$ is slow;
    \item when $u=\mathcal{O}(1)$, the dynamics may accelerate.
\end{itemize}

This scale separation defines a slow–fast structure in which
\begin{equation}
(u,v_1,v_2) \quad \text{are slow variables}, \qquad
(\mathcal{R},v_3) \quad \text{are fast variables}.
\end{equation}

The critical surface is obtained by imposing $u'=0$, which yields
\begin{equation}
v_2 u^2 = \lambda\bigl(1-(1+v_1)u\bigr)^2.
\label{eq:critical-surface}
\end{equation}

In addition, the geometric relation
\begin{equation}
u = \frac{1}{1+v_1}
\label{eq:slow-manifold}
\end{equation}
defines the \emph{slow manifold} of the system, which organizes the global dynamics and on which the reduced flow is regular.

\textbf{Slow manifold and reduced system.}  
Substituting \eqref{eq:slow-manifold} into the background equations yields the reduced system
\begin{subequations}
\label{eq:slow-syst-manifold}
\begin{align}
v_1' &= \frac{v_2}{1+v_1},\\
v_2' &= \frac{-3(v_1+1)v_2 - \zeta^2 v_1 + \mu v_1^3}{1+v_1},
\end{align}
\end{subequations}
while the equations for $v_3$ and $\mathcal{R}$ remain unchanged.  
This system describes the effective dynamics away from the singular region $u\rightarrow \infty$ ($H\rightarrow 0$).

A detailed stability analysis and the cosmological interpretation of each family of points are presented in the following subsections.

\textbf{Linear stability}.  
The Jacobian matrix of the system at a critical point 
$(u_*,v_{1*},v_{2*},v_{3*},\mathcal{R}_*)$ has a block–triangular structure:
\begin{equation}
J =
\begin{pmatrix}
u_* A & 0\\
* & B
\end{pmatrix},
\end{equation}
where:
\begin{itemize}
    \item $A$ corresponds to the slow sector $(u,v_1,v_2)$,
    \item $B$ corresponds to the fast sector $(v_3,\mathcal{R})$.
\end{itemize}

This implies that global stability is determined by combining:
\begin{itemize}
    \item the stability of the geometric background (eigenvalues of $u_*A$),
    \item the stability of the variation of $G$ (eigenvalues of $B$).
\end{itemize}

In particular:
\begin{itemize}
    \item if $u_*\approx 0$, the slow sector is nearly neutral and stability depends on $\mathcal{R}$;
    \item if $\mathcal{R}_*>0$, the mode associated with $v_3$ is unstable;
    \item if $\mathcal{R}_*<0$, $v_3$ decays and the point may be stable.
\end{itemize}

\textbf{Physical interpretation}.  
The formal analysis shows that:
\begin{itemize}
    \item The slow manifold \eqref{eq:slow-manifold} describes the regular evolution of the background.
    \item The region $u\rightarrow 0$ corresponds to a singular regime in which the equation for $\mathcal{R}$ becomes dominant.
    \item The source term $3/\mu$ and the quadratic term $\tfrac12\mathcal{R}^2$ explain the appearance of peaks in $\mathcal{R}$.
    \item The stability of the system depends critically on the sign of $\mathcal{R}$, which controls both the variation of $G$ and the evolution of $v_3$.
\end{itemize}

Taken together, the system \eqref{Aut-R} exhibits a well–defined slow–fast geometry together with a closed equation for $\mathcal{R}$, which concentrates the relevant physics: modified continuity, phantom crossing, BBN constraints, and the connection with the $S_8$ tension.

However, the geometric analysis of the system reveals the existence of a singular region at $v_1=-1$ ($u\rightarrow\infty$, $H\rightarrow 0$), which delimits the domain $v_1\leq -1$ ($u<\infty$). In this region, the variables used do not allow precise identification of the interval over which $H$ changes sign. For this reason, in Section~\ref{Sect:5.4}, an alternative formulation will be introduced that allows a consistent study of the transition.

\section{Critical points of the normalized system in \texorpdfstring{$\mathcal{R}$}{R}}
\label{Sect:5.3}

In this section, we explicitly determine the family of critical points of the autonomous system \eqref{Aut-R} in the regular sector $u>0$. The analysis is carried out systematically, combining the equilibrium conditions of each equation.

\textbf{Critical point conditions}.  
A critical point of the system \eqref{Aut-R} has coordinates
$(u_*,v_{1*},v_{2*},v_{3*},\mathcal{R}_*)$ satisfying simultaneously
\begin{equation}
u' = v_1' = v_2' = v_3' = \mathcal{R}' = 0.
\end{equation}
We focus on the regular sector $u_*>0$, leaving the case $u_*=0$ for the study of the singular region associated with the dynamics near $S_8$.

\textbf{Background equations \texorpdfstring{$(u,v_1,v_2)$}{(u,v1,v2)}.}
From \eqref{Aut:v1-R} we obtain
\begin{equation}
v_1' = u\,v_2 = 0.
\end{equation}
In the sector $u_*\neq 0$, this necessarily implies
\begin{equation}
v_{2*} = 0.
\label{eq:v2star}
\end{equation}

Equation \eqref{Aut:v2-R} then reduces to
\begin{equation}
v_2' = u\bigl[-3(v_1+1)v_2 - \zeta^2 v_1 + \mu v_1^3\bigr]
= u\bigl[- \zeta^2 v_1 + \mu v_1^3\bigr] = 0,
\end{equation}
from which
\begin{equation}
-\zeta^2 v_{1*} + \mu v_{1*}^3 = 0
\quad\Longrightarrow\quad
v_{1*}\bigl(\mu v_{1*}^2 - \zeta^2\bigr)=0.
\end{equation}
Thus,
\begin{equation}
v_{1*} = 0,
\qquad
v_{1*} = \pm \frac{\zeta}{\sqrt{\mu}}.
\label{eq:v1star-values}
\end{equation}

Finally, the condition $u'=0$ from \eqref{Aut:u-R} takes the form
\begin{equation}
u' = -u\!\left[
v_2 u^2 -\lambda
\bigl(1-(1+v_1)u\bigr)^2\right] = 0.
\end{equation}
In the sector $u_*\neq 0$, using \eqref{eq:v2star} we obtain
\begin{equation}
\bigl(1-(1+v_{1*})u_*\bigr)^2 = 0
\quad\Longrightarrow\quad
u_* = \frac{1}{1+v_{1*}},
\label{eq:ustar}
\end{equation}
provided $1+v_{1*}\neq 0$. The physical condition $u_*>0$ requires
\begin{equation}
1+v_{1*} > 0.
\label{eq:u-positive}
\end{equation}

In summary, the background sector with $u_*>0$ is determined by
\begin{equation}
v_{2*}=0,\quad
v_{1*}\in\left\{0,\ \pm\frac{\zeta}{\sqrt{\mu}}\right\},\quad
u_*=\frac{1}{1+v_{1*}},\quad 1+v_{1*}>0.
\label{eq:background-critical}
\end{equation}

\textbf{Equation for \texorpdfstring{$v_3$}{v3}.}
The condition $v_3'=0$ imposes
\begin{equation}
v_3' = \beta v_3\mathcal{R} = 0
\quad\Longrightarrow\quad
v_{3*} = 0
\quad\text{or}\quad
\mathcal{R}_* = 0.
\label{eq:v3-critical-condition}
\end{equation}

In what follows, we first analyze the case
\begin{equation}
v_{3*}=0,
\label{eq:v3zero}
\end{equation}
which partially decouples the equation for $\mathcal{R}$ and yields a discrete family of critical points with $\mathcal{R}_*\neq 0$. The case $\mathcal{R}_*=0$ will be discussed later.

\textbf{Equation for \texorpdfstring{$\mathcal{R}$}{R}.}
Under conditions \eqref{eq:v2star}, \eqref{eq:ustar} and \eqref{eq:v3zero},  
the terms proportional to $u^{2}v_{2}$, $u^{2}v_{3}$ and 
$\bigl(1-(1+v_{1})u\bigr)^{2}$ disappear from \eqref{R:normalized-R}, yielding for fixed $v_{1}$ the one–dimensional differential equation
\begin{equation}
\mathcal{R}' = f(\mathcal{R})
=
\frac{3}{\mu}
+\frac{1}{2}\mathcal{R}^{2}
-3(v_{1}+1)\mathcal{R}.
\end{equation}

Thus, the equilibrium equation $\mathcal{R}'=0$ reduces to
\begin{equation}
f(\mathcal{R})=0
\label{eq:parabola-v1}
\end{equation}
or, multiplying by $2$,
\begin{equation}
\mathcal{R}^{2}
- 6(v_{1}+1)\mathcal{R}
+ \frac{6}{\mu}
= 0.
\label{eq:R-quadratic}
\end{equation}

The discriminant
\begin{equation}
\Delta = 9(v_{1}+1)^{2}-\frac{6}{\mu}
\end{equation}
determines the number of solutions.

The solutions of \eqref{eq:R-quadratic} are
\begin{equation}
\mathcal{R}_{\pm}(v_{1})
=
3(v_{1}+1)
\pm
\sqrt{9(v_{1}+1)^{2}-\frac{6}{\mu}}.
\label{eq:Rpm}
\end{equation}
Therefore, the existence of real critical points requires the discriminant to be non–negative:
\begin{equation}
9(v_{1}+1)^{2}-\frac{6}{\mu}\ge 0
\quad\Longleftrightarrow\quad
|v_{1}+1|\ge \sqrt{\frac{2}{3\mu}}.
\label{eq:R-discriminant-condition}
\end{equation}

Expressions \eqref{eq:background-critical}, \eqref{eq:Rpm} and 
\eqref{eq:R-discriminant-condition} completely determine the regular critical points $(u_{*}>0)$ in the sector $v_{3*}=0$.

\textbf{Regime without critical points.}
The regime without critical points corresponds to
\begin{equation}
\Delta < 0,
\end{equation}
in which the polynomial has no real roots. In this situation, the function $f(\mathcal{R})$ does not vanish anywhere on the real axis, and the system has no stationary states at all.

Dynamically, the absence of critical points implies that the sign of $f(\mathcal{R})$ is constant for all $\mathcal{R}\in\mathbb{R}$. Therefore, the evolution of $\mathcal{R}(t)$ is strictly monotonic:
\begin{itemize}
    \item if $f(\mathcal{R})>0$ for all $\mathcal{R}$, then $\mathcal{R}(t)$ grows without interruption and the flow is directed towards $+\infty$;
    \item if $f(\mathcal{R})<0$ for all $\mathcal{R}$, then $\mathcal{R}(t)$ decreases uniformly and the flow is directed towards $-\infty$.
\end{itemize}

In this regime, there is no local structure that can attract or repel trajectories, nor any possibility of stabilization at a finite value. The dynamics are purely directional and completely determined by the global sign of the velocity field. This behavior characterizes an unstructured phase of the system, in which no local bifurcations or state transitions can occur.

\begin{figure}[htb]
    \centering
    \includegraphics[width=\textwidth]{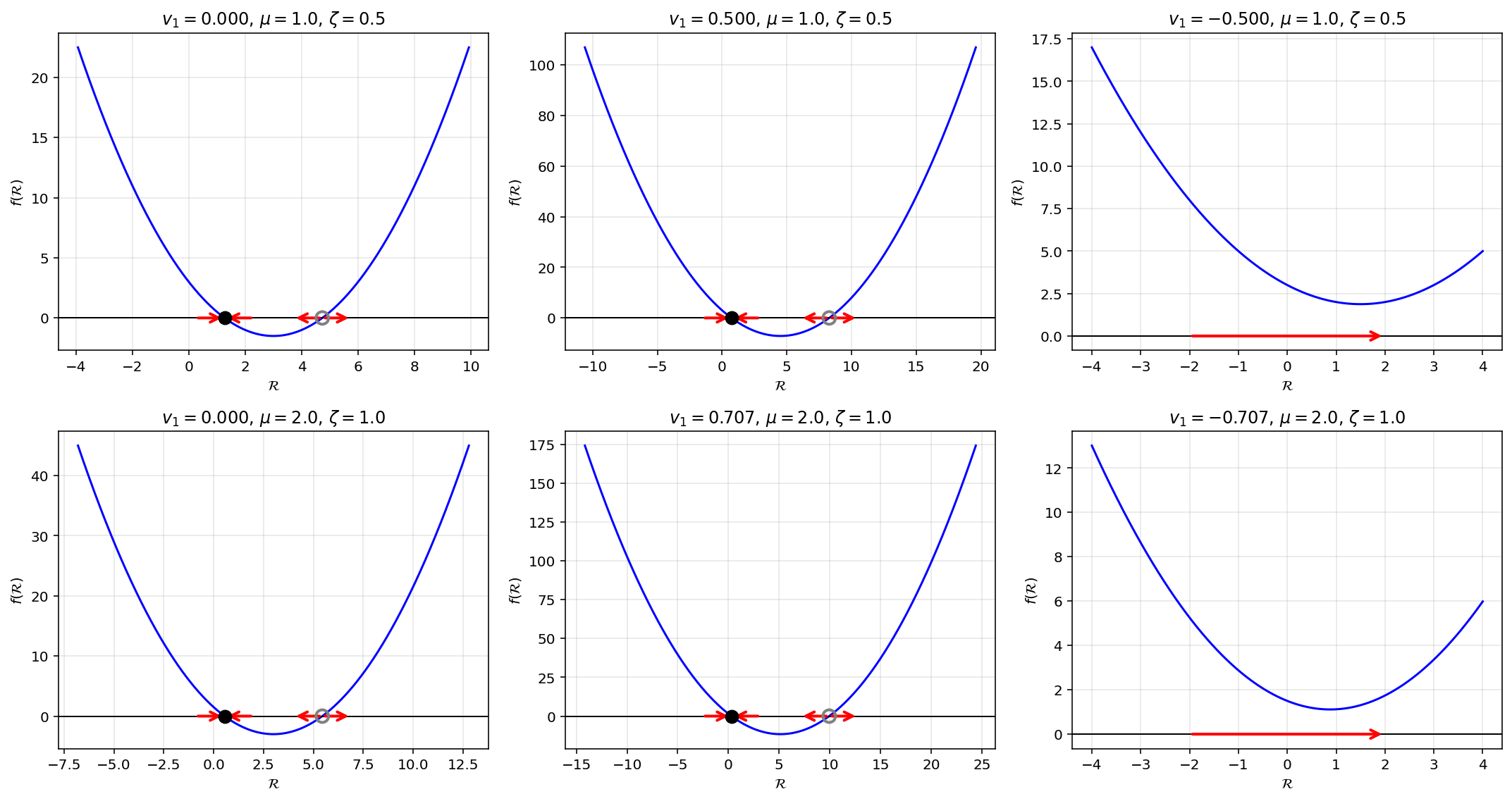}
    \caption{One-dimensional flow of equation \eqref{eq:parabola-v1} for $v_{1*}\in\{0,\pm\zeta/\sqrt{\mu}\}$ and different parameter values.}
    \label{fig:flujo-parábola-v1}
\end{figure}

Figure~\ref{fig:flujo-parábola-v1} shows the one-dimensional flow of equation \eqref{eq:parabola-v1} for $v_{1}=0$ and $\mu=1$.

\textbf{Regular critical points with \texorpdfstring{$v_{3}=0$}{v3=0}.}
The possible values of $v_{1*}$ in \eqref{eq:background-critical} are
\begin{equation}
v_{1*}=0,\qquad
v_{1*}=\frac{\zeta}{\sqrt{\mu}},\qquad
v_{1*}=-\frac{\zeta}{\sqrt{\mu}}.
\end{equation}
For each of them, we obtain:
\begin{itemize}
    \item the value of $u_{*}$ through $u_{*}=1/(1+v_{1*})$, provided that $1+v_{1*}>0$;
    \item two possible values of $\mathcal{R}_{*}$ given by \eqref{eq:Rpm}.
\end{itemize}

Thus, each admissible value of $v_{1*}$ generates, in principle, a pair of critical points, which we denote by $P^{(+)}$ and $P^{(-)}$ depending on the choice of sign in $\mathcal{R}_{\pm}$.

\begin{table}[t]
\centering
\caption{Regular critical points ($u>0$) of system \eqref{Aut-R} in the sector $v_{3}=0$. The existence of each point requires that the conditions $1+v_{1*}>0$ and $9(v_{1*}+1)^{2}-6/\mu\ge 0$ be satisfied simultaneously.}
\begin{tabular}{c c c c c c}
\hline
Label & $u_{*}$ & $v_{1*}$ & $v_{2*}$ & $v_{3*}$ & $\mathcal{R}_{*}$ \\ \hline
$P_{0}^{(\pm)}$
& $1$
& $0$
& $0$
& $0$
& $3\pm\sqrt{9-\dfrac{6}{\mu}}$ \\
$P_{+}^{(\pm)}$
& $\dfrac{1}{1+\zeta/\sqrt{\mu}}$
& $\dfrac{\zeta}{\sqrt{\mu}}$
& $0$
& $0$
& $3\!\left(1+\dfrac{\zeta}{\sqrt{\mu}}\right)
\pm
\sqrt{9\!\left(1+\dfrac{\zeta}{\sqrt{\mu}}\right)^{2}-\dfrac{6}{\mu}}$ \\
$P_{-}^{(\pm)}$
& $\dfrac{1}{1-\zeta/\sqrt{\mu}}$
& $-\dfrac{\zeta}{\sqrt{\mu}}$
& $0$
& $0$
& $3\!\left(1-\dfrac{\zeta}{\sqrt{\mu}}\right)
\pm
\sqrt{9\!\left(1-\dfrac{\zeta}{\sqrt{\mu}}\right)^{2}-\dfrac{6}{\mu}}$ \\
\hline
\end{tabular}
\label{tab:criticos-Reg}
\end{table}

Table~\ref{tab:criticos-Reg} summarizes the regular critical points with $v_{3*}=0$. The physically relevant subset will also depend on the constraints on the parameters $\mu$ and $\zeta$, as well as on the cosmological interpretation of $u_{*}$ and $v_{1*}$.

\textbf{Branch with \texorpdfstring{$\mathcal{R}_{*}=0$}{R=0} and \texorpdfstring{$v_{3*}\neq 0$}{v3≠0}.}
If instead of fixing $v_{3*}=0$ the alternative condition $\mathcal{R}_{*}=0$ is imposed in \eqref{eq:v3-critical-condition}, equation \eqref{R:normalized-R} reduces to
\begin{equation}
\mathcal{R}'\big|_{\mathcal{R}=0}
=
\frac{3}{\mu}
-
\frac{3(\beta-1)\Omega_{0}\,u^{2}v_{3}}{\mu}
=0,
\end{equation}
from which we obtain
\begin{equation}
u_{*}^{2}v_{3*}
=
\frac{1}{(\beta-1)\Omega_{0}},
\qquad
\beta\neq 1.
\label{eq:Rzero-branch}
\end{equation}

In this case, $\mathcal{R}_{*}=0$ describes a regime with constant $G$, and condition \eqref{eq:Rzero-branch} fixes a combination of $u_{*}$ and $v_{3*}$. Together with \eqref{eq:background-critical}, this defines a continuous branch of critical points, which can be written, for each admissible $v_{1*}$, as
\begin{equation}
u_{*}=\frac{1}{1+v_{1*}},\qquad
v_{2*}=0,\qquad
\mathcal{R}_{*}=0,\qquad
v_{3*}=\frac{(1+v_{1*})^{2}}{(\beta-1)\Omega_{0}},
\end{equation}
under the additional condition $1+v_{1*}>0$.

This branch is relevant for the discussion of asymptotic regimes with $\dot{G}=0$ and will be analyzed in more detail when studying the linear stability and cosmological interpretation of the critical points.

The above analysis provides the explicit location of the regular critical points of system \eqref{Aut-R}. Linear stability is obtained by evaluating the Jacobian matrix at each point and depends, in particular, on the sign of $\mathcal{R}_{*}$ and on the \emph{slow--fast} structure associated with $u_{*}$.

In physical terms:
\begin{itemize}
    \item points with $\mathcal{R}_{*}>0$ tend to reinforce the effective gravity and can be associated with regimes of enhanced structure growth;
    \item points with $\mathcal{R}_{*}<0$ correspond to an effective decrease of $G$ and are natural candidates to alleviate the $S_{8}$ tension;
    \item the branch $\mathcal{R}_{*}=0$ describes effective states with constant $G$, whose stability will determine whether the model admits asymptotic phases compatible with standard cosmology.
\end{itemize}

\section{\textit{Slow–fast} structure and desingularization of the crossing at $u=0$}
\label{Sect:5.4}
In our model, the variable $ u = {H_0}/{H}$ provides a natural normalization of the effective Hubble parameter. This choice has a direct physical interpretation: $u>0$ corresponds to expansion ($H>0$), $u<0$ implies contraction ($H<0$), and the condition $H\to 0$ (bounce or recollapse) implies $|u|\to\infty$. The crossing at $u=0$ therefore represents the crossing at $H\to\infty$, that is, an ultra-rapid regime in which expansion or contraction dominates over fractional corrections. In this sense, $u=0$ is not a special physical point, but rather a kinematic limit in which the normalization becomes infinite.

However, in the autonomous system \eqref{Aut-R}, the variable $u$ multiplies the derivatives of $v_1$ and $v_2$, introducing a differentiated time-scale structure. In particular, equations \eqref{Aut:v1-R}–\eqref{Aut:v2-R} contain a factor $u$, while the equations for $v_3$ and $\mathcal{R}$, \eqref{Aut:v3-R}–\eqref{R:normalized-R}, do not. This suggests interpreting $u$ as a parameter that controls the separation of scales: when $|u|\ll 1$ (that is, $|H|\gg H_0$), the variables $(v_1,v_2)$ evolve slowly, while $(v_3,\mathcal{R})$ evolve at an order-one rate.

\subsection{\textit{Slow–fast} structure}
\label{Sect:5.4.1}
From \eqref{Aut:v1-R}–\eqref{Aut:v2-R} we have
\begin{equation}
v_1' = u\,v_2, \qquad
v_2' = u\Bigl[-3(v_1+1)v_2 - \zeta^2 v_1 + \mu v_1^3\Bigr].
\end{equation}
For $|u|\ll 1$, we have $v_1',v_2'=\mathcal{O}(u)$, so that $(v_1,v_2)$ are \emph{slow} variables. In contrast, the equations for $v_3$ and $\mathcal{R}$,
\begin{equation}
v_3' = \beta v_3\,\mathcal{R}, \qquad
\mathcal{R}' = \frac{3}{\mu}
+ \frac{1}{2}\mathcal{R}^2
- 3(v_1+1)\mathcal{R} + \mathcal{O}(u^2),
\end{equation}
do not contain a factor $u$ and remain of order one even when $u\to0$. In particular, the constant term $3/\mu$ in \eqref{R:normalized-R} implies that $\mathcal{R}$ follows its own dynamics, \emph{time-scale decoupled} from the $(u,v_1,v_2)$ sector in the regime $|u|\ll 1$.

The equation for $u$,
\begin{equation}
u' = -u\!\left[
v_2 u^2
-\lambda\bigl(1-(1+v_1)u\bigr)^2
\right],
\label{eq:u-explicit}
\end{equation}
shows that $u$ multiplies the effective field. In particular, $u=0$ is an invariant hypersurface: if $u(n_0)=0$, then $u'(n_0)=0$ and $u(N)\equiv 0$. Moreover, if $u(n_0)\neq 0$, one can write
\begin{equation}
\frac{u'}{u}
= -\left[
v_2 u^2
-\lambda\bigl(1-(1+v_1)u\bigr)^2
\right],
\end{equation}
from which it follows that $u(N)$ preserves its sign as long as the solution exists. The original system does not allow crossing $u=0$, and the dynamics on the half-axes.

From a physical point of view, this means that the normalization $u=H_0/H$ imposes an artificial barrier: although $H$ can change sign in a cosmological bounce, the variable $u$ cannot cross $0$ within the autonomous system as formulated. To study the dynamics near $u=0$—which corresponds to an ultra-rapid expansion or contraction regime—it is necessary to desingularize the system.

\subsection{Problem when dividing by $u$ and the role of $\mathcal{R}$}
\label{Sect:5.4.2}
A natural strategy to desingularize \eqref{eq:u-explicit} is to introduce
a new parameter $\lambda$ through
\begin{equation}
\frac{d}{d\lambda} = \frac{1}{u}\frac{d}{dN},
\label{eq:lambda-local}
\end{equation}
so that
\begin{equation}
\frac{du}{d\lambda}
= \frac{1}{u}u'
= -\left[
v_2 u^2
-\lambda\bigl(1-(1+v_1)u\bigr)^2
\right],
\end{equation}
which is regular at $u=0$. However, this reparametrization presents two
important limitations:

\begin{itemize}
\item[(i)] \textbf{Sign of $u$.} The definition \eqref{eq:lambda-local} is only
locally valid on each half-axis ($u>0$ or $u<0$), since the factor
$1/u$ changes sign when $u$ changes sign. Therefore, it does not define a
global parameter and must be interpreted as a \emph{local} desingularization
in charts $u>0$ and $u<0$ separately.

\item[(ii)] \textbf{Equation for $\mathcal{R}$.} Equation
\eqref{R:normalized-R} contains a constant term $3/\mu$ and is not of the form
$u\times(\text{expression})$. If \eqref{eq:lambda-local} were applied to
$\mathcal{R}$, terms of the form $(3/\mu)/u$ would appear, singularly defined
at $u=0$. This reflects that $\mathcal{R}$ maintains its own time scale and
should not be desingularized by dividing by $u$.
\end{itemize}

Consequently, desingularization via \eqref{eq:lambda-local} must be understood
as a \emph{partial} transformation, restricted to the slow sector
$(u,v_1,v_2)$ and applied locally in each domain of sign of $u$.

\subsubsection{Partial desingularization of the sector $(u,v_1,v_2)$}
\label{Sect:5.4.3}
Applying
\begin{equation}
\frac{dx}{d\lambda}=\frac{1}{u}\frac{dx}{dN},
\end{equation}
to equations \eqref{Aut:u-R}–\eqref{Aut:v2-R}, we obtain the desingularized
system in the subspace $(u,v_1,v_2)$:
\begin{subequations}
\label{Aut-R-desing-partial}
\begin{align}
\frac{du}{d\lambda}
&= -\left[
v_2 u^2
-\lambda\bigl(1-(1+v_1)u\bigr)^2
\right],
\label{Aut:u-desing}
\\
\frac{dv_1}{d\lambda}
&= v_2,
\label{Aut:v1-desing}
\\
\frac{dv_2}{d\lambda}
&= -3(v_1+1)v_2 - \zeta^2 v_1 + \mu v_1^3.
\label{Aut:v2-desing}
\end{align}
\end{subequations}
Equation \eqref{Aut:u-desing} is regular at $u=0$, and
\begin{equation}
\left.\frac{du}{d\lambda}\right|_{u=0}
= \lambda > 0,
\end{equation}
so that, in each chart $u>0$ or $u<0$, the vector field points transversely to the hypersurface $u=0$ in the direction of $u>0$. The system \eqref{Aut-R-desing-partial} regularly describes the geometry of the flow in the sector $(u,v_1,v_2)$ and allows the crossing through $u=0$ to be analyzed as a well-defined geometric phenomenon, even though $u$ may take negative values in the complete model.

In this local analysis, $\mathcal{R}$ and $v_3$ are considered as fast variables that evolve according to their original equations in $n$ and act as external parameters (or forcings) on the reduced system \eqref{Aut-R-desing-partial}. This approximation captures the essential \textit{slow–fast} structure of the fractional model near $u=0$ without
introducing artificial singularities in the equation for $\mathcal{R}$.

\subsection{Alternative reparametrization using $T=\frac{1}{1+u}$ and slow–fast analysis}
\label{Sect:5.5}
In the autonomous system \eqref{Aut-R}, the variable $u=\frac{H_0}{H}$ can take positive values (expansion, $H>0$), negative values (contraction, $H<0$), and even diverge when $H\to 0$ (bounce or recollapse). This wide domain suggests seeking a reparametrization that allows compactification of dynamically relevant regions and, at the same time, reorganization of the slow–fast structure of the system.

A natural alternative is to introduce the variable
\begin{equation}
T=\frac{1}{1+u},\qquad u=\frac{1-T}{T},
\label{eq:T-def}
\end{equation}
valid whenever $1+u\neq 0$, that is, as long as $H\neq -H_0$. The change \eqref{eq:T-def} compactifies the infinity $|u|\to\infty$ at the boundary $T\to 0$ and shifts the region $u\simeq 0$ towards $T\simeq 1$. The sign of $T$ coincides
with that of $1+u$, so that the chart $T>0$ describes the region $H>-H_0$ and the chart $T<0$ the region $H<-H_0$. The dynamics can be analyzed locally in each chart without loss of generality.

\subsubsection{Rescaled derivative}
\label{Sect:5.5.1}
We define the rescaled derivative
\begin{equation}
\mathcal{D}X \equiv T\,X',
\label{eq:D-def}
\end{equation}
which reparametrizes the orbits without altering their geometry. When $T>0$, $\mathcal{D}$ has the same orientation as the original derivative; when $T<0$, the orientation is reversed, which does not affect the qualitative analysis.

Substituting $u=(1-T)/T$ into \eqref{Aut-R} and applying \eqref{eq:D-def}, we obtain a closed system in $(T,v_1,v_2,v_3,\mathcal{R})$.

\subsubsection{Rescaled system in $T$}
\label{Sect:5.5.2}
The resulting system is
\begin{subequations}
\label{Global-R-T-rescaled}
\begin{align}
\mathcal{D}T
&= (1-T)\left[(1-T)^2 v_2
-\lambda\big((2+v_1)T-(1+v_1)\big)^2\right],
\label{Global:T-evolution-rescaled}
\\
\mathcal{D}v_1 &= (1-T)\,v_2,
\label{Global:v1-T-rescaled}
\\
\mathcal{D}v_2 &= (1-T)\Big[-3(v_1+1)v_2 - \zeta^2 v_1 + \mu v_1^3\Big],
\label{Global:v2-T-rescaled}
\\
\mathcal{D}v_3 &= \beta\,T\,v_3\,\mathcal{R},
\label{Global:v3-T-rescaled}
\\
T\mathcal{D}\mathcal{R}
&= T^2\!\left(\frac{3}{\mu} + \frac{1}{2}\mathcal{R}^2 - 3(v_1+1)\mathcal{R}\right)
\nonumber\\
&\quad
+\left[
\lambda\, \mathcal{R}\big((2+v_1)T-(1+v_1)\big)^2
-(1-T)^2\Big(\mathcal{R}v_2 + \tfrac{3(\beta-1)\Omega_0}{\mu}v_3\Big)
\right].
\label{GlobalR:normalized-T-rescaled}
\end{align}
\end{subequations}

The curve
\begin{equation}
(2+v_1)T-(1+v_1)=0
\quad\Longleftrightarrow\quad
T=\frac{1+v_1}{2+v_1}=0,
\end{equation}
is not a singularity of the system, but rather the locus where the closure term inherited from $1-(1+v_1)u$ vanishes.

\subsubsection{Slow–fast structure of the rescaled system}
\label{Sect:5.5.3}
System \eqref{Global-R-T-rescaled} exhibits three distinct dynamical regimes, determined by the value of $T$.

\textbf{Regime I: $T\simeq 1$ (equivalent to $u\simeq 0$).}

Here $(1-T)\ll 1$, so that
\begin{equation}
\mathcal{D}T,\;\mathcal{D}v_1,\;\mathcal{D}v_2=\mathcal{O}(1-T),
\end{equation}
and the variables $(T,v_1,v_2)$ are \emph{slow}. In contrast, the equations for $v_3$ and $\mathcal{R}$ contain order-one terms:
\begin{equation}
\mathcal{D}v_3=\beta v_3\mathcal{R},\qquad
\mathcal{D}\mathcal{R}=\frac{3}{\mu}+\tfrac{1}{2}\mathcal{R}^2
-3(v_1+1)\mathcal{R}+\mathcal{O}(1-T),
\end{equation}
so that $(v_3,\mathcal{R})$ are \emph{fast}. The slow flow, with $ds=(1-T)\,d\tau$, becomes
\begin{equation}
\frac{dT}{ds}=(1-T)^2 v_2
-\lambda\big((2+v_1)T-(1+v_1)\big)^2,
\end{equation}
\begin{equation}
\frac{dv_1}{ds}=v_2,\qquad
\frac{dv_2}{ds}=-3(v_1+1)v_2-\zeta^2 v_1+\mu v_1^3.
\end{equation}

\textbf{Regime II: $T\to 0$ (equivalent to $|u|\to\infty$, bounce $H\to 0$).}

At this boundary,
\begin{equation}
\mathcal{D}v_3=\beta T v_3\mathcal{R}\to 0,
\end{equation}
so that $v_3$ becomes slow. For $\mathcal{R}$, the dominant balance in \eqref{GlobalR:normalized-T-rescaled} is
\begin{equation}
T\mathcal{D}\mathcal{R}\approx
\lambda\, \mathcal{R}(1+v_1)^2
-\Big(\mathcal{R}v_2+\tfrac{3(\beta-1)\Omega_0}{\mu}v_3\Big),
\end{equation}
which algebraically fixes the value of $\mathcal{R}$ in terms of $(v_1,v_2,v_3)$. The system here exhibits an inversion of hierarchy: the originally fast variables quickly adjust to a critical manifold.

\textbf{Regime III: $T\to\infty$ (equivalent to $u\to -1$).}

For $T\gg 1$, the dominant closure in \eqref{Global:T-evolution-rescaled} is
\begin{equation}
\mathcal{D}T\approx (1-T)\left[(1-T)^2 v_2
-\lambda(2+v_1)^2 T^2\right],
\end{equation}
which forces $\mathcal{D}T<0$ and pushes the orbits towards finite values of $T$.
In the equation for $\mathcal{R}$,
\begin{equation}
T\mathcal{D}\mathcal{R}\sim T^2\left(\frac{3}{\mu}
+\tfrac{1}{2}\mathcal{R}^2-3(v_1+1)\mathcal{R}\right)
+\lambda\, \mathcal{R}(2+v_1)^2 T^2,
\end{equation}
which implies a rapid adjustment dominated by terms $\propto T^2$.

The change of variable $T=1/(1+u)$ produces a closed and regular system in the regions $T\to 0$ and $T\to\infty$, compactifying the bounce $H\to 0$ and separating the region $u\simeq 0$ around $T\simeq 1$. The resulting slow–fast
structure presents three well-differentiated regimes, with smooth transitions between them. The fast variables $(v_3,\mathcal{R})$ adjust at both boundaries, while $(T,v_1,v_2)$ organize the global dynamics of the system.

\subsection{Alternative reparametrization using $w = u^2$ and slow–fast analysis}

Desingularizations based on dividing by $u$ or on the change $T=1/(1+u)$ allow the study of dynamics near $u=0$ or compactify regions of phase space, but both present limitations: the first is local on each half-axis $u>0$ and $u<0$, while the second does not cover the surface $u=-1$ and changes the time orientation when $1+u$ changes sign.
A conceptually more robust alternative consists of completely removing the sign of $u$ through the quadratic variable
\begin{equation}
w = u^2,\qquad u=\sigma\sqrt{w},\qquad \sigma=\mathrm{sign}(u)\in\{-1,+1\}.
\label{eq:w-def}
\end{equation}

This transformation separates the modulus $|u|$ from the sign $\sigma$, allowing the crossing through $u=0$ to be described as a crossing through $w=0$ with a regular vector field. The discrete variable $\sigma$ only changes when $w=0$, which naturally represents the change of sign of the effective Hubble parameter $H$ in a cosmological bounce.

\subsubsection{Physical motivation} Recall that $u=\frac{H_0}{H}$,
so that $u>0$ corresponds to expansion ($H>0$), $u<0$ to contraction ($H<0$), and $u\to\pm\infty$ to a bounce or recollapse ($H\to 0$). The crossing through $u=0$ represents a regime of ultra-rapid expansion or contraction ($|H|\to\infty$) and does not have a singular physical meaning. However, in the autonomous system \eqref{Aut-R}, the equation for $u$ contains a multiplicative factor $u$, which prevents $u$ from changing sign and artificially separates the dynamics on the half-axes $u>0$ and $u<0$.

The variable $w=u^2$ removes this artificial barrier: the crossing through $u=0$ is represented as a smooth crossing through $w=0$, while the physical sign of $H$ is encoded in the discrete variable $\sigma$.

\subsubsection{Autonomous system in $(w,\sigma,v_1,v_2,v_3,\mathcal{R})$} Starting from the equation for $u$,
\begin{equation}
u' = -u\!\left[
v_2 u^2
-\lambda\bigl(1-(1+v_1)u\bigr)^2
\right],
\end{equation}
we obtain
\begin{equation}
w' = 2u\,u'
= -2w\left[
v_2 u^2
-\lambda\bigl(1-(1+v_1)u\bigr)^2
\right].
\end{equation}
Substituting $u=\sigma\sqrt{w}$, the system becomes
\begin{small}
\begin{subequations}
\label{Aut-w-system}
\begin{align}
w' &= -2w\left[
v_2 w
-\lambda\bigl(1-(1+v_1)\sigma\sqrt{w}\bigr)^2
\right],\\
v_1' &= \sigma\sqrt{w}\,v_2,\\
v_2' &= \sigma\sqrt{w}\Bigl[-3(v_1+1)v_2 - \zeta^2 v_1 + \mu v_1^3\Bigr],\\
v_3' &= \beta v_3\,\mathcal{R},\\
\mathcal{R}' &= \frac{3}{\mu}
+ \frac{1}{2}\mathcal{R}^2
-\mathcal{R}\left[3(v_1+1)+ v_2 w - \lambda\bigl(1-(1+v_1)\sigma\sqrt{w}\bigr)^2\right] - \frac{3(\beta-1)\Omega_0\,w v_3}{\mu}.
\end{align}
\end{subequations}
\end{small}
The variable $\sigma$ satisfies
\begin{equation}
\sigma' = 0\quad\text{while } w>0,
\qquad
\sigma\ \text{changes sign only when } w=0.
\end{equation}

\subsubsection{Regularity at $w=0$} The system \eqref{Aut-w-system} is regular at $w=0$:

\begin{itemize}
\item The equation for $w$ contains a factor $w$ that ensures $w'\to 0$ when $w\to 0$.
\item The equations for $v_1$ and $v_2$ contain factors $\sqrt{w}$, so that $v_1',v_2'\to 0$ when $w\to 0$.
\item The equations for $v_3$ and $\mathcal{R}$ do not contain divisions by $u$ or by $1+u$, and remain regular.
\end{itemize}

In particular,
\begin{equation}
w'=0,\qquad v_1'=0,\qquad v_2'=0
\quad\text{when } w=0,
\end{equation}
which shows that $w=0$ is a critical manifold of the system, and the crossing through $u=0$ is represented as a smooth crossing through $w=0$.

\subsubsection{Slow–fast structure in the variable $w$} The slow–fast structure is naturally reorganized:

\begin{itemize}
\item For $w\ll 1$ (equivalent to $|u|\ll 1$), the variables $(v_1,v_2)$ are slow, since $v_1',v_2'=\mathcal{O}(\sqrt{w})$.
\item The variables $(v_3,\mathcal{R})$ remain fast, with dynamics dominated by terms independent of $w$.
\item The variable $w$ is slow, since $w'=\mathcal{O}(w)$.
\end{itemize}

The critical manifold $w=0$ describes the regime of ultra-rapid expansion or contraction ($|H|\to\infty$), where the slow variables freeze, and the fast variables adjust to their internal dynamics.
 
The crossing through $u=0$ corresponds to a crossing through $w=0$ with a regular vector field. Physically, this represents passage through a regime of extremely rapid expansion or contraction, in which $|H|\gg H_0$. The sign of $H$ is encoded in $\sigma$, which changes only when $w=0$, naturally reproducing the change of sign of the Hubble parameter in a cosmological bounce.

\subsubsection{Advantages and limitations}

The desingularization via $w=u^2$ presents several advantages:
\begin{itemize}
\item[(i)] It is global in $u\in\mathbb{R}$ and does not require dividing by $u$ or by $1+u$.
\item[(ii)] It keeps the equation for $\mathcal{R}$ in its original form, without introducing singular terms.
\item[(iii)] The crossing through $u=0$ is represented as a smooth crossing through $w=0$.
\item[(iv)] The slow–fast structure is naturally preserved.
\end{itemize}

Its only disadvantage is the introduction of the discrete variable $\sigma$, but this is standard in blow-up techniques and does not affect the geometric structure of the system.

The transformation $w=u^2$ provides a global, regular, and physically interpretable desingularization of the autonomous system \eqref{Aut-R}. It allows the study of the crossing through $u=0$ without introducing artificial singularities and preserves the slow–fast structure of the fractional model.

\subsection{Comparison of desingularizations and representation of the slow manifold}
\label{Sect:5.6}
The autonomous system \eqref{Aut-R} presents a \textit{slow–fast} structure induced by the multiplicative factor $u$ in the equations for $(v_1,v_2)$.
This structure becomes degenerate on the hypersurface $u=0$, where the slow variables freeze and the fast ones $(v_3,\mathcal{R})$ maintain order-one dynamics. To study the geometry of the flow near $u=0$, as well as the global dynamics associated with expansion, contraction, and cosmological bounces, we consider three alternative desingularizations:
(i) a local reparametrization in $u$, (ii) a global reparametrization in $T=1/(1+u)$, and (iii) a quadratic transformation $w=u^2$. Each one regularizes a different aspect of the system and represents the slow manifold in a
different way.

\subsubsection{Local desingularization via $\lambda$}
\label{Sect:5.6.1}
The transformation
\begin{equation}
\frac{d}{d\lambda}=\frac{1}{u}\frac{d}{dN}
\end{equation}
removes the multiplicative factor $u$ in the equation for $u$ and regularizes the slow sector $(u,v_1,v_2)$ in a neighborhood of $u=0$. In this chart, the slow manifold appears as
\begin{equation}
\mathcal{S}_0^{(\lambda)}=\{u=0\},
\end{equation}
where $v_1'=v_2'=0$ and the transversal flow is described by a regular equation in $\lambda$. This desingularization is suitable for studying the transversality of the flow at $u=0$, but it is \emph{local}: it is only valid on each half-axis $u>0$ or $u<0$, and cannot be applied to the equation for $\mathcal{R}$, which contains terms independent of $u$.

\subsubsection{Global reparametrization via $T=\frac{1}{1+u}$}
\label{Sect:5.6.2}
The transformation
\begin{equation}
T=\frac{1}{1+u},\qquad u=\frac{1-T}{T},
\end{equation}
compactifies the infinity $|u|\to\infty$ at the boundary $T\to 0$ and shifts the region $u\simeq 0$ towards $T\simeq 1$. With the rescaled derivative $\mathcal{D}X=T X'$, the slow equations contain a factor $(1-T)$, so that the slow manifold is represented as
\begin{equation}
\mathcal{S}_0^{(T)}=\{T=1\}.
\end{equation}
This chart is global in the region $H\neq -H_0$ and allows a uniform description of the dynamics near the cosmological bounce $H\to 0$ ($T\to 0$). The equation for $\mathcal{R}$ remains regular in $T$, which facilitates global analysis.
However, the chart does not cover the surface $u=-1$ ($T=\infty$), and the time orientation changes when $1+u$ changes sign.

\subsubsection{Global desingularization via $w=u^2$}
\label{Sect:5.6.3}
The quadratic transformation
\begin{equation}
w=u^2,\qquad u=\sigma\sqrt{w},\qquad \sigma=\mathrm{sign}(u),
\end{equation}
completely removes the sign of $u$ and turns the crossing through $u=0$ into a smooth crossing through $w=0$. The equation for $w$ contains a factor $w$, while the equations for $(v_1,v_2)$ contain factors $\sqrt{w}$, so that the slow
manifold is represented as
\begin{equation}
\mathcal{S}_0^{(w)}=\{w=0\}.
\end{equation}
This desingularization is global in $u\in\mathbb{R}$, does not require dividing by $u$ or by $1+u$, and keeps the equation for $\mathcal{R}$ in its original form. The discrete variable $\sigma$ encodes the physical sign of $H$ and only
changes when $w=0$, naturally reproducing the change of sign of the Hubble parameter in a cosmological bounce.

\subsection{Comparative synthesis}
\label{Sect:5.6.4}
The three desingularizations represent the same slow manifold—the regime $|H|\gg H_0$ where the slow variables freeze—but they do so in different charts:
\begin{itemize}
\item The transformation in $\lambda$ is \emph{local} and geometrically direct:
\begin{equation}
\mathcal{S}_0^{(\lambda)}=\{u=0\}.
\end{equation}

\item The transformation in $T$ is \emph{global} and compactifying:
\begin{equation}
\mathcal{S}_0^{(T)}=\{T=1\}.
\end{equation}

\item The transformation in $w=u^2$ is \emph{global and regular} without singular divisions:
\begin{equation}
\mathcal{S}_0^{(w)}=\{w=0\}.
\end{equation}
\end{itemize}

Each representation is useful for a different purpose: local analysis of transversality ($\lambda$), global dynamics and bounces ($T$), or complete regularity of the system without orientation inversions ($w$). Taken together, these three charts provide a complete and complementary description of the slow–fast structure of the fractional system.

\section{Variation of \texorpdfstring{$G$}{G} and growth of perturbations}
\label{Sect:5.7}

\subsection{Fractional evolution of the density} 

We start from the barotropic equation of state
\begin{equation}
p = \gamma \rho,
\end{equation}
where $\gamma$ characterizes the type of fluid ($\gamma=0$ for cold matter, $\gamma=1/3$ for radiation, etc.).

In the fractional framework, the energy conservation equation is modified by the presence of the exponent $\alpha$, which leads to a matter density evolution of the form
\begin{equation}
\rho(t) = \rho_0\,a^{-3(1+\gamma)}\,t^{-(\alpha-1)(1+\gamma)},
\label{eq:rho-frac}
\end{equation}
where $\rho_0$ is the initial density.   
The first factor $a^{-3(1+\gamma)}$ corresponds to the standard dilution due to cosmic expansion, while the second factor
$t^{-(\alpha-1)(1+\gamma)}$ introduces a fractional time correction that modifies the dilution rate. In particular:
\begin{itemize}
    \item If $\alpha=1$, the standard behavior of general relativity is recovered.
    \item If $\alpha<1$, the density decays faster with time, reinforcing dilution.
    \item If $\alpha>1$, the density decays more slowly, prolonging the contribution of matter.
\end{itemize}

To analyze the evolution of the matter density fraction
\begin{equation}
\Omega_m(N) \equiv \frac{\rho_m}{\rho_{\rm crit}}, \qquad N \equiv \ln a,
\end{equation}
it is useful to derive its autonomous equation. Using the definition of $\Omega_m$ and effective energy conservation, we obtain
\begin{equation}
\Omega_m'(N) 
= -3\,\Omega_m(N)\,\bigl(1-\Omega_m(N)\bigr)\,\bigl(1+w_{\rm eff}(N)\bigr),
\label{eq:Omega-prime}
\end{equation}
where $w_{\rm eff}(N)$ is the effective equation-of-state parameter of the cosmological background. This expression shows that the variation of $\Omega_m$ depends both on the fraction of matter present and on the remaining energy
content ($1-\Omega_m$), modulated by the effective equation of state.

Taken together, equations \eqref{eq:rho-frac} and \eqref{eq:Omega-prime} describe how fractional corrections alter the dilution of matter and the evolution of its relative fraction in the universe. These modifications directly impact the growth of perturbations and cosmological observables such as $f\sigma_8$.

\subsection{Effective equation-of-state parameter} 

The effective equation-of-state parameter of the cosmological background is denoted by $w_{\mathrm{eff}}(N)$ and is introduced to condense into a single function how fractional corrections modify the dilution of matter and the dynamics of the total energy.

In physical terms:
\begin{itemize}
    \item In standard cosmology, the equation-of-state parameter $w$ is defined as the ratio between pressure and energy density,
    \begin{equation}
    w = \frac{p}{\rho}.
    \end{equation}
    \item In the fractional formalism, $w_{\mathrm{eff}}(N)$ plays the same role but applied to the effective fluid that results from fractional corrections.
    \item Therefore, it appears in the autonomous equation for $\Omega_m(N)$, indicating that the variation of the matter fraction depends on how much the rest of the universe’s energy content “weighs,” modulated by the effective equation of state.
\end{itemize}

In practice, $w_{\mathrm{eff}}(N)$ can be deduced from the dynamics of $H(N)$ and its derivatives, since it is directly related to the evolution of the scale factor. In standard cosmology, one uses the identity
\begin{equation}
w_{\mathrm{eff}}(N) = -1 - \frac{2}{3}\,\frac{H'(N)}{H(N)},
\end{equation}
whereas in the fractional model, this relation is modified by additional terms, such as the effective friction
\begin{equation}
\frac{1-\alpha}{tH},
\end{equation}
and the gravitational source
\begin{equation}
t^{-(\alpha -1)(1+\gamma)}.
\end{equation}

In summary, $w_{\mathrm{eff}}(N)$ is the equation-of-state parameter of the effective cosmological fluid, which incorporates fractional corrections. Operationally, it is obtained from the energy conservation equation together with the evolution of $H(N)$. Physically, it plays a central role by controlling how matter dilutes relative to other components and by determining the growth rate of perturbations.

\subsection{Fractional growth equation in autonomous form}

From the fractional field equations 
\eqref{tag4}, \eqref{tag5}, \eqref{tag6}, and using the transformations
\begin{equation}
\frac{d}{dt}=H\frac{d}{dN},\quad
3H+\frac{1-\alpha}{t}=H\!\left(3+\frac{1-\alpha}{tH}\right),
\end{equation}
together with the definition of the effective gravity
\begin{equation}
G_{\rm eff}(N)=G_0\,e^{\int \mathcal{R}\,dN},
\end{equation}
the linear growth equation in the sub–Hubble regime takes the autonomous form
\begin{equation}
\delta'' 
+ \left(2+\frac{H'}{H}+\frac{1-\alpha}{tH}\right)\delta'
- \frac{3}{2}\,G_{\rm eff}(N)\,\Omega_m(N)\,
t^{-(\alpha-1)(1+\gamma)}\,\delta = 0.
\label{eq:delta-frac}
\end{equation}

Fractional effects manifest themselves in the modified effective friction $(1-\alpha)/(tH)$ and in the corrected gravitational source $t^{-(\alpha-1)(1+\gamma)}$: 
\begin{itemize}
    \item \textbf{Cosmological friction:} $(1-\alpha)/(tH)$ increases damping if $\alpha<1$ and reduces it if $\alpha>1$.
    \item \textbf{Gravitational source:} $t^{-(\alpha-1)(1+\gamma)}$ strengthens or weakens the matter contribution depending on the sign of $\alpha-1$.
\end{itemize}

The sign of $\mathcal{R}$ determines how the effective gravity evolves and, therefore, the growth of structure. If $\mathcal{R}<0$, effective gravity decreases and the growth of perturbations is weakened. If $\mathcal{R}>0$, effective gravity increases and perturbations grow more rapidly.

Fractional terms can amplify or attenuate this effect, modifying the prediction of $f\sigma_8$ and, therefore, the extent to which the model can alleviate the $S_8$ tension.

\subsubsection{Growth factor and observable \texorpdfstring{$f\sigma_8$}{f sigma8}}
\label{Sect:5.7.1}

We define the logarithmic growth rate as
\begin{equation}
f \equiv \frac{d\ln\delta}{d\ln a}
= \frac{1}{\delta}\frac{d\delta}{dN},
\qquad N \equiv \ln a .
\label{eq:def-f}
\end{equation}

From this, we obtain
\begin{equation}
\delta' = f\,\delta, \qquad
\delta'' = f'\delta + f^2\delta, 
\label{eq:delta-primes}
\end{equation}
where the prime denotes derivative with respect to $N$.

We assume that the fractional growth equation for $\delta$ is given by \eqref{eq:delta-frac}. 
Substituting the relations \eqref{eq:delta-primes} into \eqref{eq:delta-frac} we obtain
\begin{equation}
\left(f'\delta + f^2\delta\right)
+ \left(2+\frac{H'}{H}+\frac{1-\alpha}{tH}\right)\left(f\delta\right)
- \frac{3}{2}\,G_{\rm eff}\,\Omega_m\,t^{-(\alpha-1)(1+\gamma)}\,\delta = 0.
\end{equation}

Dividing by $\delta$, we arrive at the autonomous equation for the growth rate:
\begin{equation}
f' + f^2
+ \left(2+\frac{H'}{H}+\frac{1-\alpha}{tH}\right)f
- \frac{3}{2}\,G_{\rm eff}\,\Omega_m\,t^{-(\alpha-1)(1+\gamma)} = 0.
\label{eq:f-frac}
\end{equation}

The relevant observable is
\begin{equation}
F(N) = f(N)\,\sigma_8(N),
\label{eq:def-F}
\end{equation}
where $\sigma_8(N)$ scales with the linear growth factor:
\begin{equation}
\sigma_8' = f\,\sigma_8.
\label{eq:sigma8prime}
\end{equation}

Differentiating \eqref{eq:def-F} we obtain
\begin{equation}
F' = f'\sigma_8 + f\sigma_8' = (f' + f^2)\sigma_8.
\label{eq:Fprime1}
\end{equation}

Using \eqref{eq:f-frac}, the combination $f'+f^2$ can be written as
\begin{equation}
f' + f^2 =
- \left(2+\frac{H'}{H}+\frac{1-\alpha}{tH}\right)f
+ \frac{3}{2}\,G_{\rm eff}\,\Omega_m\,t^{-(\alpha-1)(1+\gamma)}.
\label{eq:fplusf2}
\end{equation}

Inserting \eqref{eq:fplusf2} into \eqref{eq:Fprime1} yields
\begin{equation}
F' =
\left[
- \left(2+\frac{H'}{H}+\frac{1-\alpha}{tH}\right)f
+ \frac{3}{2}\,G_{\rm eff}\,\Omega_m\,t^{-(\alpha-1)(1+\gamma)}
\right]\sigma_8.
\end{equation}

Finally, using $\sigma_8 = F/f$, we obtain the autonomous equation for $F(N)$:
\begin{equation}
F' =
- \left(2+\frac{H'}{H}+\frac{1-\alpha}{tH}\right)F
+ \frac{3}{2}\,G_{\rm eff}\,\Omega_m\,t^{-(\alpha-1)(1+\gamma)}\,\frac{F}{f}.
\label{eq:F-frac}
\end{equation}

\subsubsection{Fractional closure relation: $f \simeq \Omega_m^{\gamma_{\rm frac}}$}
\label{Sect:5.7.2}

In the fractional framework, we generalize the standard approximation by means of an effective growth index $\gamma_{\rm frac}$, which may depend on the fractional parameters (for example, $\alpha$ and $\gamma$):
\begin{equation}
f(N)\simeq\Omega_m(N)^{\gamma_{\rm frac}}.
\label{eq:fgamma-frac-improved}
\end{equation}
The consistency condition requires recovering the GR limit when $\alpha\rightarrow 1$:
\begin{equation}
\gamma_{\rm frac}\longrightarrow\gamma_{\Lambda{\rm CDM}}\quad(\alpha\rightarrow 1).
\end{equation}

Substituting \eqref{eq:fgamma-frac-improved} into the definition $F=f\sigma_8$ and using \eqref{eq:F-frac}, we obtain the closed equation for $F$:
\begin{equation}
F' =
- \left( 2 + \frac{H'}{H} + \frac{1-\alpha}{tH} \right) F
+ \frac{3}{2}\,G_{\rm eff}\,\Omega_m\,t^{-(\alpha-1)(1+\gamma)}\,\sigma_8,
\label{eq:F-frac-closure-improved}
\end{equation}
where it is emphasized that both $H=H(N)$ and $t=t(N)$ are dynamical variables.

The observable $\sigma_8(N)$ is reconstructed from $f$ as
\begin{equation}
\sigma_8(N)=\sigma_{8,0}\,
\exp\!\left(\int_0^N \Omega_m(\tilde N)^{\gamma_{\rm frac}}\,d\tilde N\right),
\label{eq:sigma8-frac-improved}
\end{equation}
with $\sigma_{8,0}$ the present value. In the limit $\alpha\rightarrow 1$ and $G_{\rm eff}\rightarrow 1$ the standard GR expressions are recovered.

\subsubsection{Determination of the fractional growth index}
\label{Sect:5.7.3}

To characterize the growth of perturbations in the fractional framework, we adopt the ansatz
\begin{equation}
\gamma_{\rm frac}(\alpha)=\gamma_{\Lambda{\rm CDM}}+c_1(1-\alpha)+c_2(1-\alpha)^2+\mathcal{O}\!\big((1-\alpha)^3\big),
\end{equation}
and, under the hypothesis of slow variation of $\gamma_{\rm frac}$, we obtain
\begin{equation}
f'=\gamma_{\rm frac}\,\Omega_m^{\gamma_{\rm frac}-1}\,\Omega_m',
\end{equation}
where the prime denotes derivative with respect to $N$.

The consistency condition can be written as
\begin{equation}
\begin{aligned}
&-\tfrac{3}{2}\, G_{\rm eff}\, t^{-(\alpha-1)(1+\gamma)}\,\Omega_m
+ 3\,\gamma_{\rm frac}\,(1+w_{\rm eff})\,(1-\Omega_m)\,\Omega_m^{\gamma_{\rm frac}}
+ \Omega_m^{2\gamma_{\rm frac}}
\\& + \frac{\Omega_m^{\gamma_{\rm frac}}\left(1-\alpha+2tH+tH'\right)}{tH}=0,
\end{aligned}
\end{equation}
or, equivalently,
\begin{equation}
\Omega_m^{\gamma_{\rm frac}}\!\left[\Big(2\gamma_{\rm frac}(1-\Omega_m)+1\Big)\frac{H'}{H}
+2+\frac{1-\alpha}{tH}\right]
+\Omega_m^{2\gamma_{\rm frac}}
-\frac{3}{2}\,G_{\rm eff}\,\Omega_m\,t^{-(\alpha-1)(1+\gamma)}=0.
\end{equation}

Expanding around
\begin{equation}
\Omega_m(N) = 1 + \Delta\Omega_m + \mathcal{O}\!\big((\Delta\Omega_m)^2\big),
\qquad
\alpha = 1 + \Delta_\alpha + \mathcal{O}\!\big((\Delta_\alpha)^2\big),
\end{equation}
we obtain
\begin{align}
&\Bigg[
-\tfrac{3}{2}\,G_{\rm eff} + \frac{H'(N)}{H(N)} + 3
+ \Delta_\alpha\left(
\tfrac{3}{2}(\gamma+1)\,G_{\rm eff}\,\ln t
- \frac{1}{tH(N)}
\right)
+ \mathcal{O}\!\big(\Delta_\alpha^2\big)
\Bigg] \nonumber \\
&\quad + \Delta\Omega_m \Bigg[
-\tfrac{3}{2}\,G_{\rm eff}
+ 4\,\gamma_{\rm frac}
- \frac{\gamma_{\rm frac}\,H'(N)}{H(N)}
+ \left(
\tfrac{3}{2}\,G_{\rm eff}(\gamma+1)\,\ln t
- \frac{\gamma_{\rm frac}}{tH(N)}
\right)\Delta_\alpha
+ \mathcal{O}\!\big(\Delta_\alpha^2\big)
\Bigg] \nonumber \\
&\quad + \mathcal{O}\!\big((\Delta\Omega_m)^2\big).
\end{align}

Collecting the terms of order $\mathcal{O}(\Delta\Omega_m)^0$ and $\mathcal{O}(\Delta\Omega_m)^1$, and discarding higher-order terms in $\Delta_\alpha$, we obtain the system
\begin{align}
-\tfrac{3}{2}\,G_{\rm eff}(N) + \frac{H'(N)}{H(N)} + 3
+ \Delta_\alpha\left(
\tfrac{3}{2}(\gamma+1)\,G_{\rm eff}(N)\,\ln t
- \frac{1}{tH(N)}
\right) &= 0, \\
-\tfrac{3}{2}\,G_{\rm eff}(N)
+ 4\,\gamma_{\rm frac}
- \frac{\gamma_{\rm frac}\,H'(N)}{H(N)}
+ \left(
\tfrac{3}{2}\,G_{\rm eff}(N)(\gamma+1)\,\ln t
- \frac{\gamma_{\rm frac}}{tH(N)}
\right)\Delta_\alpha &= 0.
\end{align}

We define
\begin{equation}
X\equiv\frac{\dot H}{H^2}, \qquad A(t)\equiv tH(t).
\end{equation}
The expansion around $\Omega_m\simeq1$, $\alpha\simeq 1$ leads to the reduced system
\begin{align}
3 - \tfrac{3}{2}\,G_{\rm eff} + X - \dfrac{\Delta_\alpha}{A}
+ \tfrac{3}{2}\,G_{\rm eff}(1+\gamma)\,\Delta_\alpha\,\ln t &= 0, \\
-\tfrac{3}{2}\,G_{\rm eff} + 4\,\gamma_{\rm frac}
- X\,\gamma_{\rm frac}
- \dfrac{\gamma_{\rm frac}\,\Delta_\alpha}{A}
+ \tfrac{3}{2}\,G_{\rm eff}(1+\gamma)\,\Delta_\alpha\,\ln t &= 0.
\end{align}

Solving, we obtain
\begin{align}
\label{eq:XY-sol}
\frac{\dot H}{H^2} &=
\frac{1}{2} 
+ G_{\rm eff}\left(
\frac{3(\gamma_{\rm frac}-1)}{4\,\gamma_{\rm frac}}
- \frac{3(1+\gamma)(\gamma_{\rm frac}-1)\,\Delta\alpha\,\ln t}{4\,\gamma_{\rm frac}}
\right), \\
t H(t) &=
\frac{4\,\gamma_{\rm frac}\,\Delta\alpha}{
14\,\gamma_{\rm frac}
- 3\,G_{\rm eff}(1+\gamma_{\rm frac})
+ 3\,G_{\rm eff}(1+\gamma)(1+\gamma_{\rm frac})\,\Delta\alpha\,\ln t}.
\end{align}

Finally, the evolution of $G_{\rm eff}$ and of $\mathcal{R}$ is described by
\begin{equation}
\dot{G_{\rm eff}}=G_{\rm eff} H\mathcal{R},
\end{equation}
and, around $\Omega_m\simeq1$, $\alpha\simeq 1$,
\begin{equation}
\frac{\dot{\mathcal{R}}}{H} \simeq \frac{3}{\mu}
+ \frac{1}{2}\mathcal{R}^2
- 3\mathcal{R}
+ \lambda\mathcal{R}\left(1-\frac{H_0}{H}\right)^2
- \frac{3(\beta-1)\Omega_0}{\mu}\left(\frac{H_0}{H}\right)^2 G_{\rm eff}^\beta.
\end{equation}

The fractional growth index $\gamma_{\rm frac}$ provides a compact way to encode how deviations from $\alpha=1$ modify the growth of matter perturbations. Its determination requires consistency between the autonomous growth equation, the background dynamics of $H(N)$, and the effective gravity $G_{\rm eff}(N)$. Expansions around $\Omega_m\simeq 1$ and $\alpha\simeq 1$ yield analytic constraints linking $\gamma_{\rm frac}$ to the fractional parameters $(\alpha,\gamma)$ and to the evolution of $\mathcal{R}$. This framework allows one to quantify how fractional corrections alter the standard closure relation $f\simeq \Omega_m^\gamma$ and to assess their impact on cosmological observables such as $f\sigma_8$.

\subsubsection{Asymptotic solutions in terms of $t$}
\label{Sect:5.7.4}
We consider the regime close to matter domination and a small fractional correction:
\begin{equation}
\Omega_m\simeq 1, \qquad \alpha\simeq 1, \qquad |\Delta\alpha|\ll 1,
\end{equation}
with $G_{\rm eff}$ varying slowly (that is, bounded $\mathcal{R}$).
This allows us to work at leading order and first order in $\Delta\alpha$, retaining logarithmic terms in $t$. The hypothesis of slow evolution of $G_{\rm eff}$ is crucial: it ensures that fractional corrections accumulate smoothly, and that integrations can be carried out by treating $G_{\rm eff}$ as approximately constant at each step.

\textbf{Asymptotic solution of $H(t)$ from $\dot H/H^2$.}
We define effective coefficients that depend on $G_{\rm eff}$, which is assumed constant at leading order:
\begin{equation}
A_0 = \tfrac{1}{2} +  G_{\rm eff}\,\frac{3(\gamma_{\rm frac}-1)}{4\,\gamma_{\rm frac}}, 
\qquad
A_1(t) = -\,G_{\rm eff}\,\frac{3(1+\gamma)(\gamma_{\rm frac}-1)}{4\,\gamma_{\rm frac}}\;\Delta\alpha\,\ln t.
\end{equation}

The Hubble evolution equation is written as
\begin{equation}
\frac{\dot H}{H^2} = A_0 + A_1(t),
\end{equation}
which implies
\begin{equation}
\frac{d}{dt}\!\left(\frac{1}{H}\right) = A_0 + A_1(t).
\end{equation}

Integrating asymptotically and retaining logarithmic terms in $t$:
\begin{equation}
\frac{1}{H(t)} \;\simeq\; A_0\,t +  \Delta\alpha\,\tilde A\,\bigl(t\ln t - t\bigr),
\quad
\tilde A = -\,G_{\rm eff}\,\frac{3(1+\gamma)(\gamma_{\rm frac}-1)}{4\,\gamma_{\rm frac}}.
\end{equation}

From this, we obtain
\begin{equation}
H(t) \;\simeq\; \frac{1}{\,t\!\left[A_0 +  \Delta\alpha\,\tilde A\,(\ln t - 1)\right]\,}.
\end{equation}

\textit{Interpretation:} $H(t)$ decays as $1/t$ with logarithmic corrections due to the fractional part. The slow variation of $G_{\rm eff}$ allows these corrections to accumulate smoothly.

\textbf{Asymptotic solution of $H(t)$ from $tH(t)$.}
The same result can be derived from the closed relation for $tH(t)$,
where the denominator separates into a constant term and a logarithmic one:
\begin{equation}
\begin{aligned}
D(t) & = D_0 +  D_1\,\ln t,  \\ 
D_0 &= 14\,\gamma_{\rm frac} - 3\,G_{\rm eff}(1+\gamma_{\rm frac}), \quad  
D_1 = 3\,G_{\rm eff}(1+\gamma)(1+\gamma_{\rm frac})\,\Delta\alpha.
\end{aligned}
\end{equation}

The relation
\begin{equation}
t H(t) = \frac{4\,\gamma_{\rm frac}\,\Delta\alpha}{D(t)}
\end{equation}
produces
\begin{equation}
H(t) \;\simeq\; \frac{4\,\gamma_{\rm frac}\,\Delta\alpha}{\,t\,[D_0 + D_1\ln t]\,}.
\end{equation}

\textit{Interpretation:} Both expressions for $H(t)$ are consistent if the coefficients are identified at the same order in $\Delta\alpha$, taking $G_{\rm eff}$ as quasi-constant due to its slow evolution.

\textbf{Asymptotic evolution of $G_{\rm eff}(t)$.} The variation of $G_{\rm eff}$ is governed by
\begin{equation}
\dot{G}_{\rm eff} = G_{\rm eff}\,H\,\mathcal{R}.
\end{equation}

If $\mathcal{R}\approx \mathcal{R}_0$ on the asymptotic scale (or varies slowly), integration gives
\begin{equation}
\ln\!\frac{G_{\rm eff}(t)}{G_{\rm eff}(t_0)} \;\simeq\; \mathcal{R}_0 \int_{t_0}^{t} H(\tilde t)\,d\tilde t.
\end{equation}

Using the asymptotic form of $H(t)$:
\begin{equation}
H(t)\simeq \dfrac{C}{t\,[D_0 + D_1\ln t]}, \quad C=4\,\gamma_{\rm frac}\,\Delta\alpha,
\end{equation}
we obtain
\begin{equation}
\int \frac{dt}{t\,[D_0 + D_1\ln t]} = \frac{1}{D_1}\,\ln\!\bigl(D_0 + D_1\ln t\bigr) + \text{const},
\end{equation}
and therefore
\begin{equation}
G_{\rm eff}(t) \;\simeq\; G_{\rm eff}(t_0)\,\left[\frac{D_0 + D_1\ln t}{D_0 + D_1\ln t_0}\right]^{\frac{\mathcal{R}_0\,C}{D_1}}.
\end{equation}

\textit{Interpretation:} $G_{\rm eff}(t)$ follows a slow power law in terms of $\ln t$.
This means its variation is very smooth, which justifies treating it as constant in the previous steps. If $\mathcal{R}_0<0$, $G_{\rm eff}$ decreases; if $\mathcal{R}_0>0$, it increases.

\textbf{Asymptotic behavior of $\mathcal{R}(t)$.} The dynamics of $\mathcal{R}(t)$ are crucial because they directly control the variation of $G_{\rm eff}(t)$ through the relation $\dot G_{\rm eff} = G_{\rm eff} H \mathcal{R}$.
Therefore, understanding its fixed points and stability allows us to anticipate whether effective gravity stabilizes or deviates significantly from the GR limit.

The general evolution equation is
\begin{equation}
\frac{\dot{\mathcal{R}}}{H} \;\simeq\; \frac{3}{\mu} + \frac{1}{2}\mathcal{R}^2 - 3\mathcal{R} 
+ \lambda\,\mathcal{R}\left(1-\frac{H_0}{H}\right)^2 
- \frac{3(\beta-1)\Omega_0}{\mu}\left(\frac{H_0}{H}\right)^2 G_{\rm eff}^{\beta}.
\end{equation}

\noindent
\textbf{Assumption $H\gg H_0$ and slow $G_{\rm eff}(t)$:} in the early regime, the terms proportional to $(H_0/H)^2$ are suppressed and $\left(1-\frac{H_0}{H}\right)^2 \to 1$. This simplifies the equation
\begin{equation}
\frac{d\mathcal{R}}{d N}=\frac{\dot{\mathcal{R}}}{H} \;\approx\; \frac{3}{\mu} + (\lambda - 3)\,\mathcal{R} + \tfrac{1}{2}\,\mathcal{R}^2.
\label{eq:parabola-lambda}
\end{equation}

\begin{figure}[htb]
    \centering
    \includegraphics[width=1\textwidth]{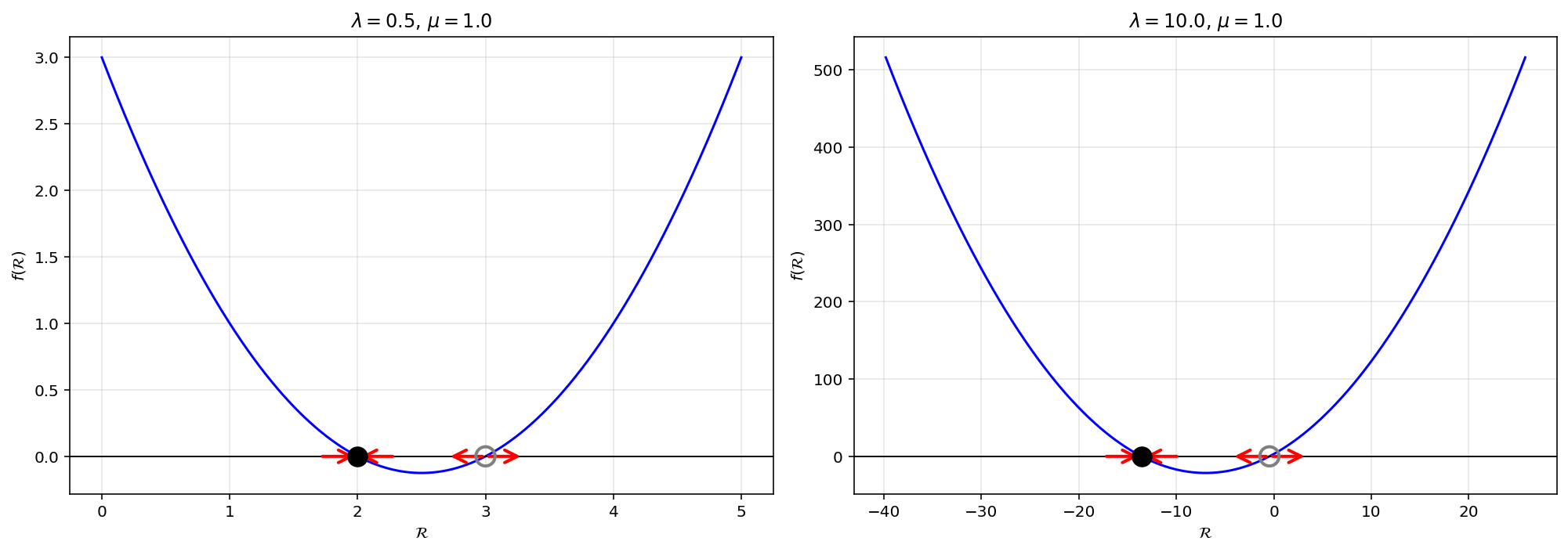}
    \caption{Phase portraits of equation \eqref{eq:parabola-lambda} for $\mu = 1, \lambda = 0.5$ and $\mu=1, \lambda = 10$.}
    \label{fig:flujo-parábola-lambda}
\end{figure}

This reduced form shows that the competition between the constant term $3/\mu$, the linear term $(\lambda-3)\mathcal{R}$, and the quadratic term $\mathcal{R}^2/2$ determines the existence and stability of the fixed points. In Figure \ref{fig:flujo-parábola-lambda}, the one-dimensional flow of equation \eqref{eq:parabola-lambda} is compared for values $\lambda \ll 1$ and $\lambda \gg 1$.

\noindent
\textbf{Fixed points:} obtained by solving
\begin{equation}
\tfrac{1}{2}\,\mathcal{R}^2 + (\lambda-3)\,\mathcal{R} + \frac{3}{\mu}=0,
\end{equation}
with roots
\begin{equation}
\mathcal{R}_\pm \;=\; -(\lambda-3) \;\pm\; \sqrt{(\lambda-3)^2 - \frac{6}{\mu}}.
\end{equation}
The condition for real existence is $(\lambda-3)^2 \ge 6/\mu$. The stable fixed point
is the smaller value, which for $\lambda\ll 3$ reproduces the standard result:
\begin{equation}
\mathcal{R}_* \;=\; 3\left(1 - \sqrt{1 - \frac{2}{3\mu}}\right), \qquad \mu> \tfrac{2}{3}.
\end{equation}

\noindent
\textbf{Linearization and relaxation:} we define
$f(\mathcal{R})=\frac{3}{\mu}+(\lambda-3)\mathcal{R}+\tfrac{1}{2}\mathcal{R}^2$,
with derivative $f'(\mathcal{R}_*)=(\lambda-3)+\mathcal{R}_*$. The linearized dynamics are
\begin{equation}
\dot{\mathcal{R}} \;\approx\; -\,\kappa\,H\,\bigl(\mathcal{R}-\mathcal{R}_*\bigr),
\qquad
\kappa \;=\; -\,f'(\mathcal{R}_*) \;=\; 3 - \lambda - \mathcal{R}_*.
\end{equation}
Thus, the relaxation rate toward the fixed point is controlled by $\kappa$ and modulated by the integral of $H$.

Integrating,
\begin{equation}
\mathcal{R}(t) \;\simeq\; \mathcal{R}_* +  \bigl(\mathcal{R}(t_0)-\mathcal{R}_*\bigr)\,
\exp\!\left(-\kappa\!\int_{t_0}^{t} H(\tilde t)\,d\tilde t\right).
\end{equation}
Using the asymptotic form $H(t)\simeq \dfrac{C}{t\,[D_0 + D_1\ln t]}$ with $C=4\,\gamma_{\rm frac}\,\Delta\alpha$,
\begin{equation}
\mathcal{R}(t) \;\simeq\; \mathcal{R}_* +  \bigl(\mathcal{R}(t_0)-\mathcal{R}_*\bigr)\,
\left[\frac{D_0 + D_1\ln t}{D_0 + D_1\ln t_0}\right]^{-\kappa\,C/D_1}.
\end{equation}

\noindent
\textbf{Physical interpretation:} relaxation toward $\mathcal{R}_*$ is controlled by $\kappa$ and modulated by the integral of $H$, producing dependencies in $\ln t$. The slow evolution of $G_{\rm eff}(t)$ justifies treating it as quasi-constant in the integrations, introducing only mild logarithmic modulations on cosmological scales.

\subsubsection{Asymptotic case: $\lambda \gg 3$}

When $\lambda$ is much greater than 3, the linear term dominates the dynamics.
The fixed points are obtained from
\begin{equation}
\tfrac{1}{2}\,\mathcal{R}^2 + (\lambda-3)\,\mathcal{R} + \frac{3}{\mu}=0,
\end{equation}
with roots
\begin{equation}
\mathcal{R}_\pm = -(\lambda-3) \;\pm\; \sqrt{(\lambda-3)^2 - \tfrac{6}{\mu}}.
\end{equation}

In the limit $\lambda \gg 3$, they approximate to
\begin{equation}
\mathcal{R}_* \;\simeq\; -\,\frac{3}{\mu(\lambda-3)} \;\to\; 0^-,
\qquad
\mathcal{R}_{\rm other} \;\simeq\; -\,2(\lambda-3).
\end{equation}

\noindent
The stable fixed point is the one close to zero, $\mathcal{R}_*\approx 0^-$. Relaxation toward this value is rapid, with linearization coefficient
\begin{equation}
\kappa \;=\; 3 - \lambda - \mathcal{R}_* \;\simeq\; -\,\lambda,
\end{equation}
which implies strong attraction toward $\mathcal{R}_*$. Consequently,
\begin{equation}
\dot G_{\rm eff} = G_{\rm eff} H \mathcal{R} \;\to\; 0,
\end{equation}
and $G_{\rm eff}(t)$ stabilizes in cosmic time, reproducing the GR limit.
This result is important: a large value of $\lambda$ acts as a self-regulation mechanism that suppresses any deviation
of the effective gravity, ensuring that the fractional model does not depart from the standard prediction in matter-dominated epochs.

\subsection{Primordial nucleosynthesis (BBN) limits on the temporal variation of the gravitational constant}
\label{Sect:6.2}
Earth measurements of Newton’s gravitational constant show small-amplitude oscillations, with values between
$6.6721\times 10^{-11}$ and $6.6725\times 10^{-11}$ N m$^{2}$ kg$^{-2}$, with a periodicity close to $5.9$ years \cite{133,134}. In the fractional framework, the temporal variation of $G$ is obtained asymptotically as
\begin{equation}
\frac{\dot{G}}{G}\Big|_{t=t_0}
\approx \frac{(2+\alpha)(1+\gamma)}{1-\beta}\,H_0,
\label{tag16}
\end{equation}
which allows exploration of regimes where $\dot{G}/G$ is comparable to or even much larger than $H_0$.

Most analyses of primordial nucleosynthesis (BBN) assume the limit $\dot{G}/G\ll H$ \cite{135,136}, while the opposite case, $\dot{G}/G\gg H$ has been studied in \cite{137,138}. A rapidly varying gravitational constant significantly modifies the predicted abundances of light elements, since it alters the expansion history during the early stages of the universe. Equation \eqref{tag10} generalizes the rapid variation law used in \cite{137,138}, where
$G=G_N(1+A\cos(\omega t+\phi))$.

Models of this type have also been considered in the context of dark energy \cite{139}, showing relevant effects on BBN and on the production of primordial elements. Typical observational limits on the variation of $G$ are in the range
$\dot{G}/G\sim 10^{-12}$--$10^{-14}$ yr$^{-1}$, coming from: BBN abundances \cite{140}, lunar laser ranging measurements \cite{141}, quasar spectra \cite{142}, and gravitational-wave observations of binary neutron stars \cite{143}.

\begin{itemize}
    \item For $\beta=0.95$ and $\alpha=0.5$ one obtains $\dot{G}/G\approx 50(1+\gamma)H_0$ yr$^{-1}$. For $\gamma=-0.9$ (compatible with $\gamma=-1.013^{+0.038}_{-0.043}$ \cite{144}), it follows that $\dot{G}/G\gg H_0$, in agreement with the rapid variation models of \cite{137,138}.

    \item For $\beta=0.1$ and $\alpha=2$ one obtains $\dot{G}/G\approx 4.44(1+\gamma)H_0$ yr$^{-1}$. For $\gamma=-0.96$, this yields $\dot{G}/G\approx 2.4\times 10^{-12}$ yr$^{-1}$, consistent with BBN limits \cite{140}.

    \item For $\beta=0.0001$ and $\alpha=1.5$ one obtains $\dot{G}/G\approx 1.96\times 10^{-8}$ yr$^{-1}$, in agreement with gravitational-wave observations of neutron star binaries \cite{143}.
\end{itemize}

\subsection{Physical interpretation and connection with $\mathcal{R}$.}
\label{Sect:6.2.1}
The temporal variation of $G$ is directly related to the quantity
\begin{equation}
\mathcal{R}=\frac{\dot{G}/G}{H},
\end{equation}
which controls the modified continuity and the fractional part of the effective equation-of-state parameter,
\begin{equation}
w_{\mathrm{eff}}
= w + \frac{\mathcal{R}}{3}
+ \frac{\alpha-1}{3tH}(1+w).
\end{equation}
During BBN, the standard condition $|\mathcal{R}|<\varepsilon\ll1$ guarantees that the expansion does not deviate significantly from the standard scenario. However, in the presence of fractional antifriction ($\alpha>1$), the evolution of $\mathcal{R}$ can be amplified, requiring even stricter constraints on the model parameters.

An increasing gravitational constant ($\mathcal{R}>0$) accelerates expansion, reduces the time available for neutron decay, and tends to increase helium abundance. Conversely, $\mathcal{R}<0$ suppresses expansion and favors the formation of heavier elements. These trends are reinforced or attenuated by the fractional friction terms $(1-\alpha)/(tH)$ and by the modified gravitational source $t^{-(\alpha-1)(1+\gamma)}$.

\subsection{Constraints and consistency with the model.}
\label{Sect:6.2.2}

The model allows both $\dot{G}/G>0$ and $\dot{G}/G<0$, depending on the parameters $(\alpha,\beta,\gamma)$. The negative case does not necessarily imply a decrease in $H$, since the fractional Friedmann equation includes additional terms. For $\dot{G}/G\gg H$ and $\gamma<-1$, equation \eqref{tag6} is approximated by
\begin{equation}
\mu(2+\alpha)(1+\gamma)^2 H_0^2/(1-\beta)^2=16\pi G\rho_0,
\end{equation}
which, compared with the standard model $\rho_0=3H_0^2/(8\pi G)$, leads to the condition
\begin{equation}
\mu(2+\alpha)(1+\gamma)^2/(1-\beta)^2=6.
\end{equation}

The previously obtained constraints,
\begin{equation}
(\alpha-1)(\gamma+1)<0,
\end{equation}
that is,
\begin{equation}
\alpha>1,\;\gamma<-1
\quad\text{or}\quad
0<\alpha<1,\;\gamma>-1,
\end{equation}
remain necessary to guarantee dynamical consistency.

The model allows crossing the phantom divide, thereby generating cosmic acceleration without introducing exotic fields. The energy density decreases for $\gamma<-1$ with $0<\alpha<1$ (phantom-dominated universe) or for $\gamma>-1$ with $1<\alpha<2$ (quintessence-dominated universe). For $\alpha=1$, the standard law is recovered $\rho=\rho_0 a^{-3(1+\gamma)}$.

Taken together, BBN imposes strong constraints on the temporal variation of $G$, on the magnitude of $\mathcal{R}$, and on the fractional parameters of the model. These constraints are essential to ensure that the fractional dynamics do not unacceptably alter the early expansion history of the universe.

\subsection{Cosmological tensions}
\label{Sect:6.3}

Current estimates of the Hubble parameter show significant dispersion: the Hubble Space Telescope Key Project obtains
$H_0=72\pm 8$ km s$^{-1}$ Mpc$^{-1}$ \cite{7}, while the CMB yields $72.2\pm 4$ km s$^{-1}$ Mpc$^{-1}$ for non-flat $\Lambda$CDM and $68.7\pm 3.1$ km s$^{-1}$ Mpc$^{-1}$ for flat $\Lambda$CDM \cite{10,11}. The SH0ES program reports $H_0=74.03\pm 1.42$ km s$^{-1}$ Mpc$^{-1}$ \cite{9}. In this context, we analyze the temporal variations of the effective gravitational constant $G_{\rm eff}(t)$ shown in Figure~\ref{Fig.6a}, for different values of $H_0$, the EoS parameter $\gamma$, and the fractional parameters $\zeta$, $\mu$, $\alpha$, and $\beta$.

\begin{figure*}[htbp]
\centering
\includegraphics[width=0.31\linewidth]{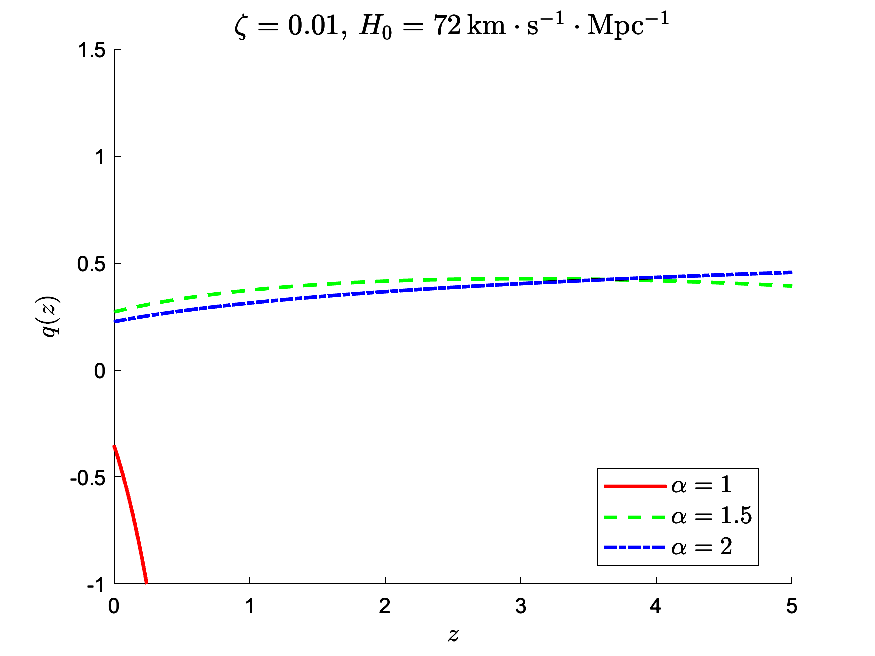}\hspace{0.01\linewidth}
\includegraphics[width=0.31\linewidth]{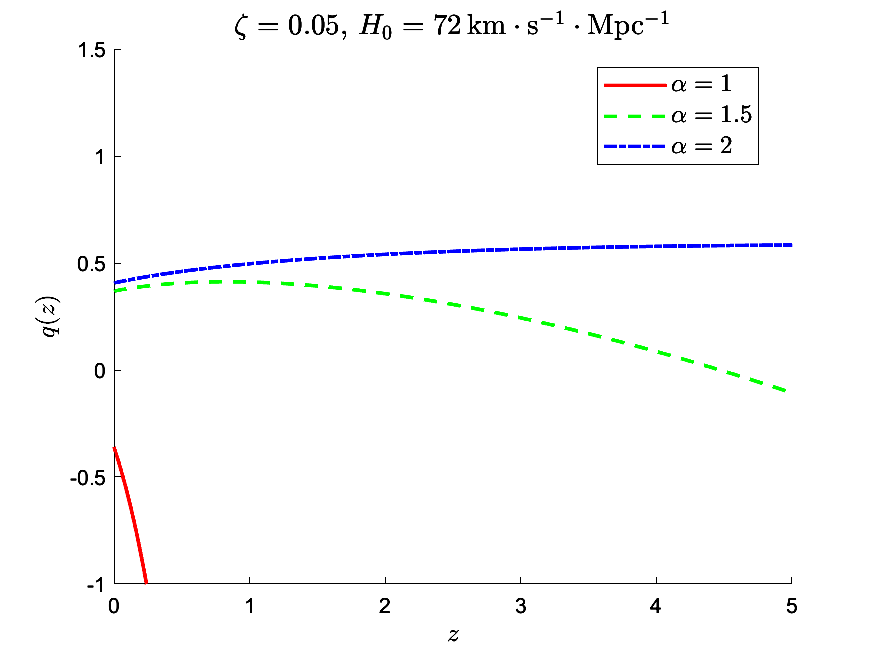}\hspace{0.01\linewidth}
\includegraphics[width=0.31\linewidth]{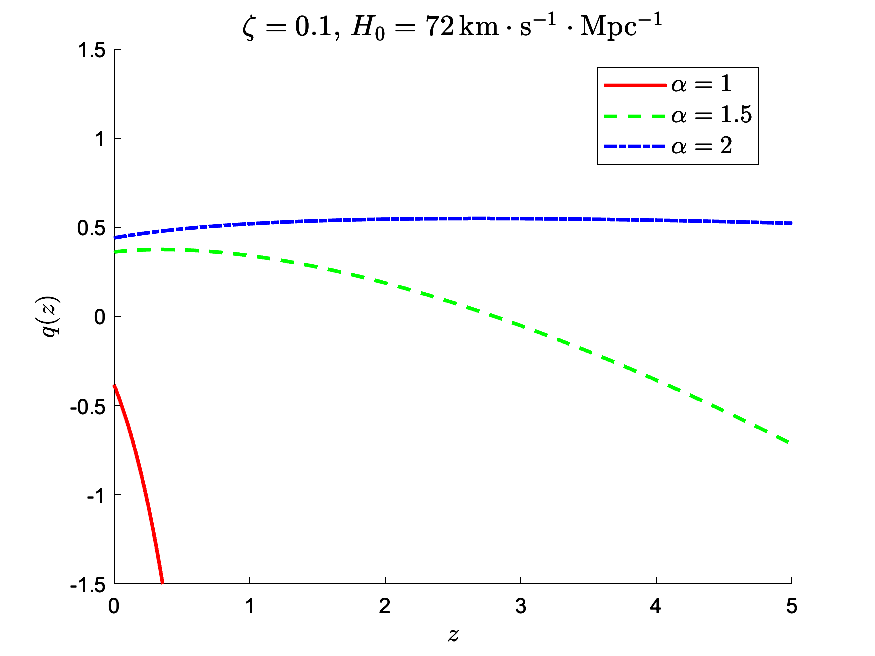}\\

\includegraphics[width=0.31\linewidth]{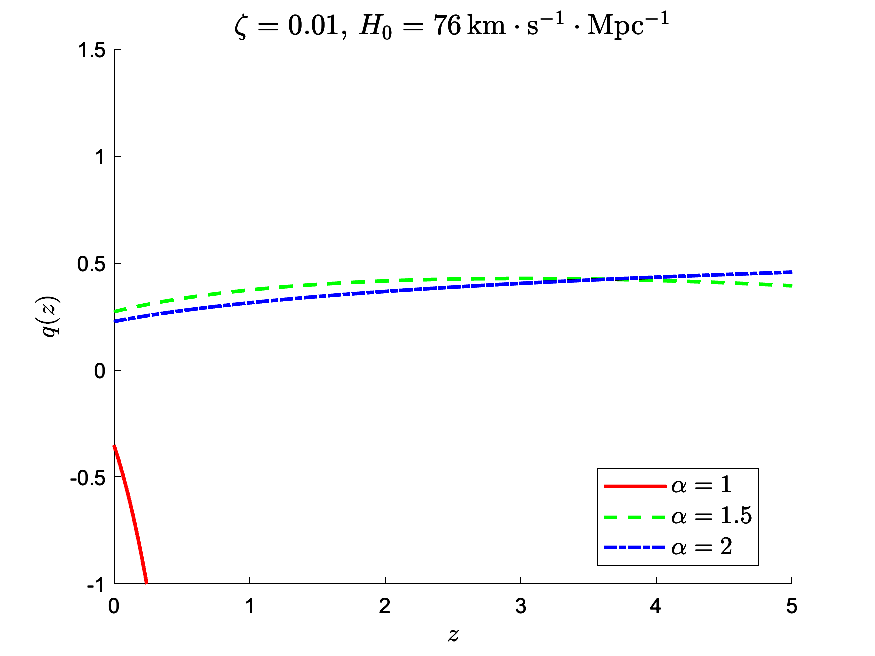}\hspace{0.01\linewidth}
\includegraphics[width=0.31\linewidth]{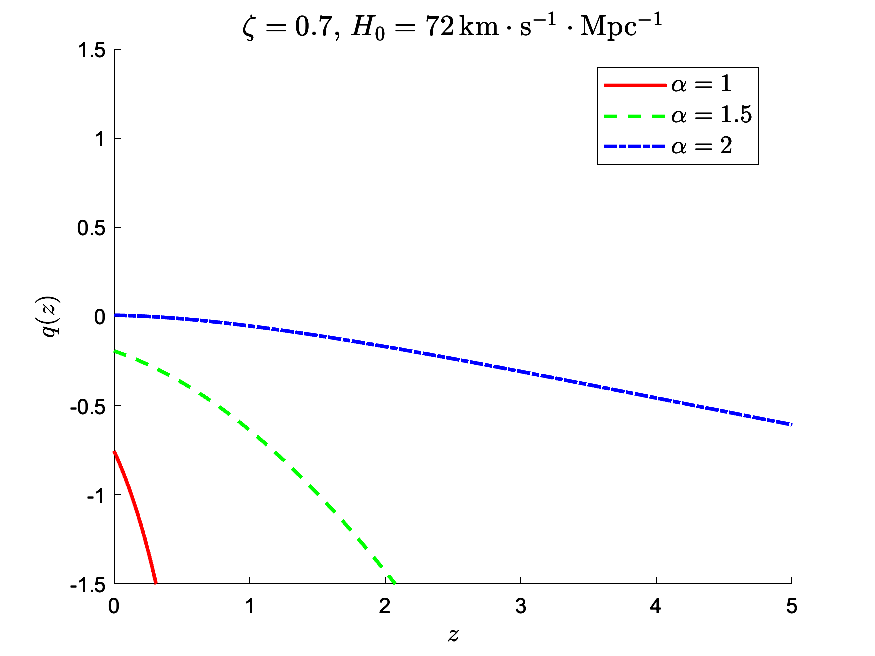}\hspace{0.01\linewidth}
\includegraphics[width=0.31\linewidth]{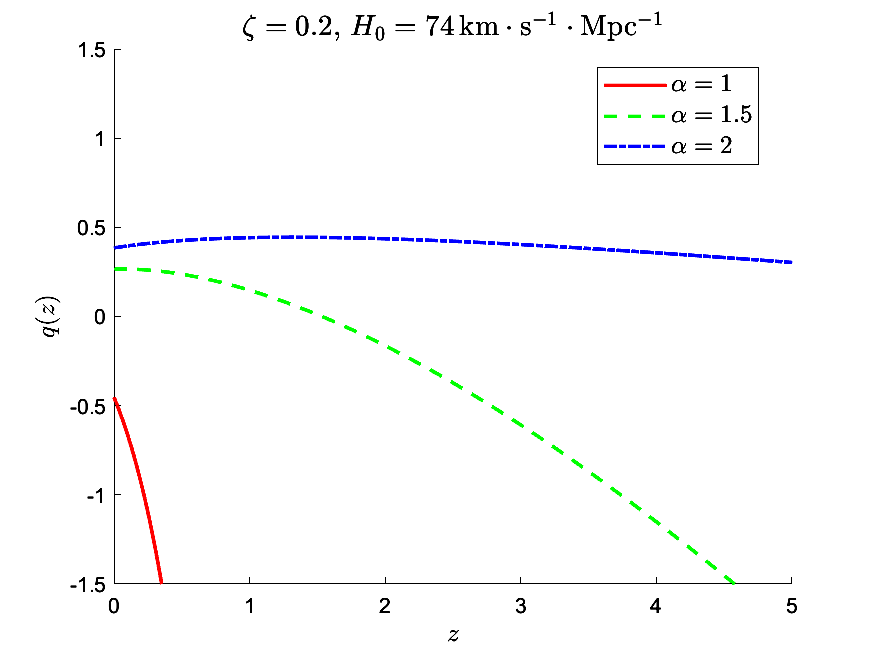}\\

\includegraphics[width=0.31\linewidth]{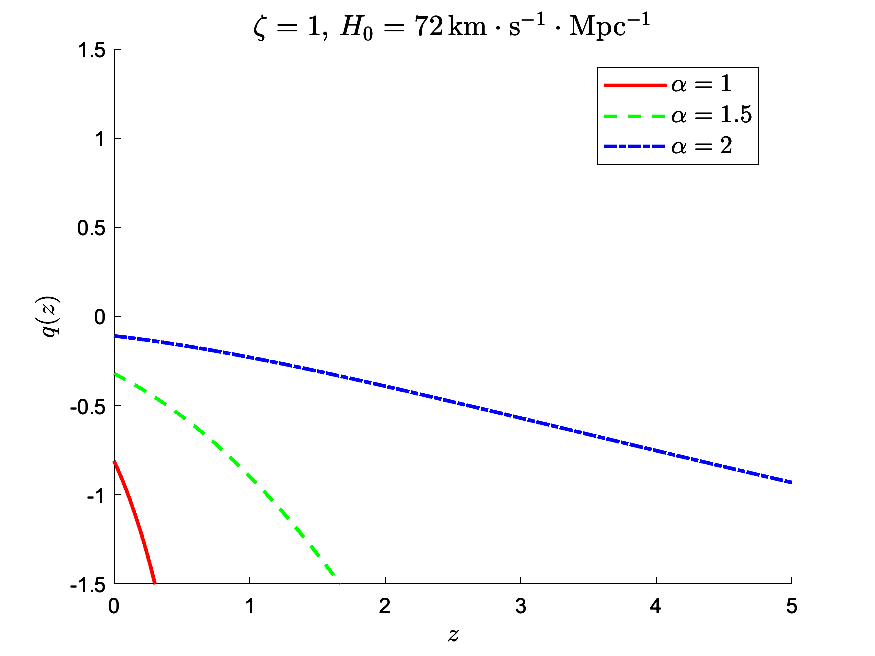}\hspace{0.01\linewidth}
\includegraphics[width=0.31\linewidth]{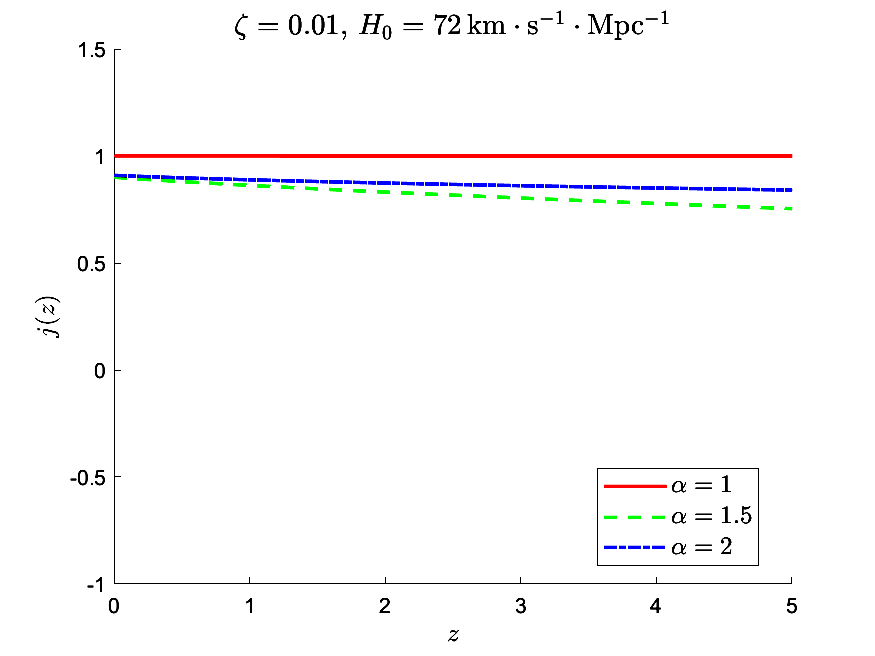}\hspace{0.01\linewidth}
\includegraphics[width=0.31\linewidth]{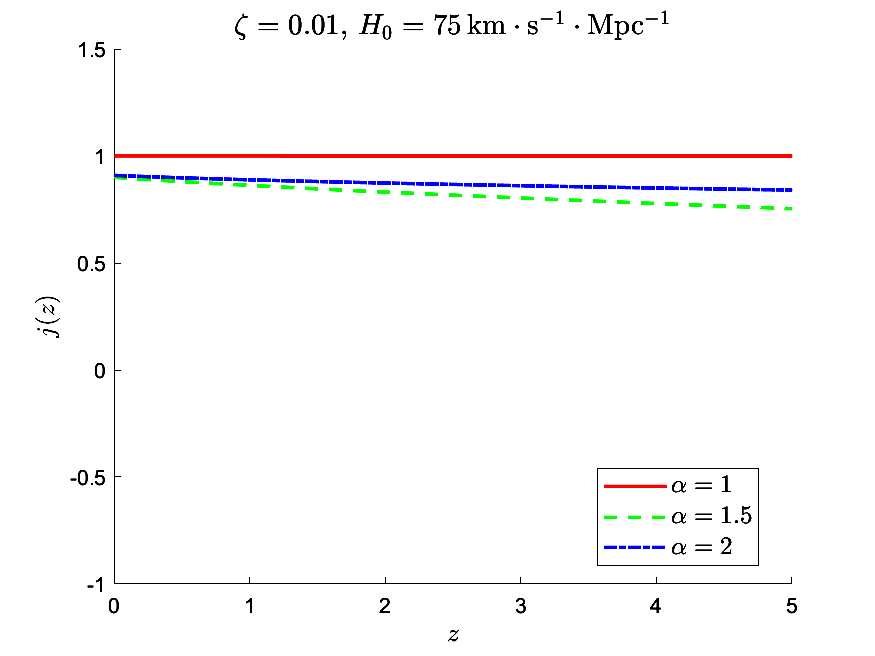}\\

\includegraphics[width=0.31\linewidth]{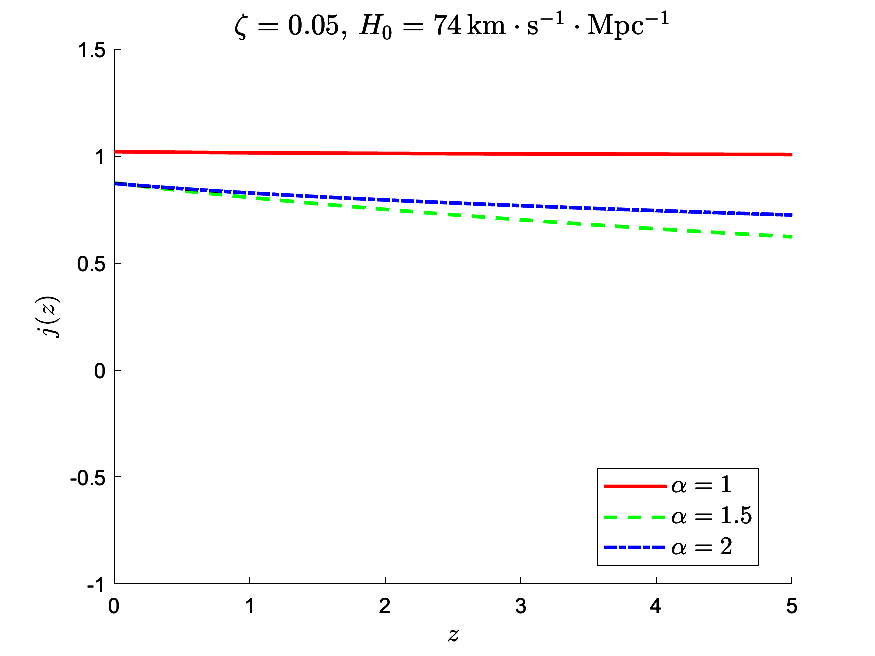}\hspace{0.01\linewidth}
\includegraphics[width=0.31\linewidth]{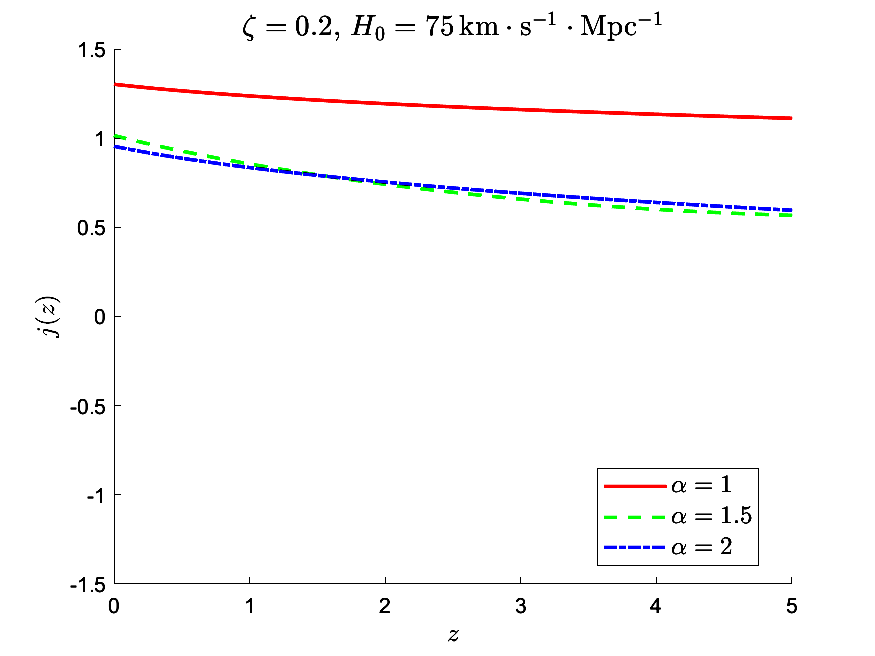}\hspace{0.01\linewidth}
\includegraphics[width=0.31\linewidth]{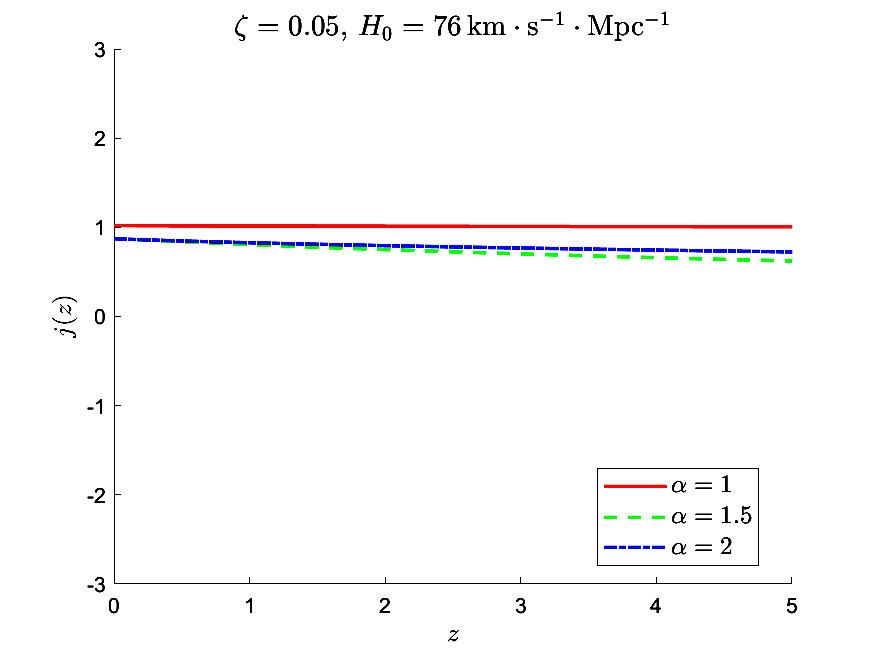}\\

\includegraphics[width=0.31\linewidth]{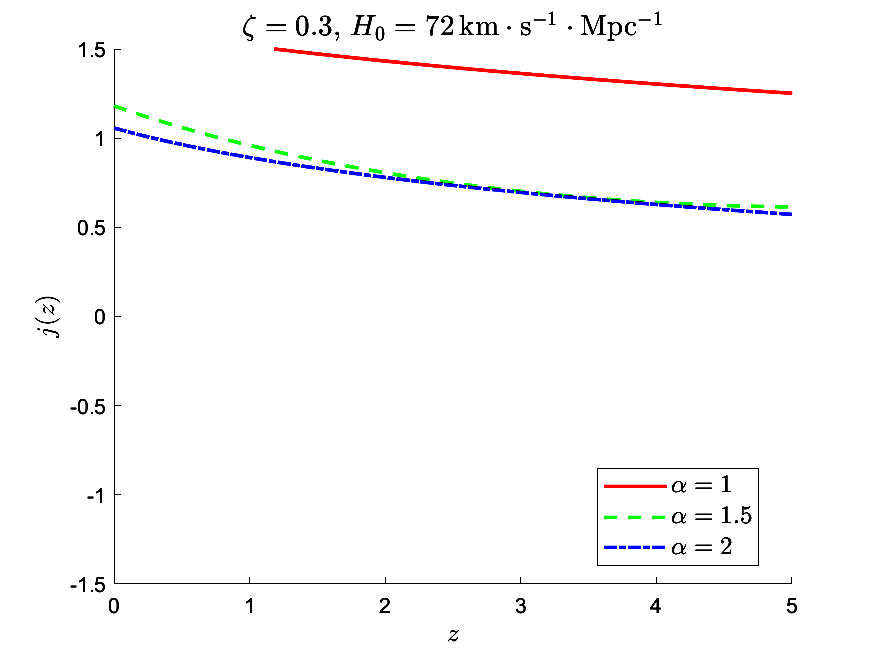}\hspace{0.01\linewidth}
\includegraphics[width=0.31\linewidth]{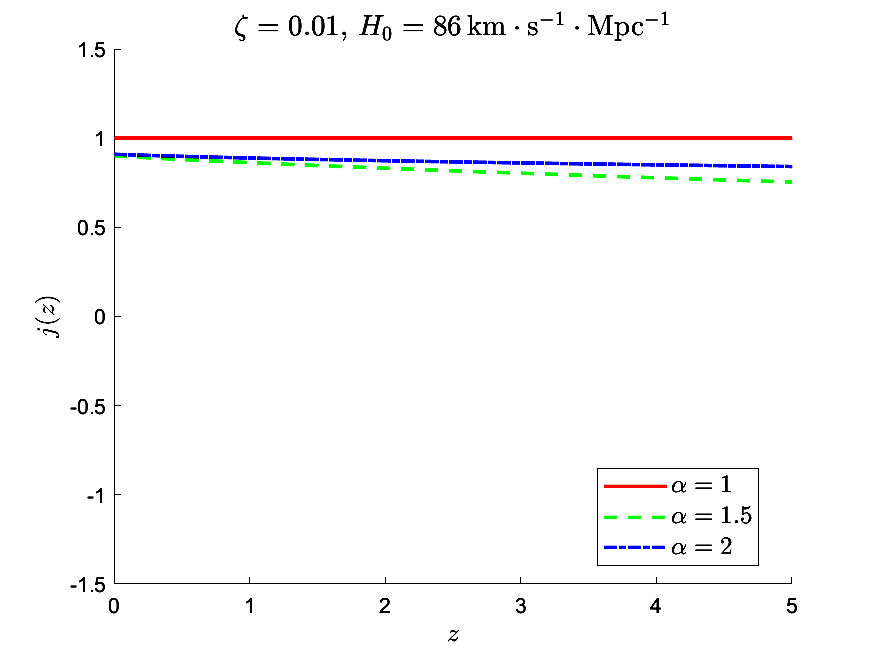}\hspace{0.01\linewidth}
\includegraphics[width=0.31\linewidth]{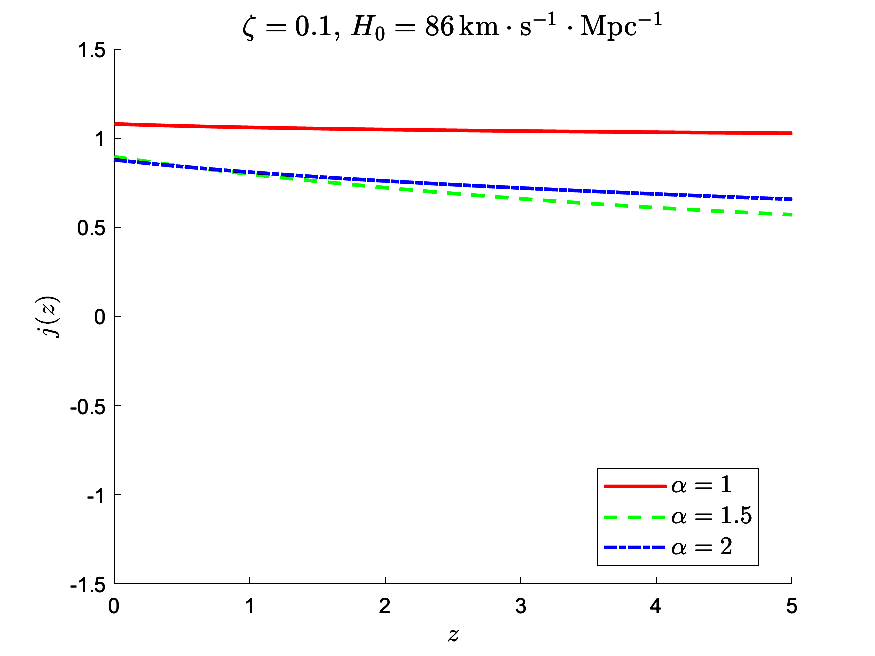}

\caption{\label{Fig.5} Evolution of the \emph{jerk} parameter and the deceleration parameter as a function of redshift.}
\end{figure*}

\begin{figure*}[htbp]
\centering
\includegraphics[width=\linewidth]{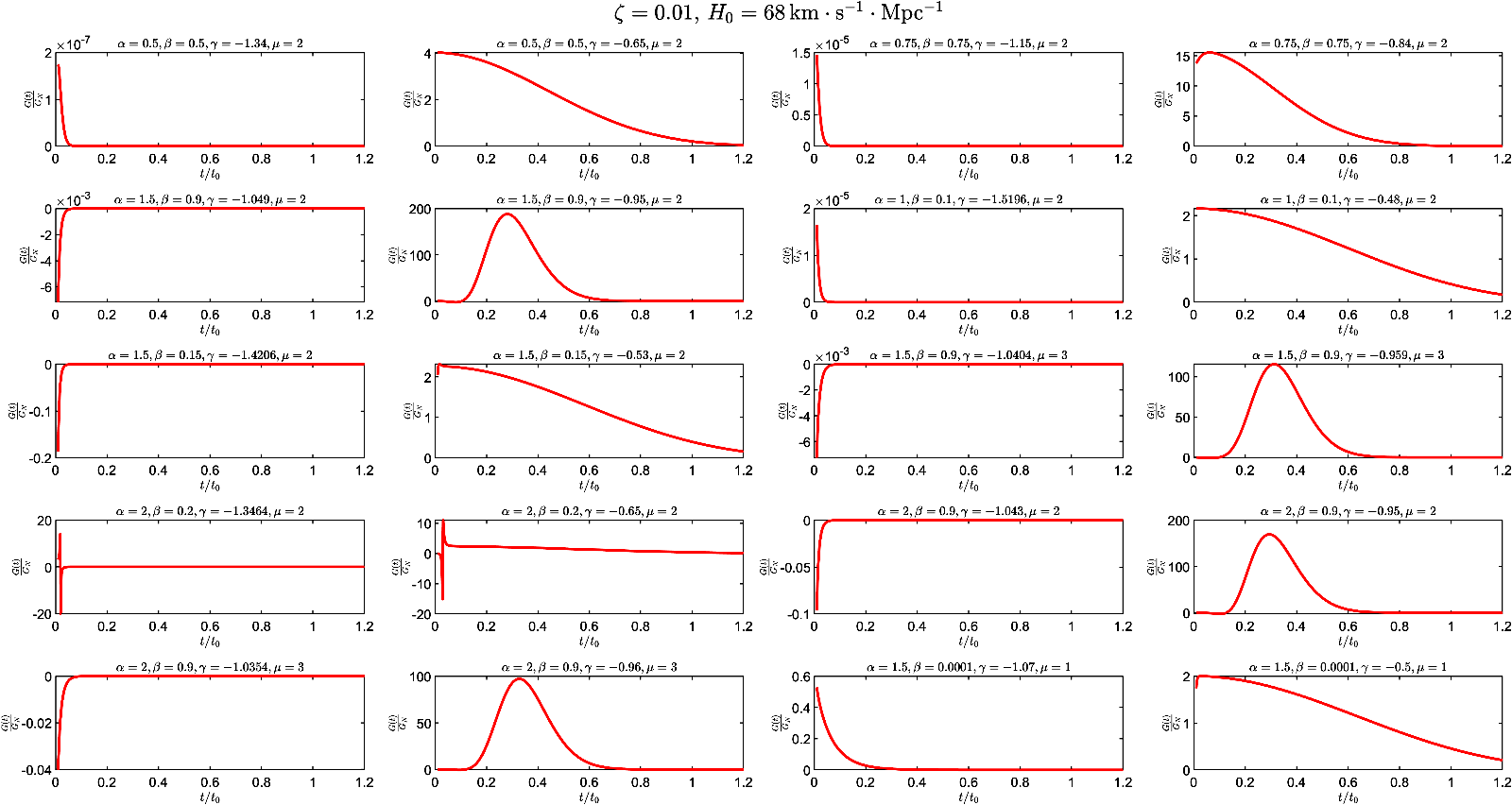}
\caption{\label{Fig.6a} Variation of the gravitational constant as a function of cosmic time (cont.).}
\end{figure*}

\begin{figure*}[htbp]\ContinuedFloat
\centering
\includegraphics[width=\linewidth]{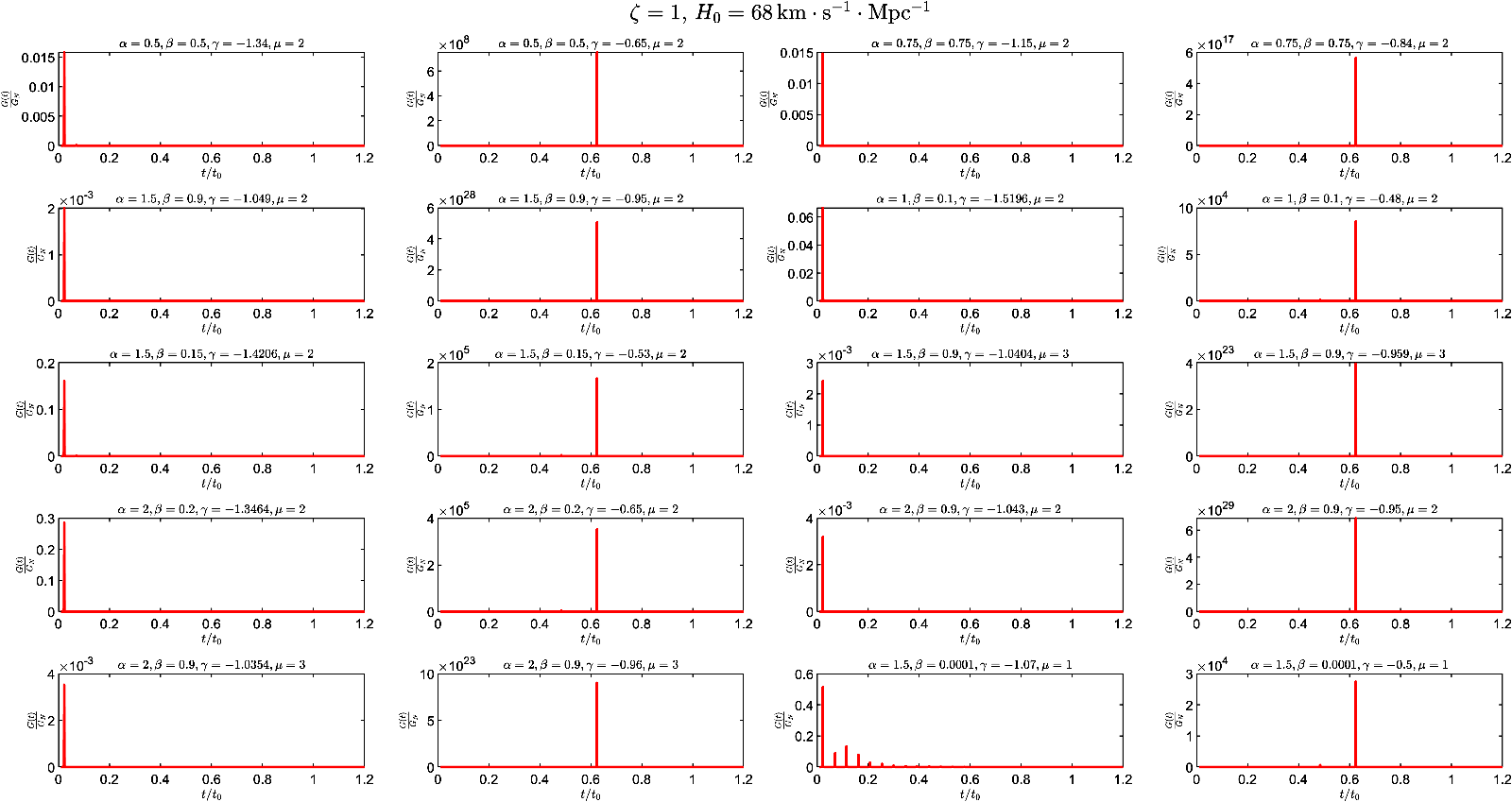}
\caption{\label{Fig.6b} Variation of the gravitational constant as a function of cosmic time (cont.).}
\end{figure*}

\begin{figure*}[htbp]\ContinuedFloat
\centering
\includegraphics[width=\linewidth]{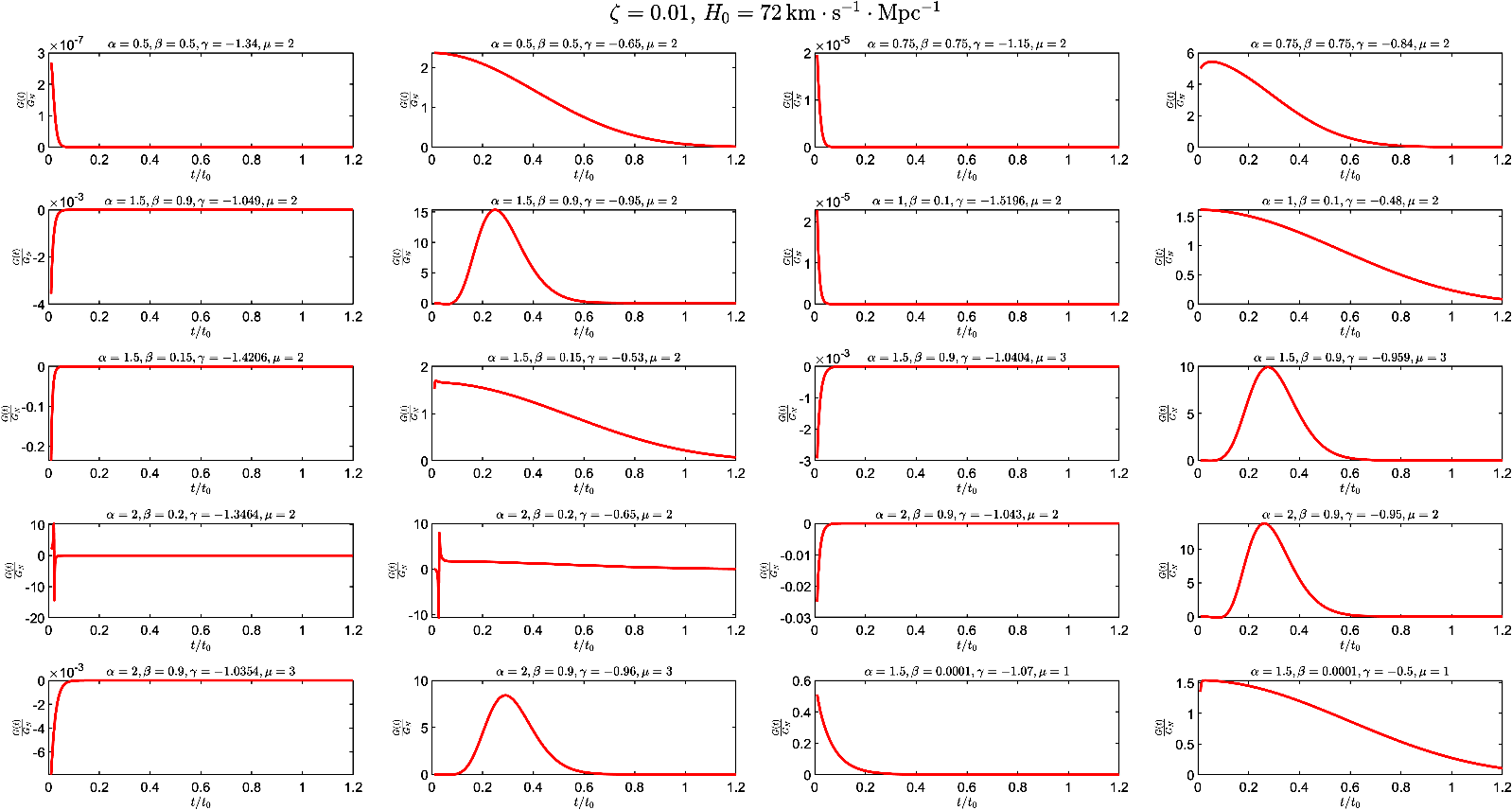}
\caption{\label{Fig.6c} Variation of the gravitational constant as a function of cosmic time (cont.).}
\end{figure*}

\begin{figure*}[htbp]\ContinuedFloat
\centering
\includegraphics[width=\linewidth]{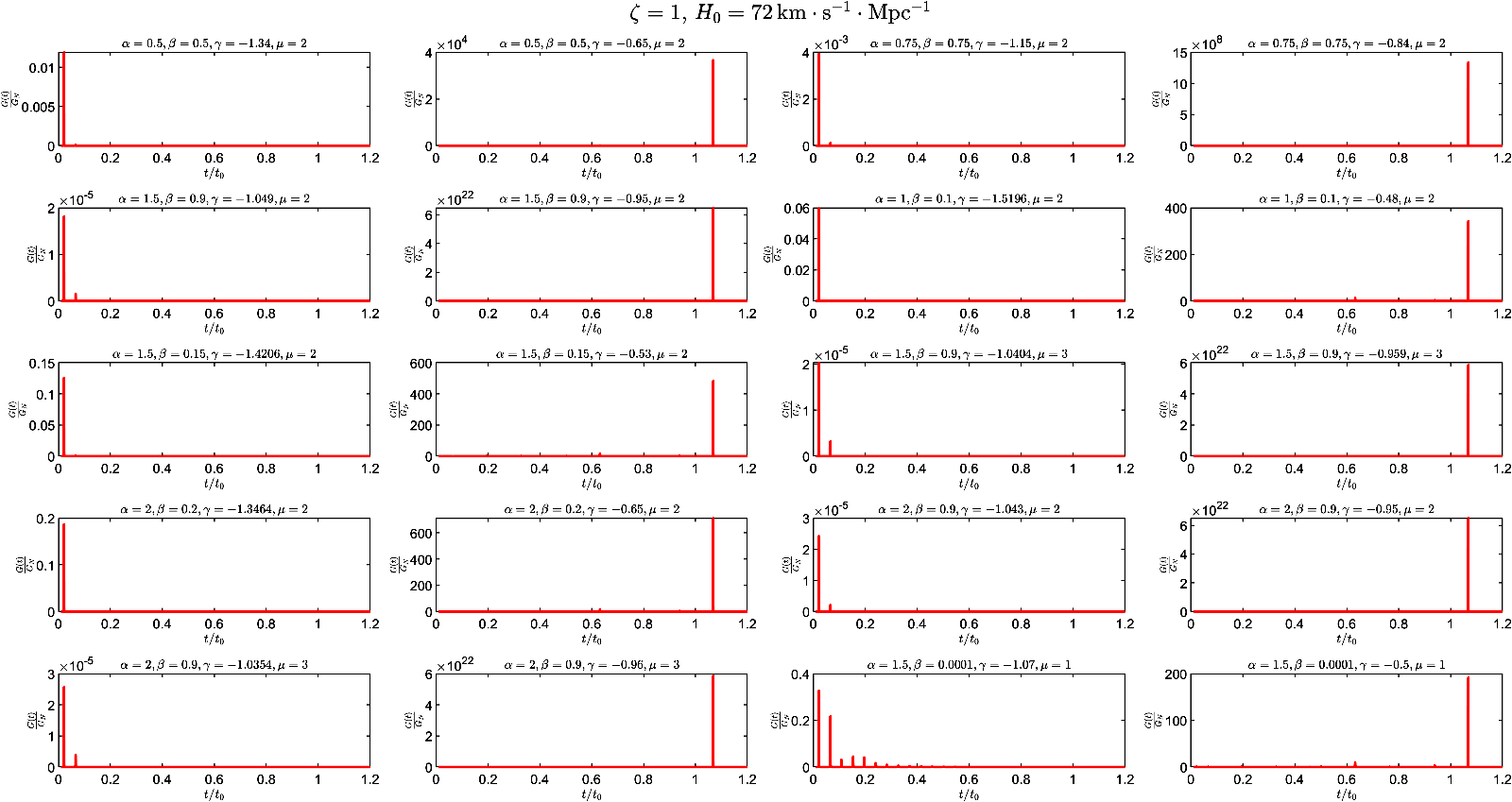}
\caption{\label{Fig.6d} Variation of the gravitational constant as a function of cosmic time (cont.).}
\end{figure*}

\begin{figure*}[htbp]\ContinuedFloat
\centering
\includegraphics[width=\linewidth]{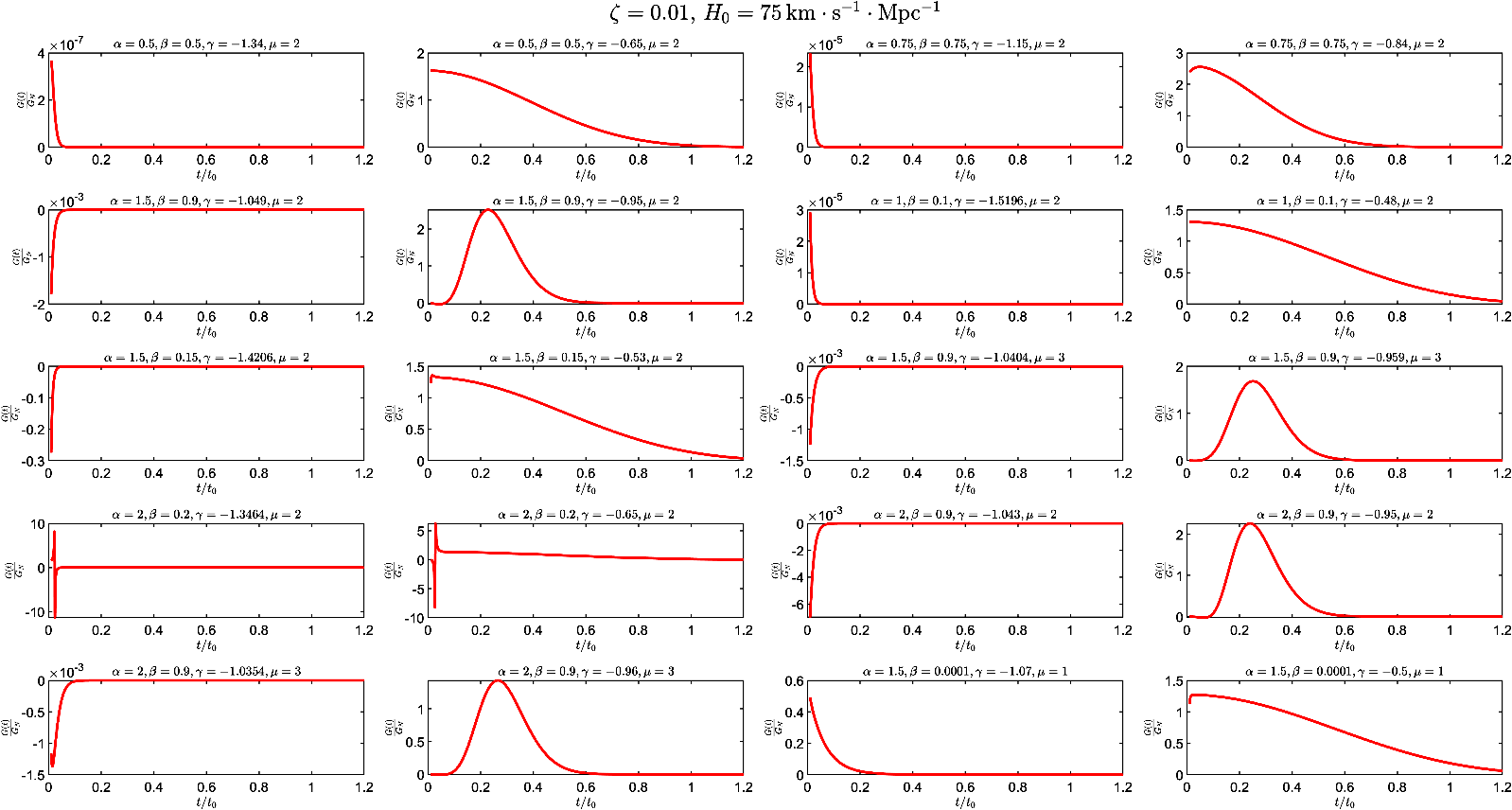}
\caption{\label{Fig.6e} Variation of the gravitational constant as a function of cosmic time (cont.).}
\end{figure*}

\begin{figure*}[htbp]\ContinuedFloat
\centering
\includegraphics[width=\linewidth]{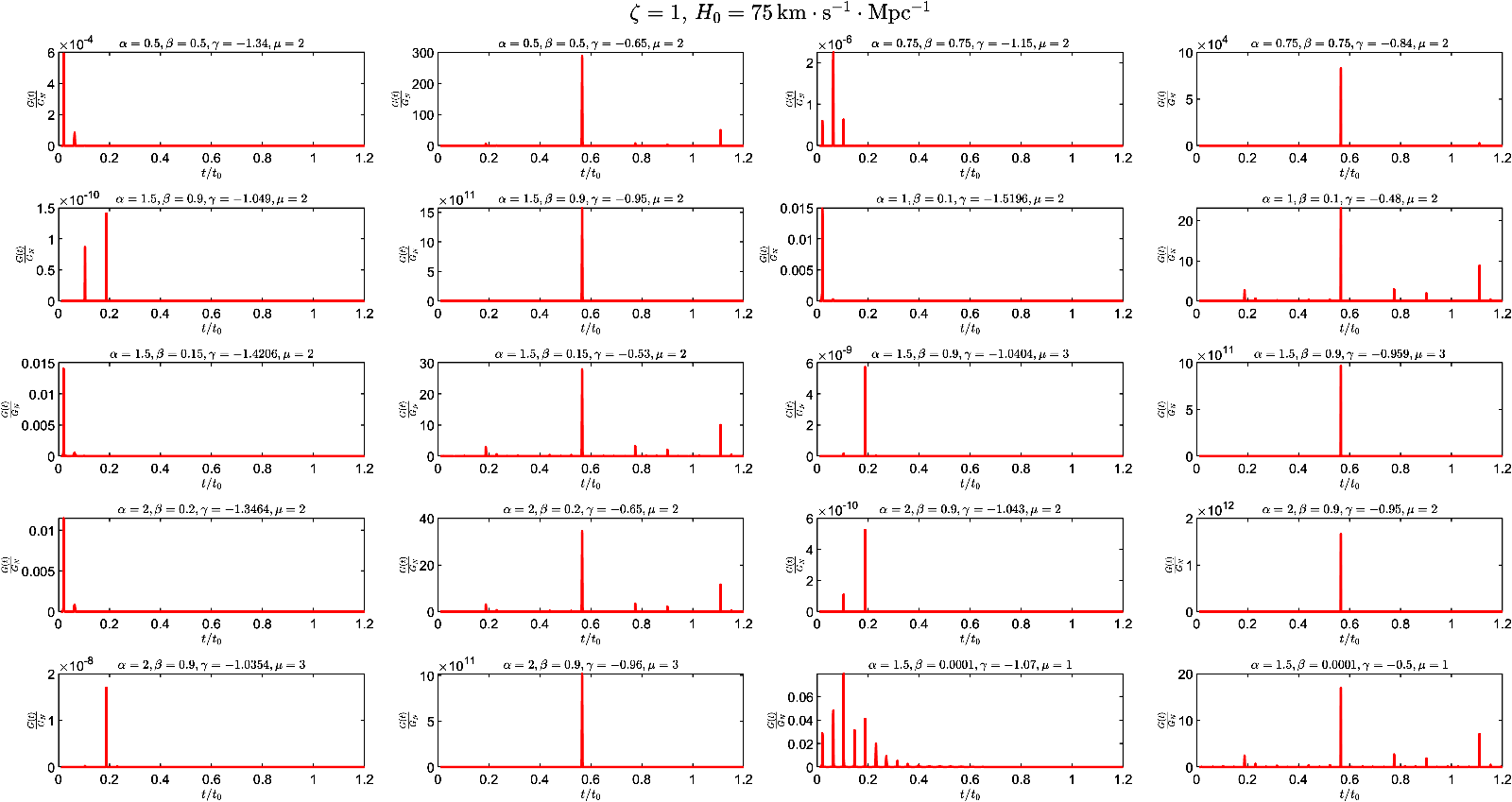}
\caption{\label{Fig.6f} Variation of the gravitational constant as a function of cosmic time.}
\end{figure*}

These variations allow us to evaluate how the model parameters affect the fluctuations of $G_{\rm eff}$ throughout cosmic time. We observe that both $\zeta$ and the equation-of-state parameter $\gamma$ strongly influence the
dynamics of the effective gravitational constant. In particular, the phantom-dominated phase ($\gamma=-1.1$ and $\zeta>1$) exhibits recurrent fluctuations of $G_{\rm eff}$ with small amplitudes that decay over time, eventually tending to $G_N$.

\begin{figure*}[htbp]
\centering
\includegraphics[width=0.31\linewidth]{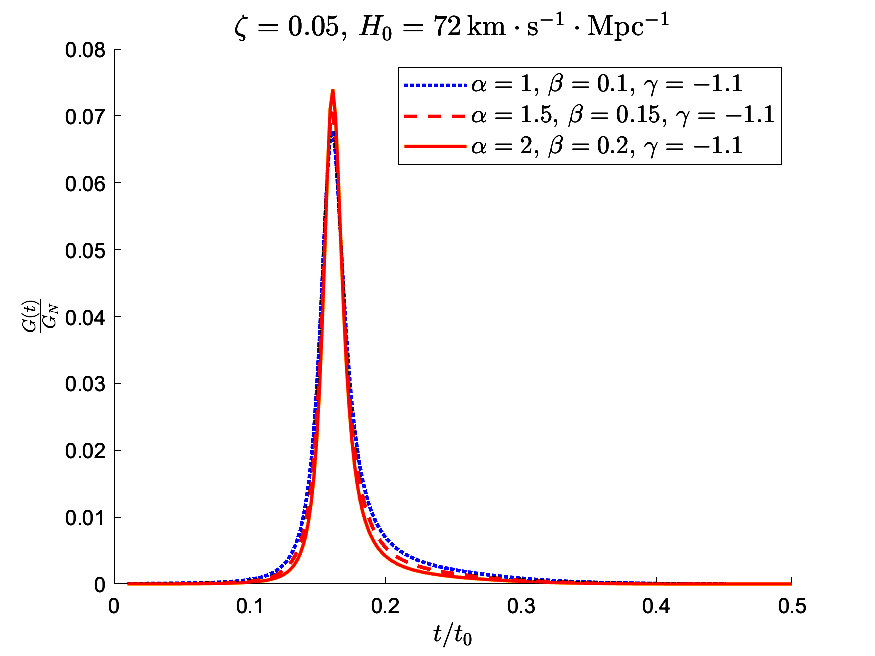}\hspace{0.01\linewidth}
\includegraphics[width=0.31\linewidth]{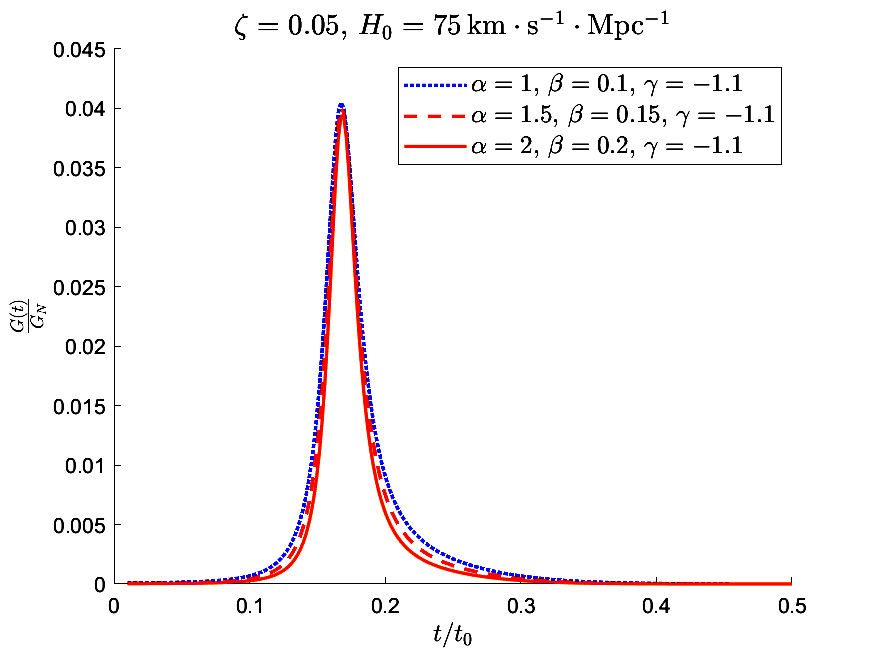}\hspace{0.01\linewidth}
\includegraphics[width=0.31\linewidth]{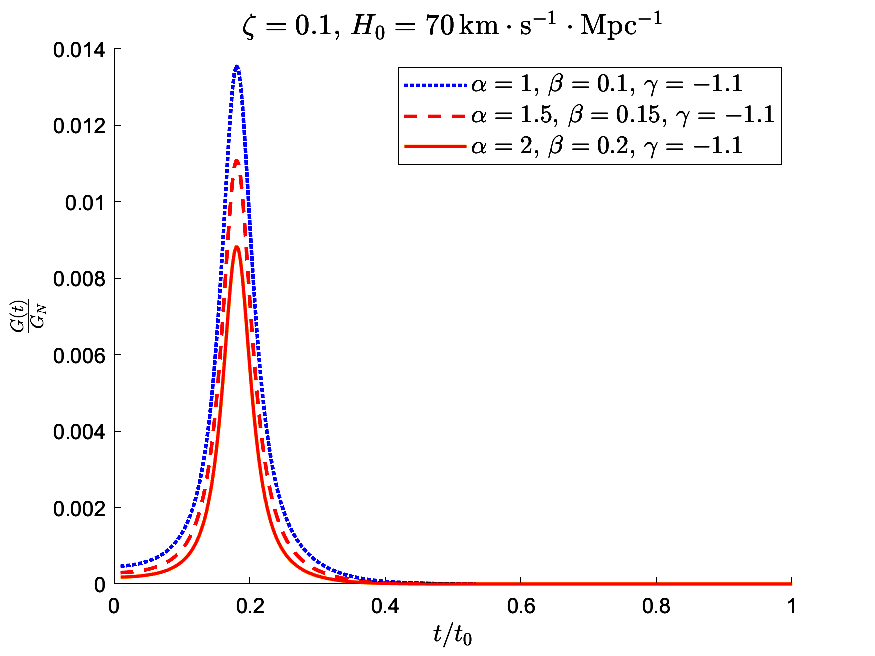}\\
\includegraphics[width=0.31\linewidth]{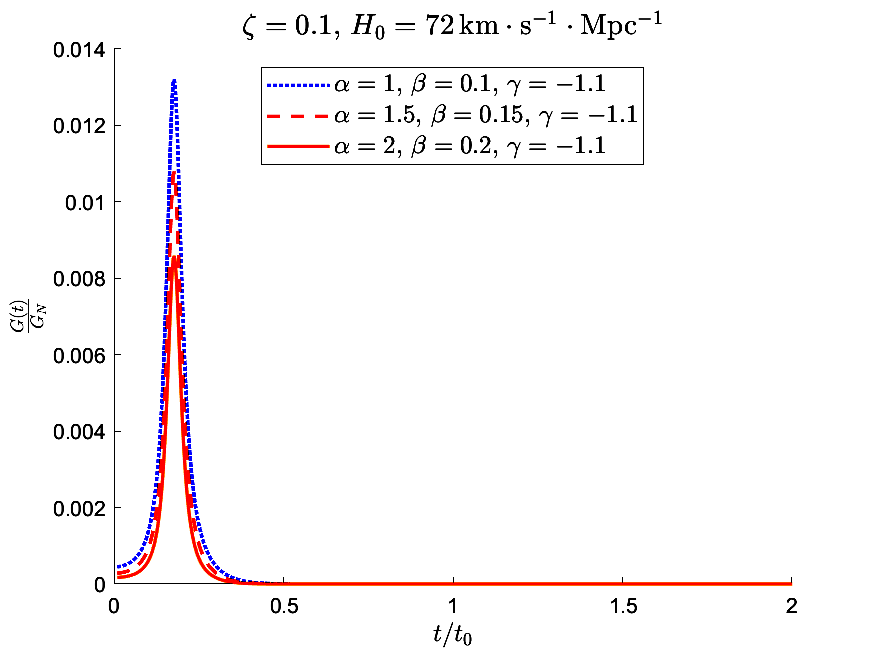}\hspace{0.01\linewidth}
\includegraphics[width=0.31\linewidth]{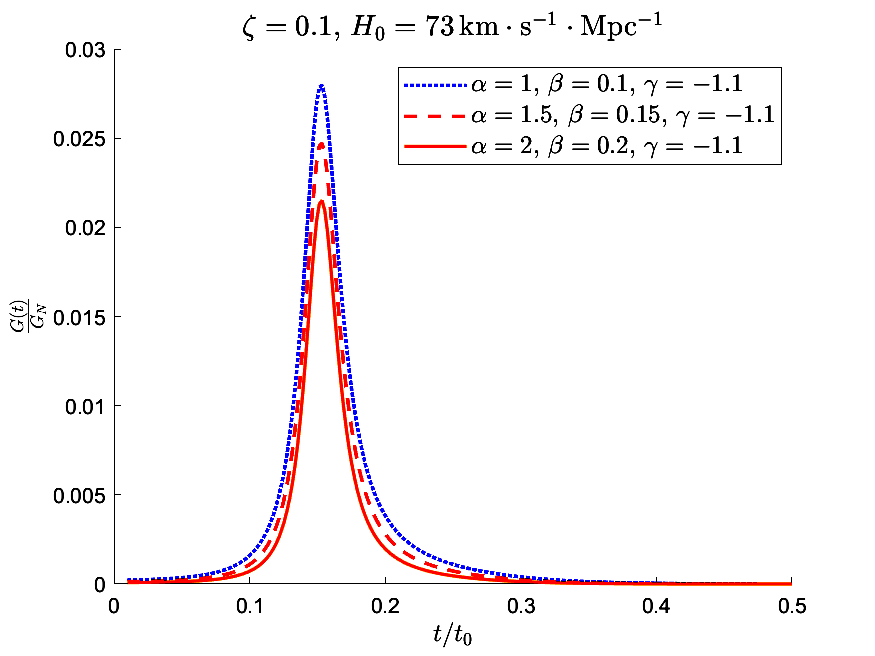}\hspace{0.01\linewidth}
\includegraphics[width=0.31\linewidth]{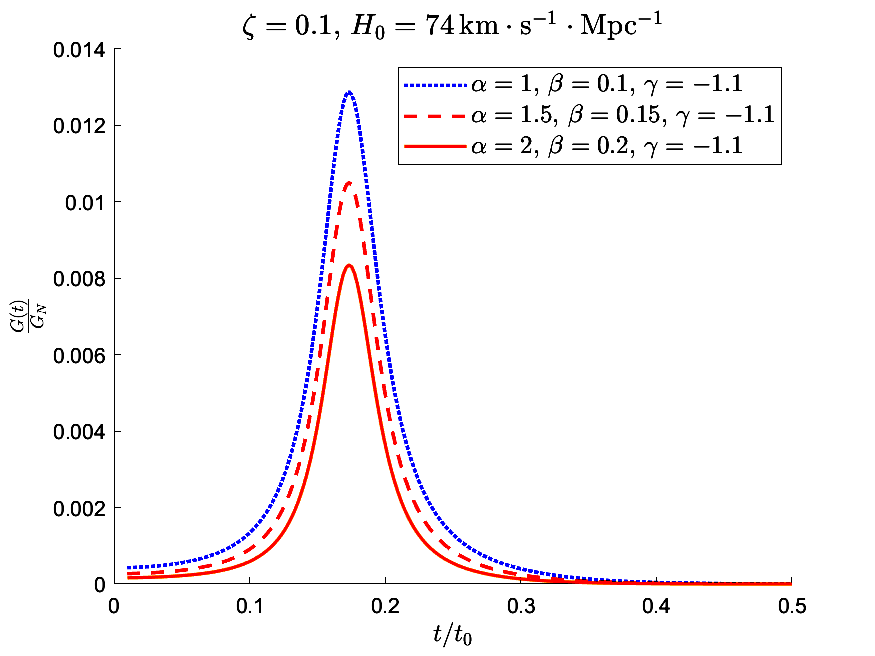}\\
\includegraphics[width=0.31\linewidth]{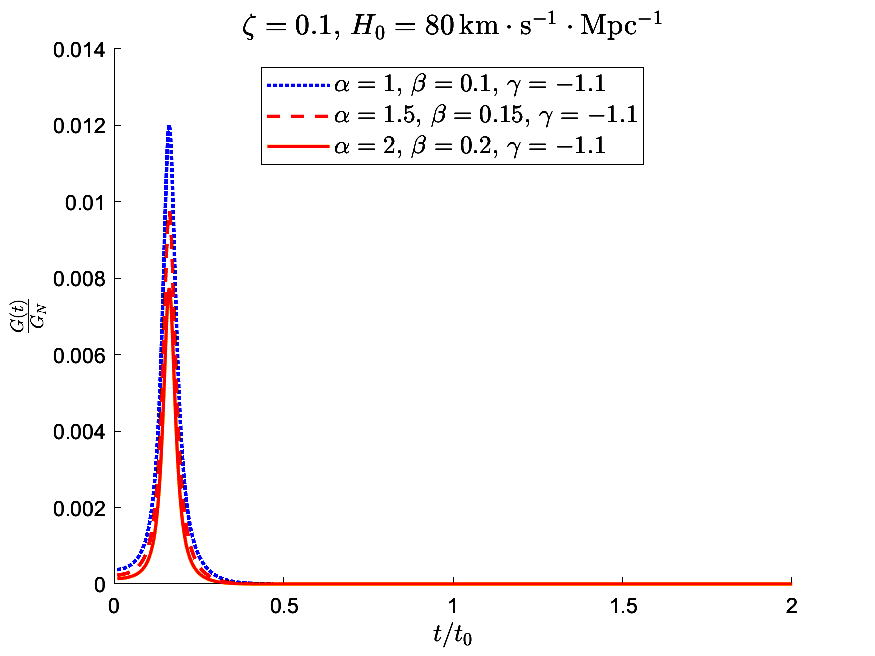}\hspace{0.01\linewidth}
\includegraphics[width=0.31\linewidth]{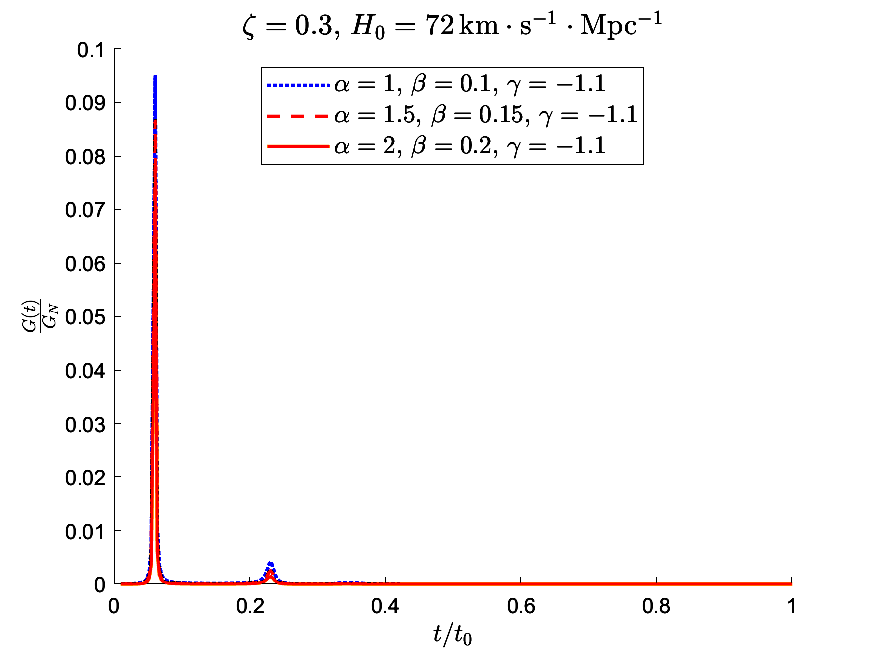}\hspace{0.01\linewidth}
\includegraphics[width=0.31\linewidth]{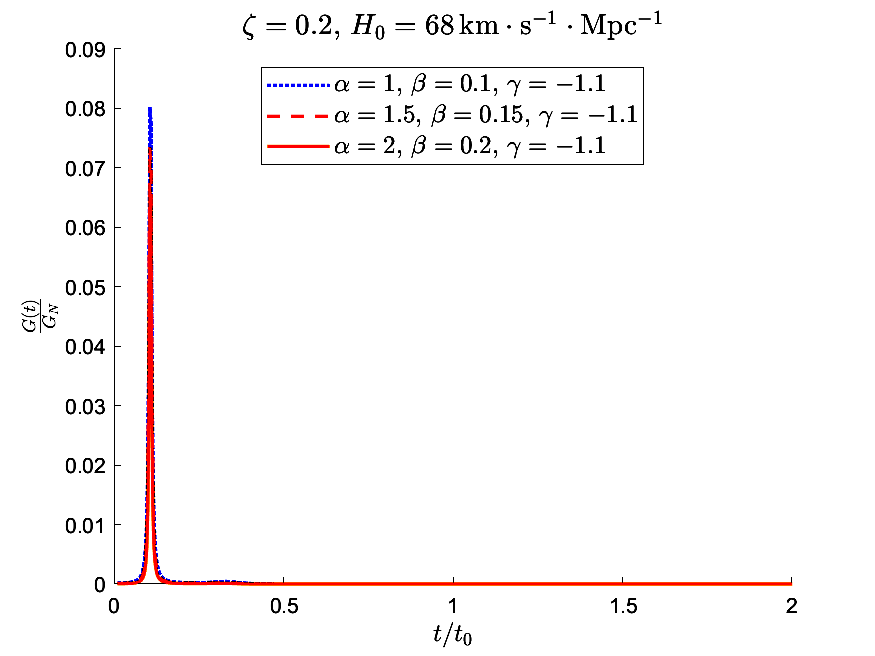}\\
\includegraphics[width=0.31\linewidth]{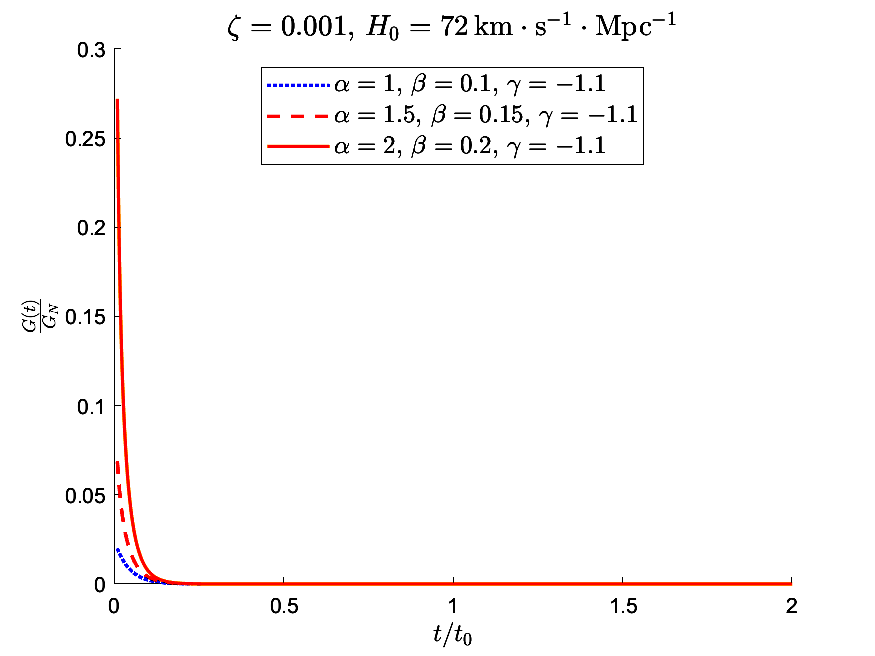}\hspace{0.01\linewidth}
\includegraphics[width=0.31\linewidth]{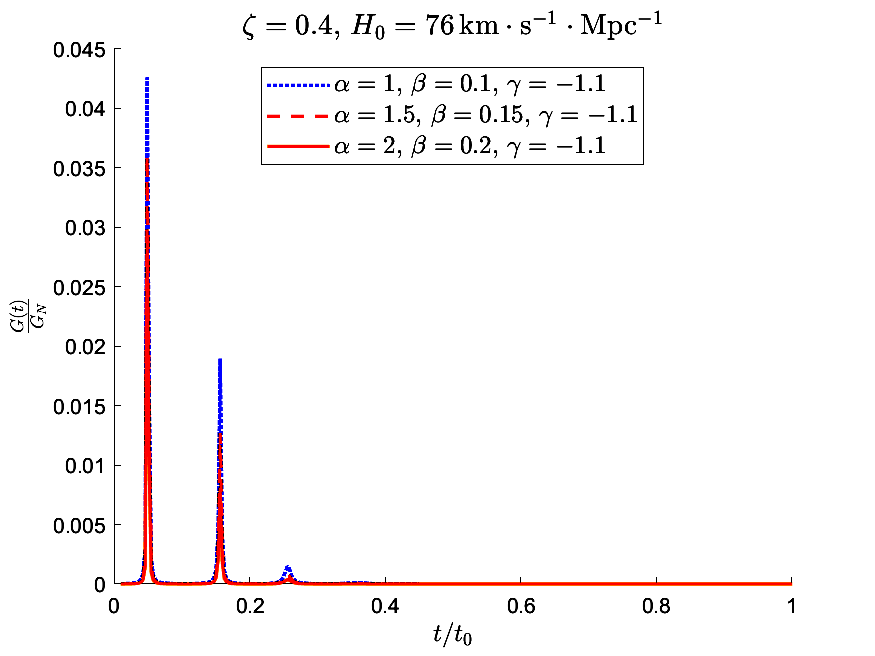}\hspace{0.01\linewidth}
\includegraphics[width=0.31\linewidth]{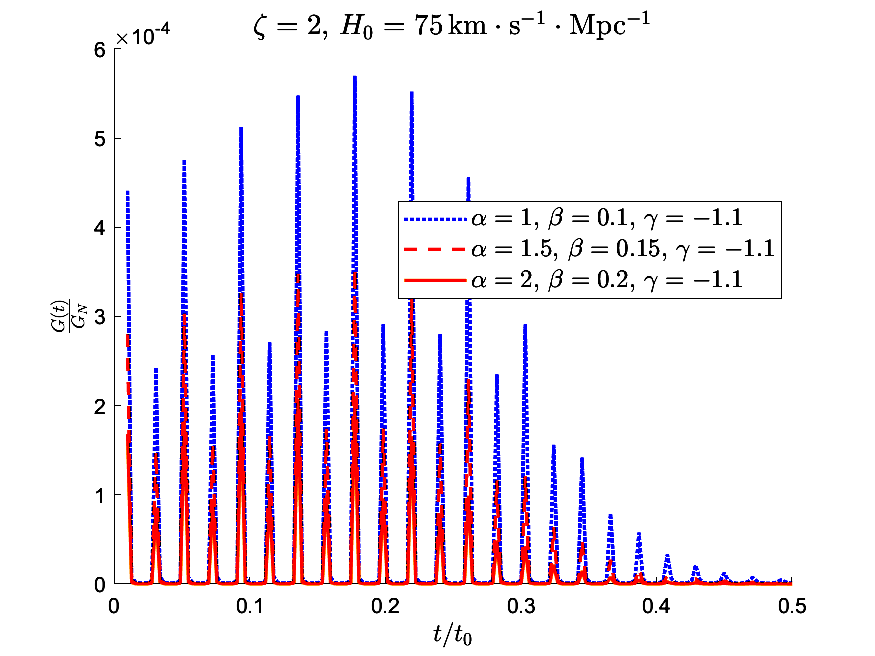}\\
\includegraphics[width=0.31\linewidth]{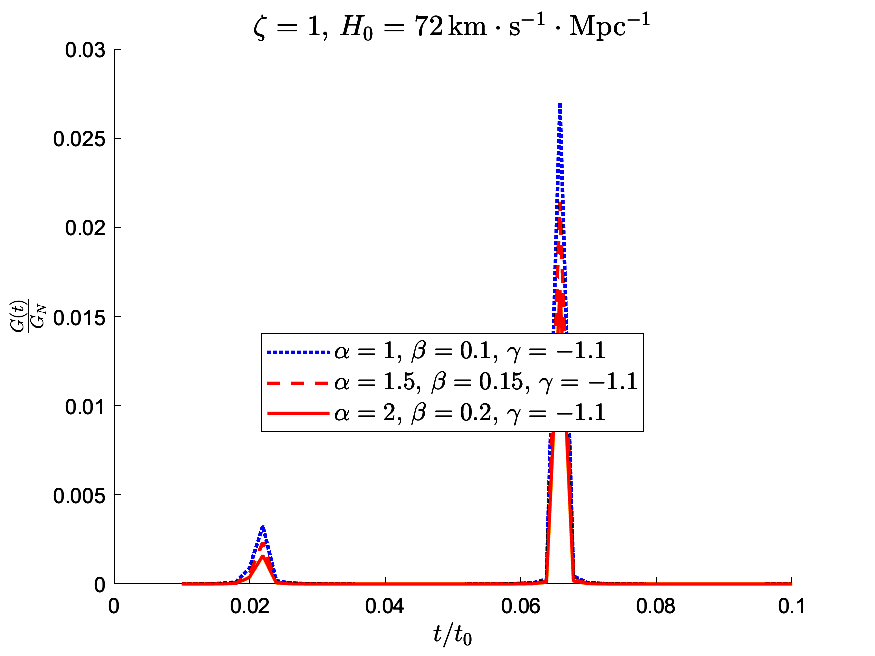}\hspace{0.01\linewidth}
\includegraphics[width=0.31\linewidth]{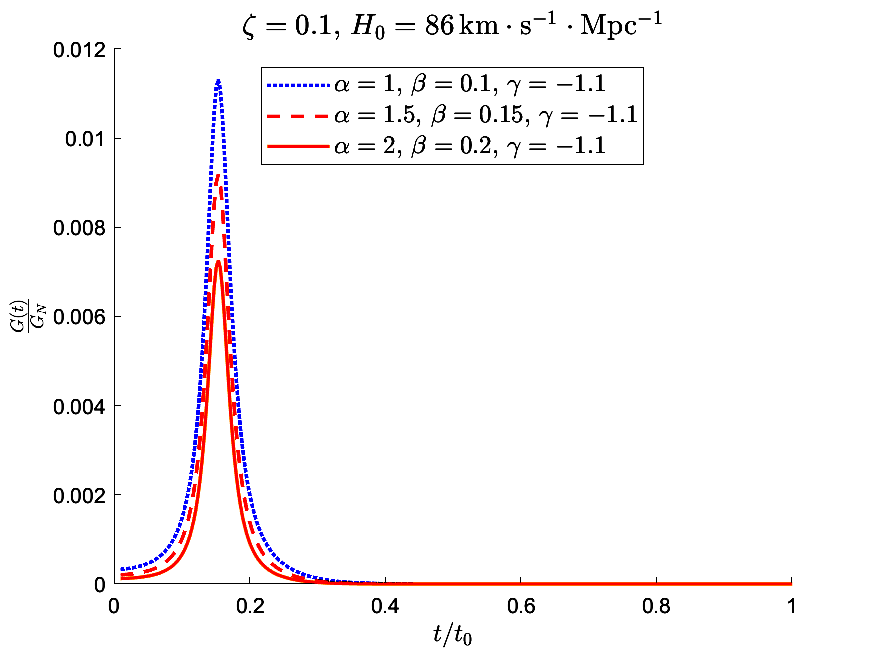}\hspace{0.01\linewidth}
\includegraphics[width=0.31\linewidth]{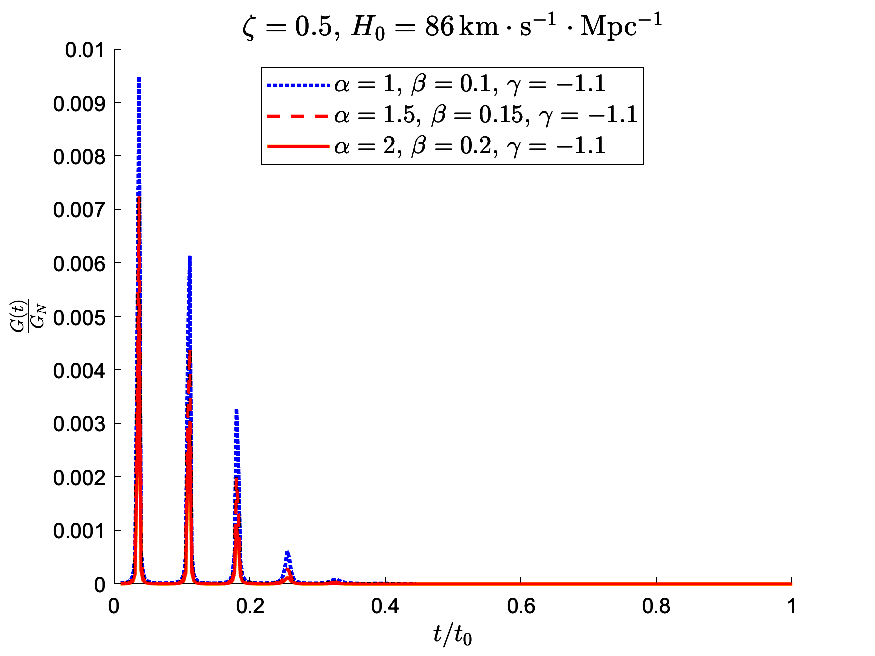}
\caption{\label{Fig.7} Variations of the gravitational parameter as a function of cosmic time for $ \gamma = -1.1 $.}
\end{figure*}

\begin{figure*}[htbp]
\centering
\includegraphics[width=0.31\linewidth]{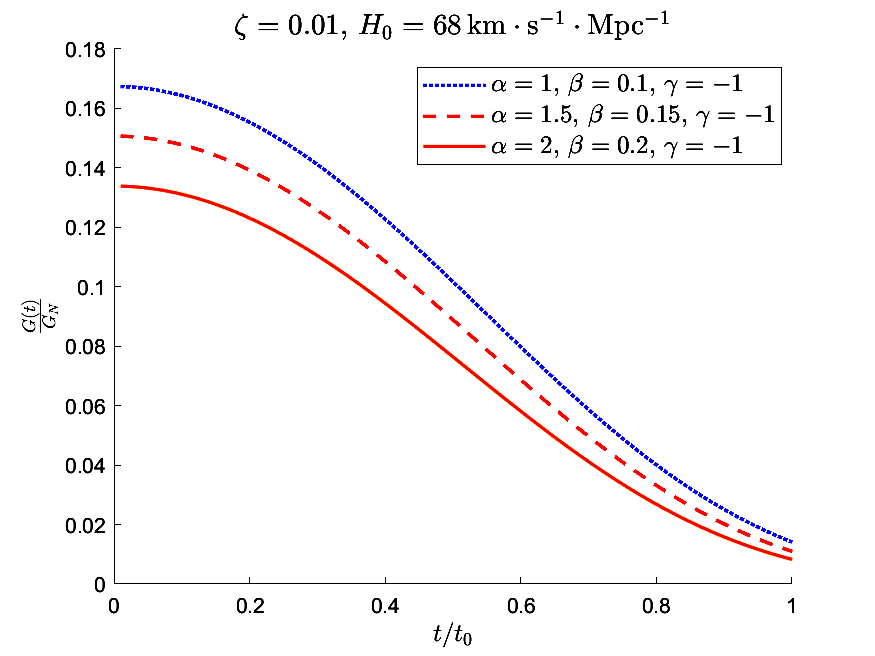}\hspace{0.01\linewidth}
\includegraphics[width=0.31\linewidth]{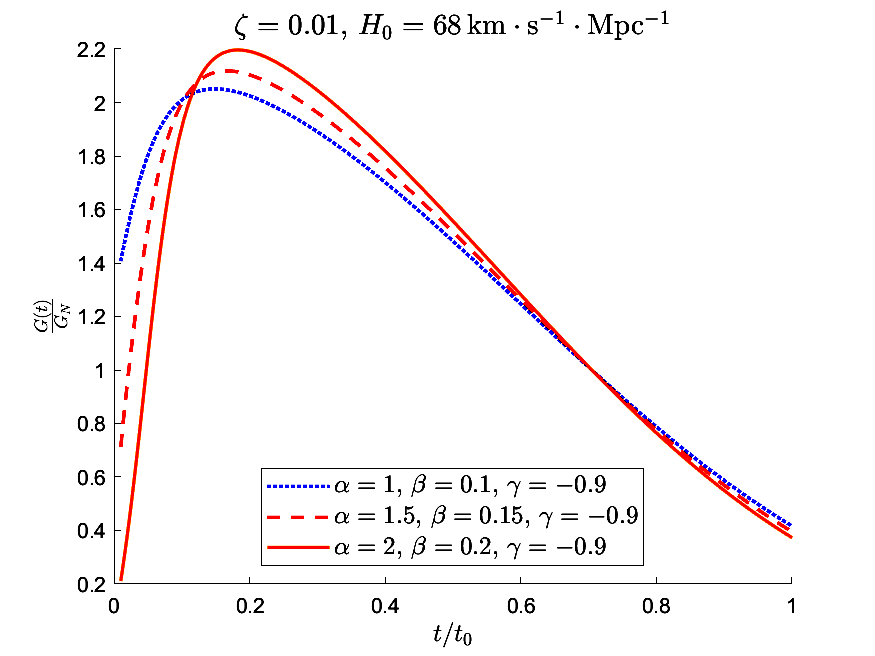}\hspace{0.01\linewidth}
\includegraphics[width=0.31\linewidth]{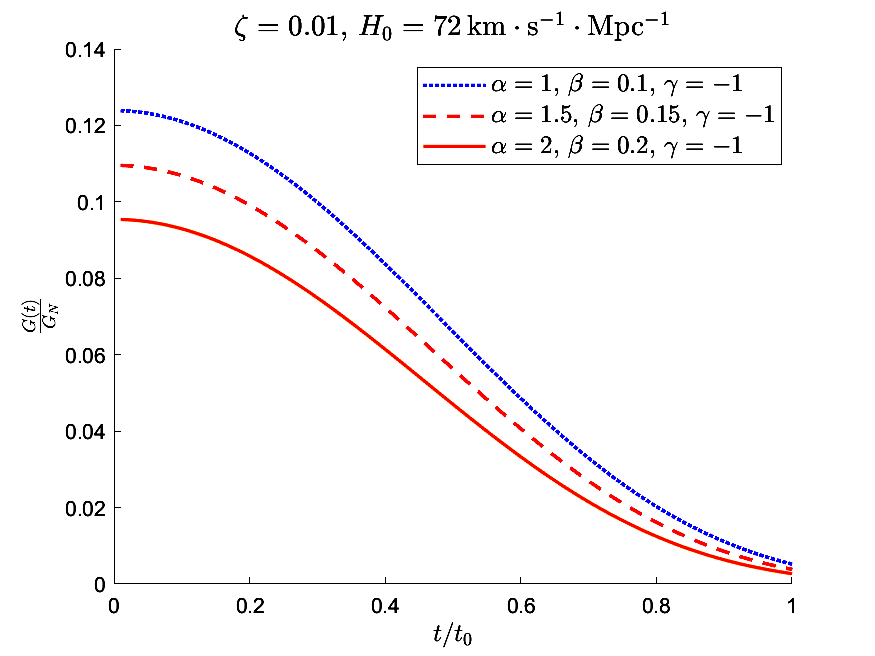}\\
\includegraphics[width=0.31\linewidth]{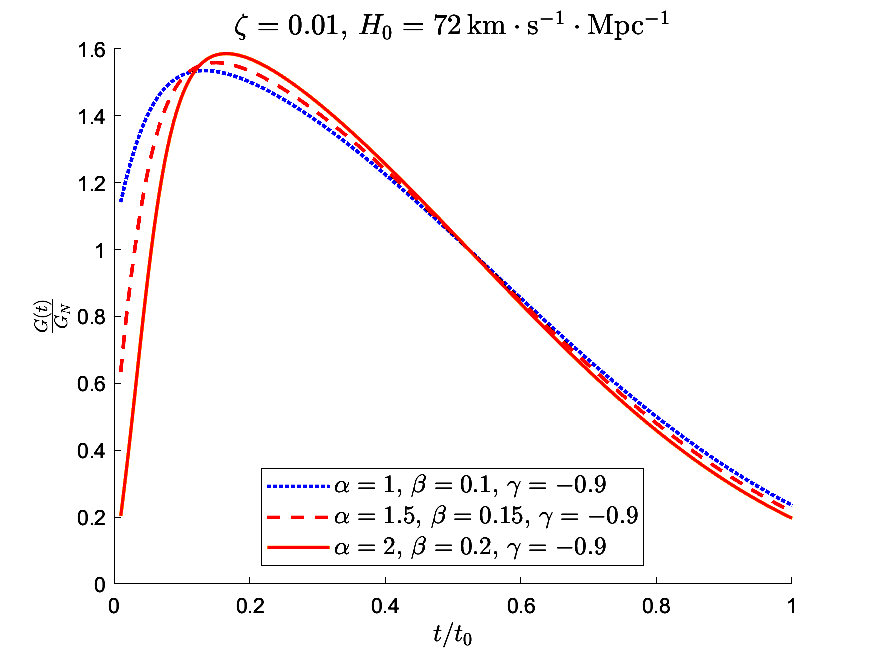}\hspace{0.01\linewidth}
\includegraphics[width=0.31\linewidth]{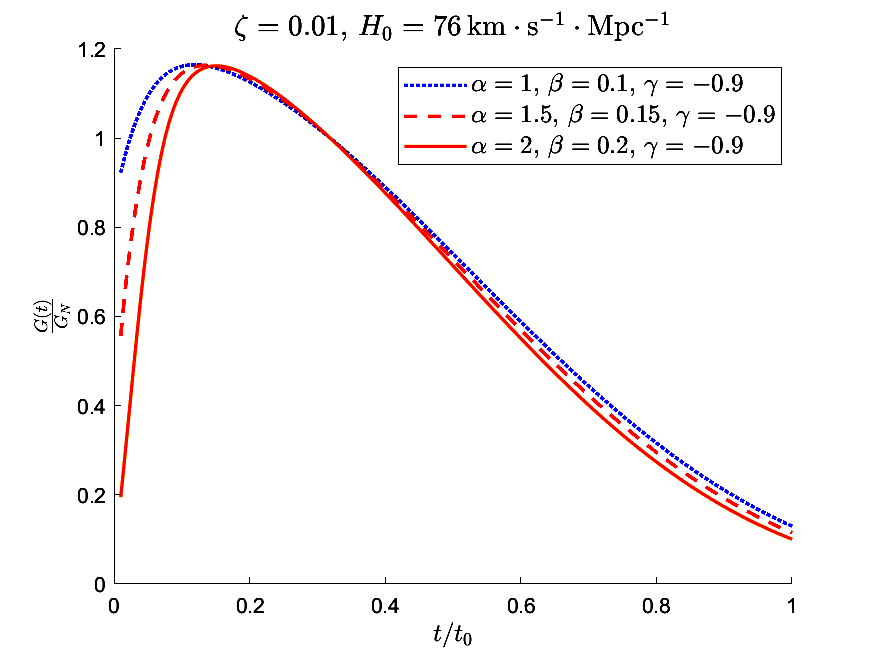}\hspace{0.01\linewidth}
\includegraphics[width=0.31\linewidth]{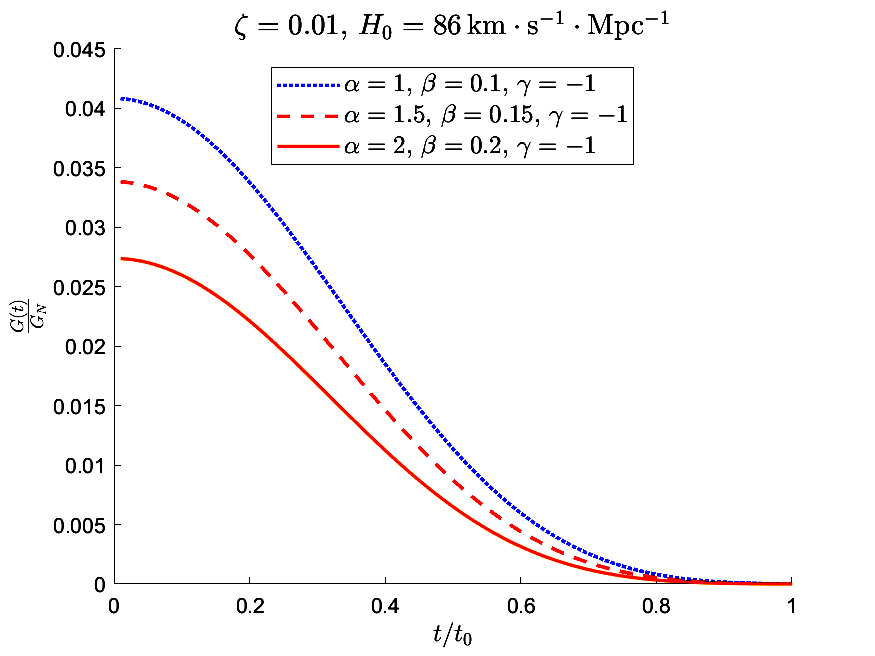}\\
\includegraphics[width=0.31\linewidth]{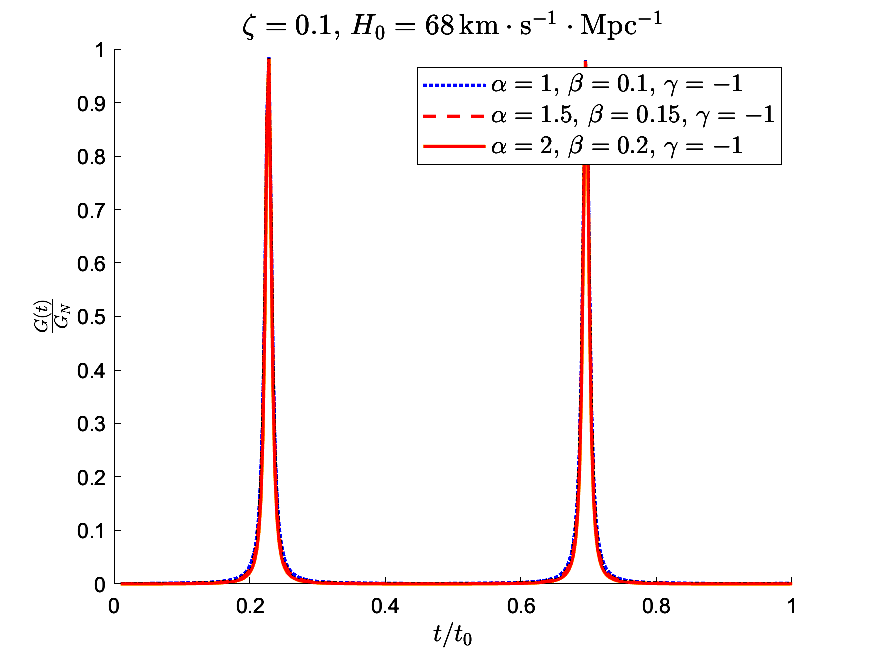}\hspace{0.01\linewidth}
\includegraphics[width=0.31\linewidth]{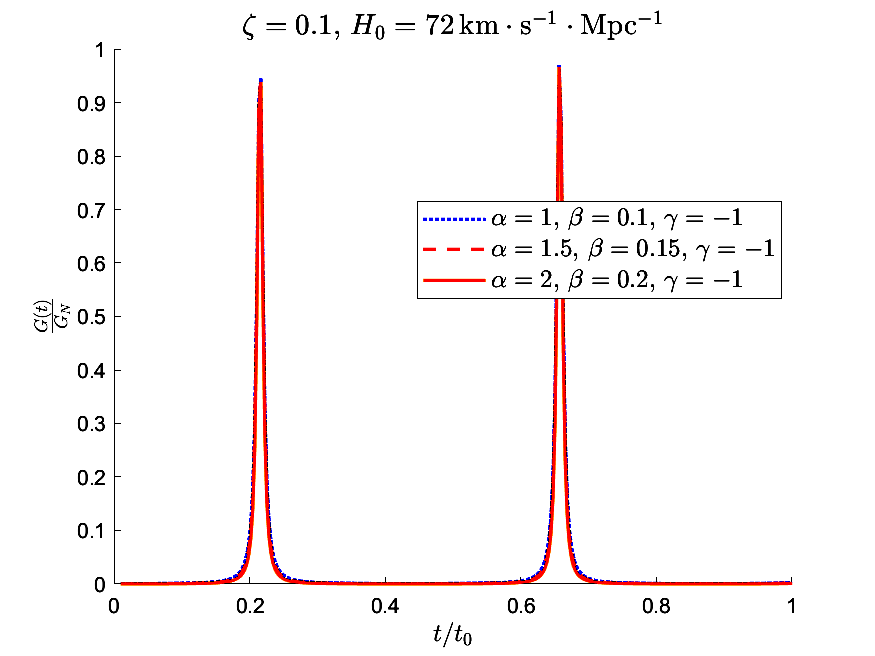}\hspace{0.01\linewidth}
\includegraphics[width=0.31\linewidth]{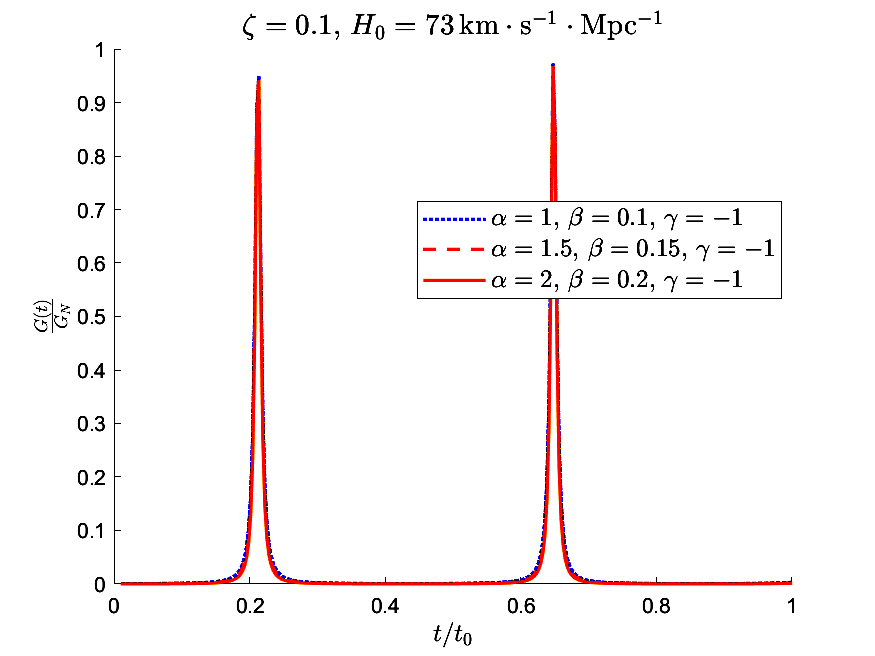}\\
\includegraphics[width=0.31\linewidth]{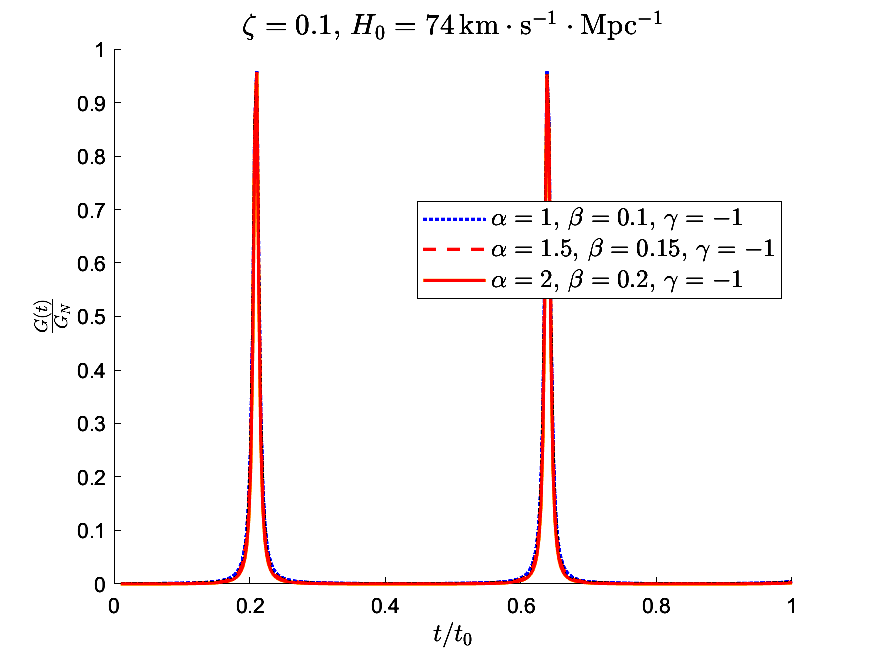}\hspace{0.01\linewidth}
\includegraphics[width=0.31\linewidth]{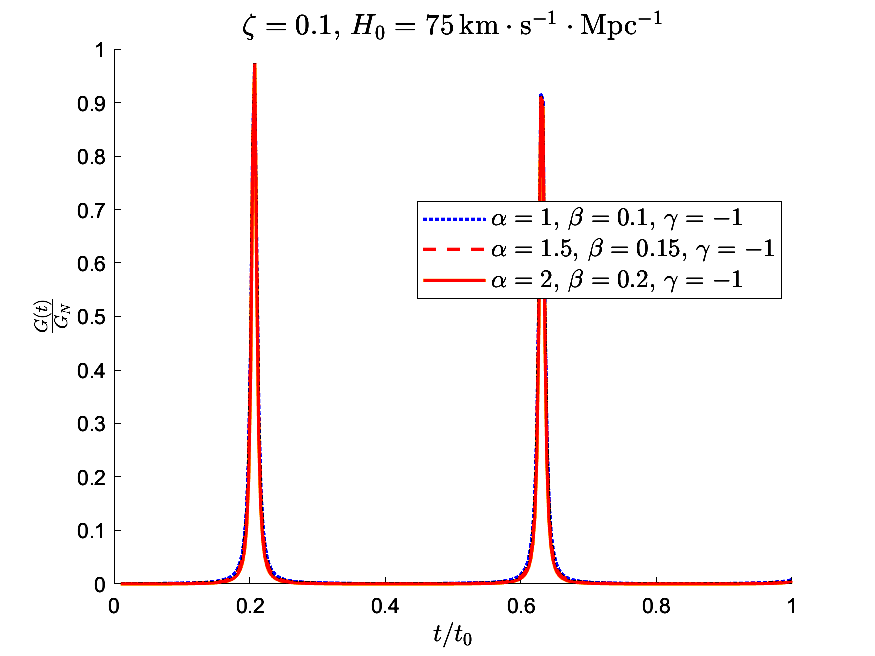}\hspace{0.01\linewidth}
\includegraphics[width=0.31\linewidth]{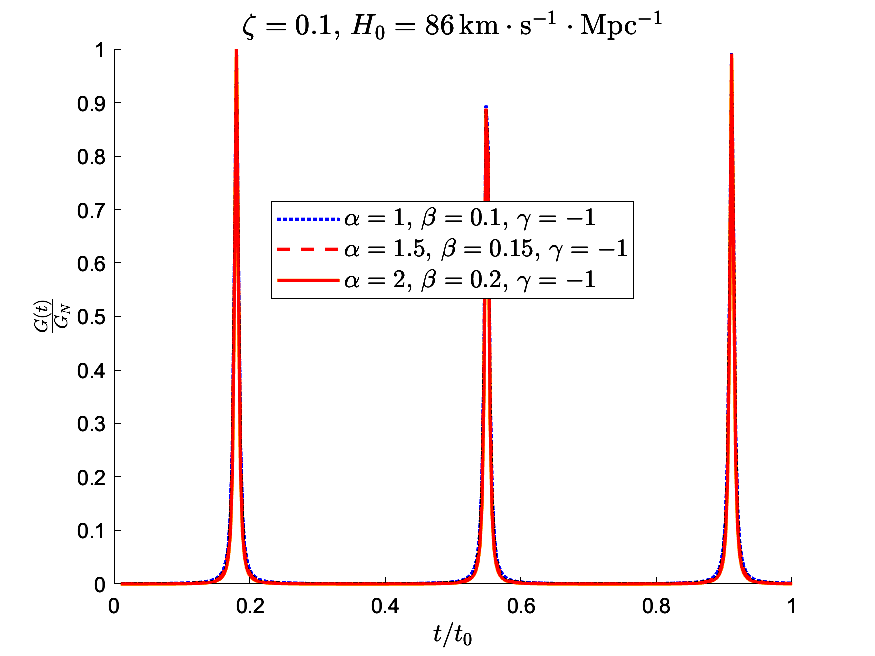}\\
\includegraphics[width=0.31\linewidth]{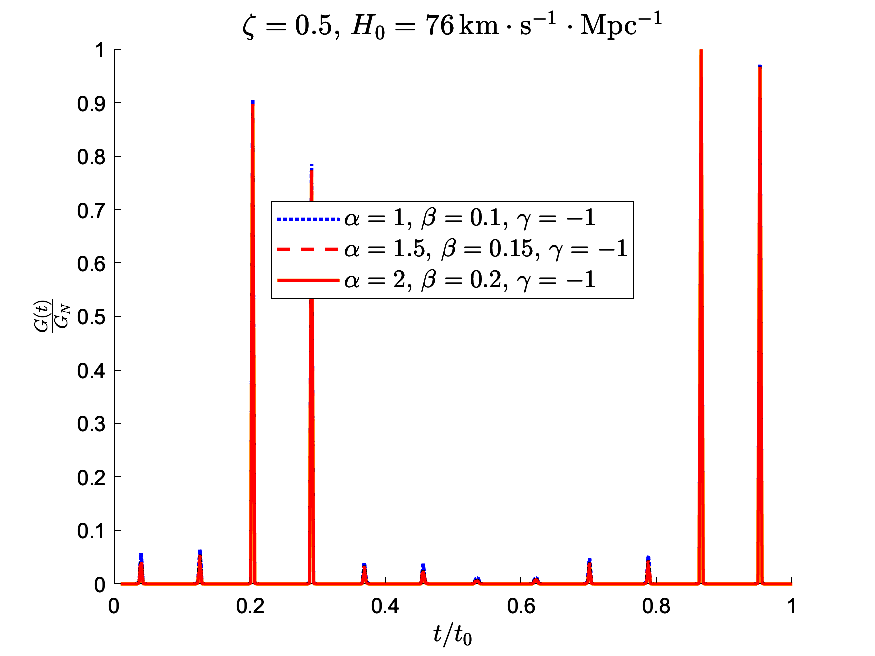}\hspace{0.01\linewidth}
\includegraphics[width=0.31\linewidth]{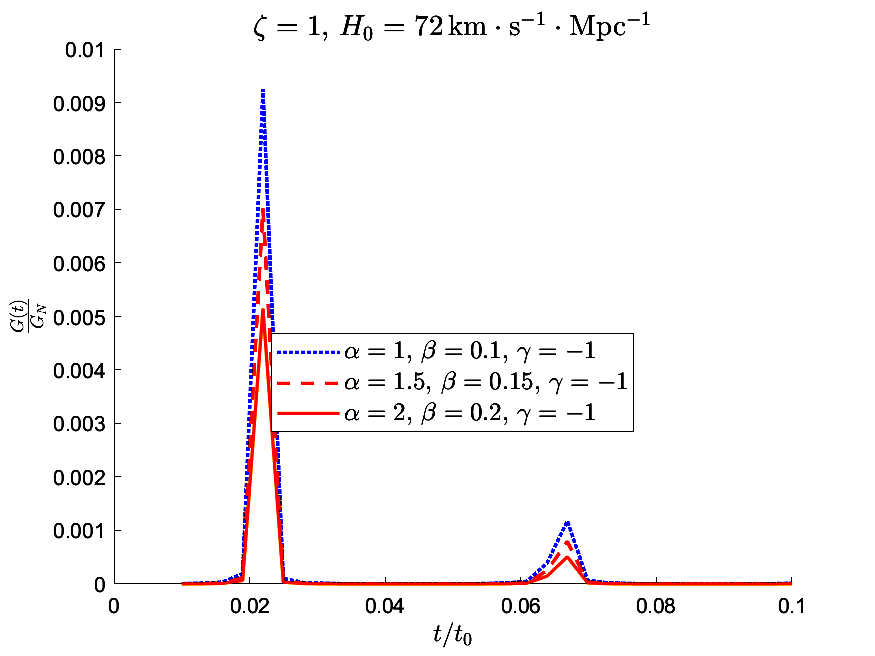}\hspace{0.01\linewidth}
\includegraphics[width=0.31\linewidth]{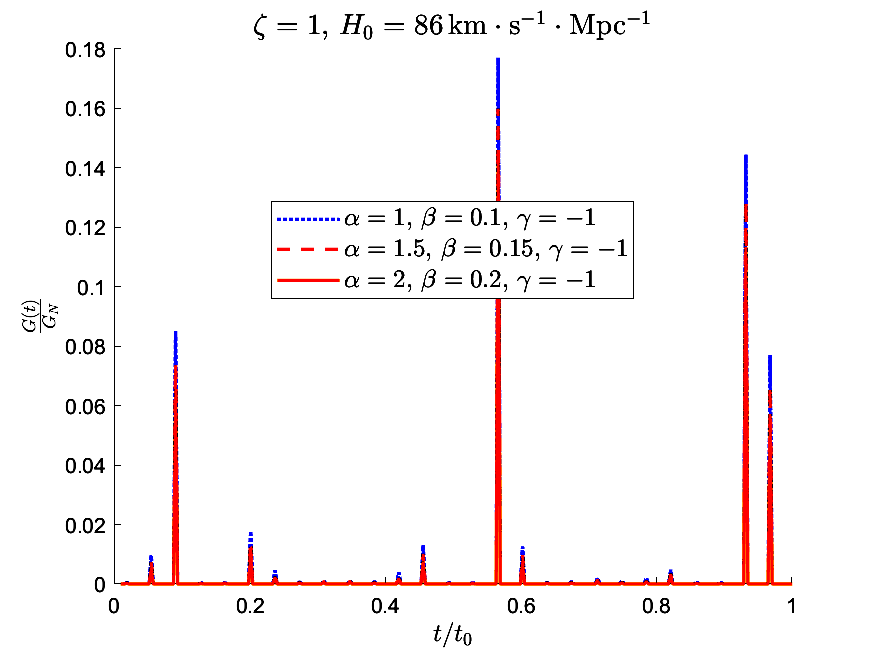}
\caption{\label{Fig.8} Variations of the gravitational parameter as a function of cosmic time for $ \gamma = -1 $.}
\end{figure*}

\begin{figure*}[htbp]
\centering
\includegraphics[width=0.31\linewidth]{figures/image344.png}\hspace{0.01\linewidth}
\includegraphics[width=0.31\linewidth]{figures/image346.png}\hspace{0.01\linewidth}
\includegraphics[width=0.31\linewidth]{figures/image347.png}\\
\includegraphics[width=0.31\linewidth]{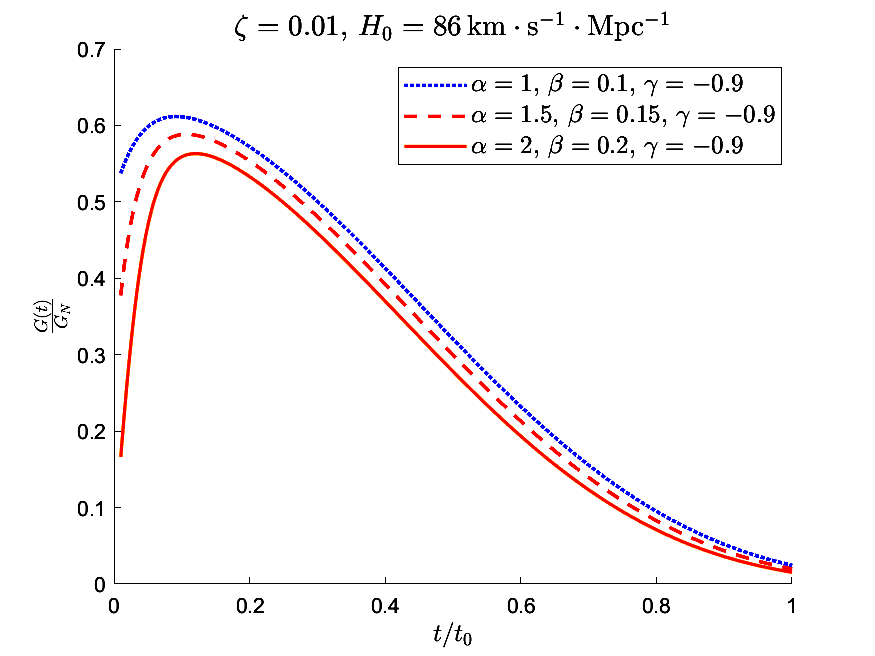}\hspace{0.01\linewidth}
\includegraphics[width=0.31\linewidth]{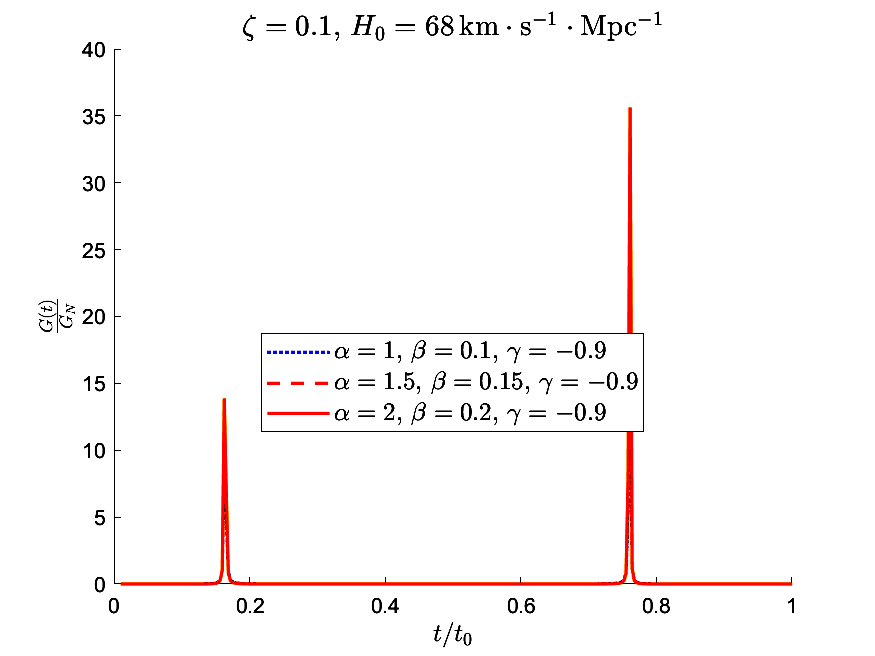}\hspace{0.01\linewidth}
\includegraphics[width=0.31\linewidth]{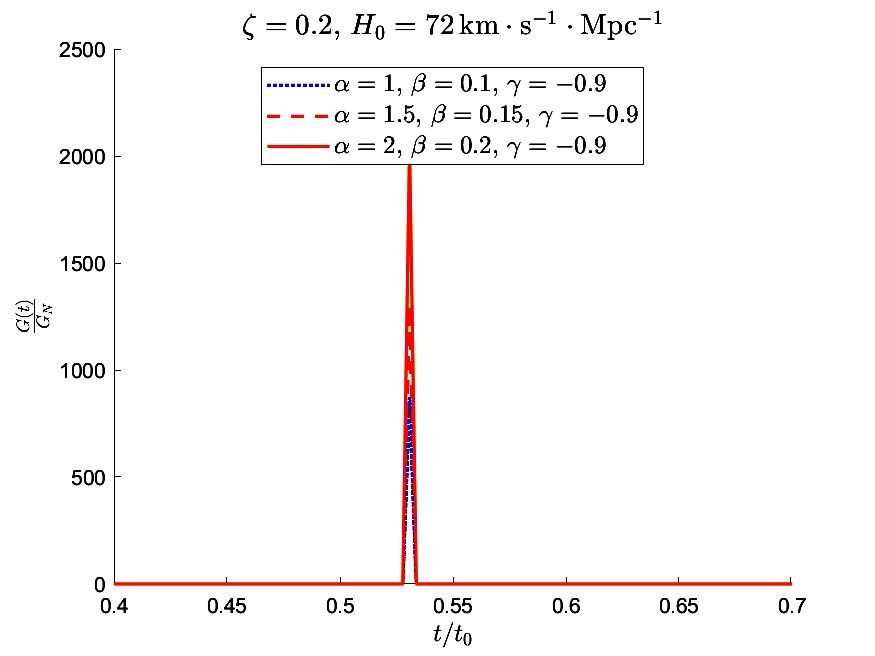}\\
\includegraphics[width=0.31\linewidth]{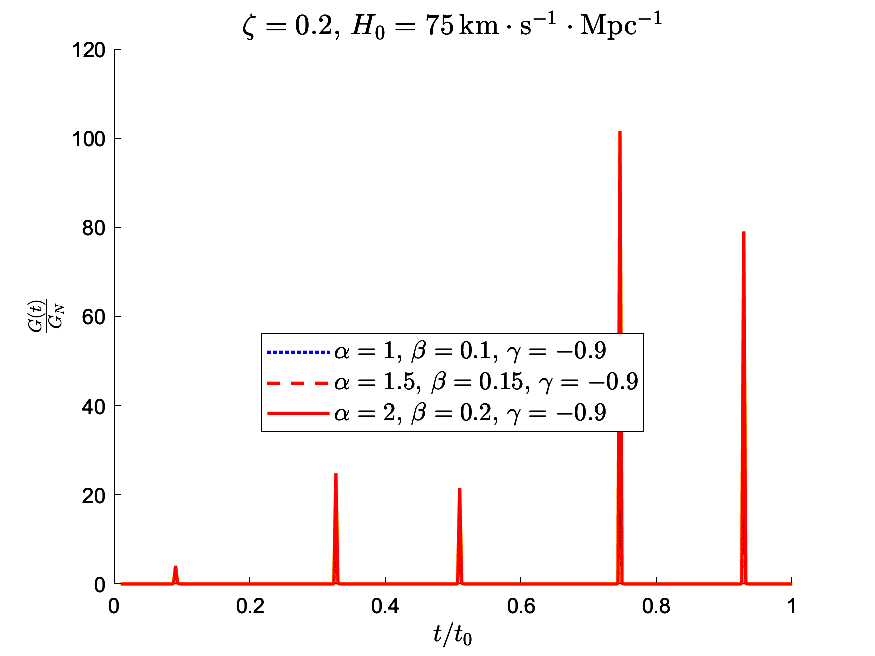}\hspace{0.01\linewidth}
\includegraphics[width=0.31\linewidth]{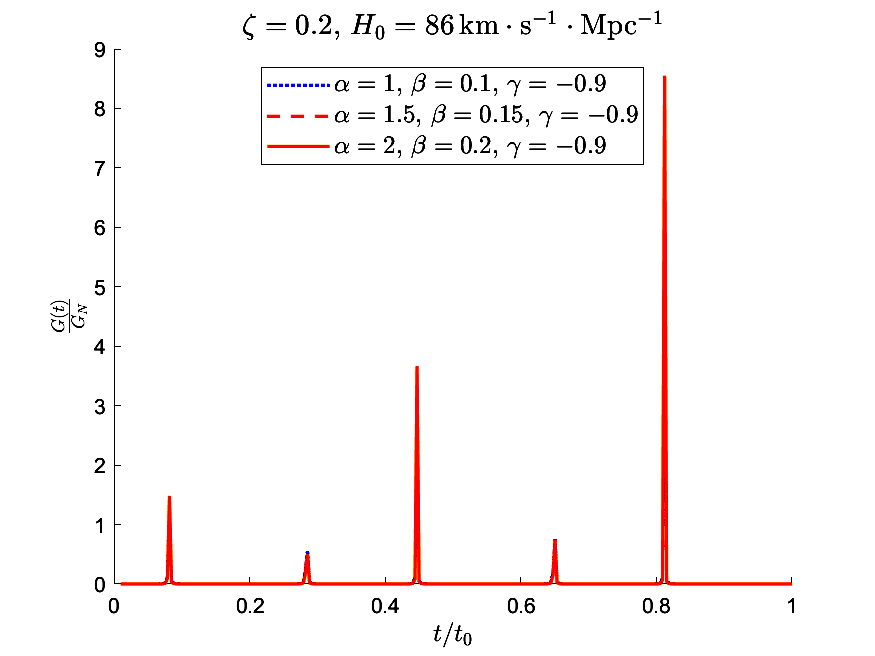}\hspace{0.01\linewidth}
\includegraphics[width=0.31\linewidth]{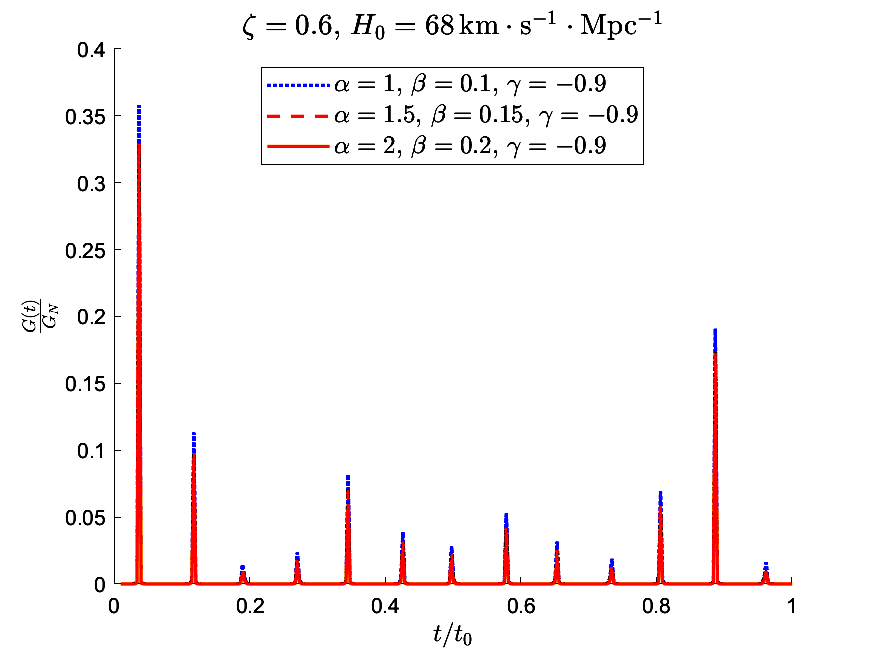}\\
\includegraphics[width=0.31\linewidth]{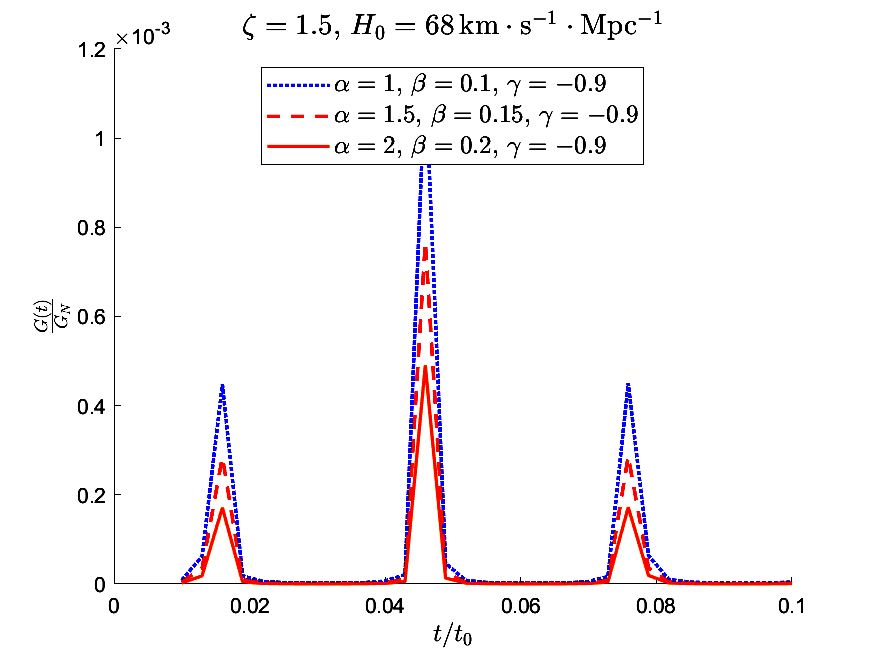}\hspace{0.01\linewidth}
\includegraphics[width=0.31\linewidth]{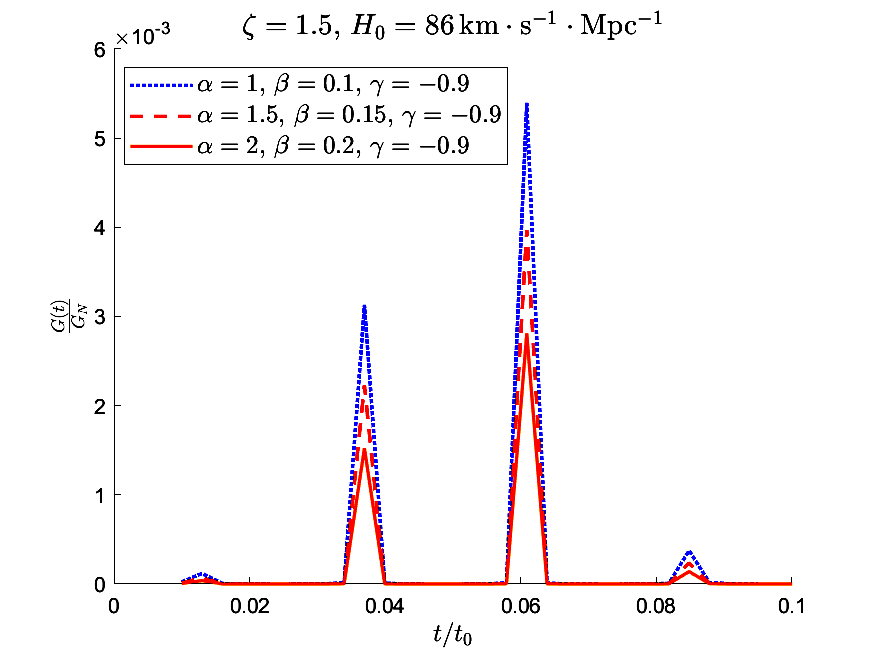}\hspace{0.01\linewidth}
\includegraphics[width=0.31\linewidth]{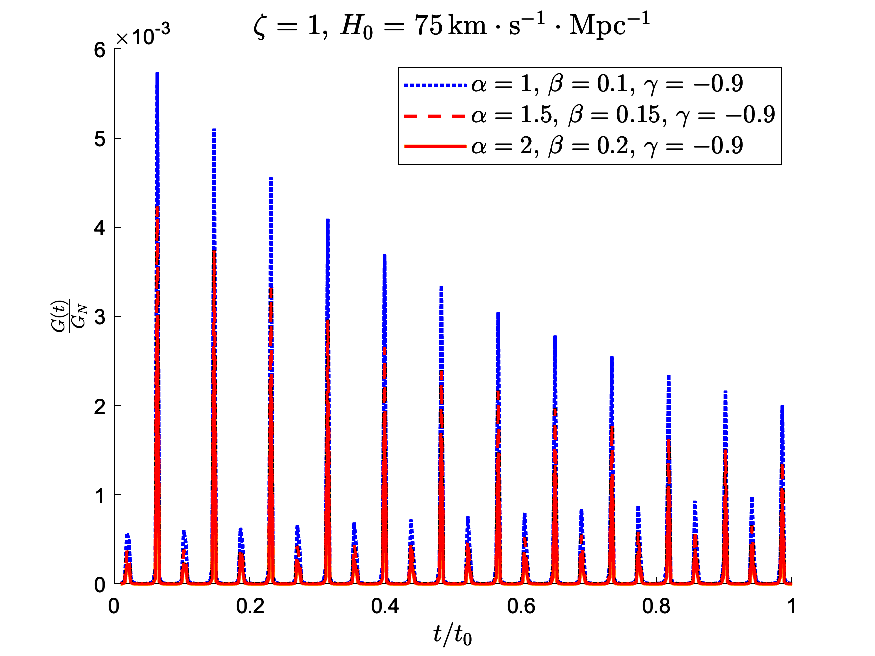}
\caption{\label{Fig.9} Variations of the gravitational parameter as a function of cosmic time for $ \gamma = -0.9 $.}
\end{figure*}

For $\gamma=-0.9$, the fluctuations gradually decay for $\zeta\geq 1$, while for $\gamma=-1$ irregular oscillations are observed. Figures~\ref{Fig.7}, \ref{Fig.8}, and \ref{Fig.9} compare these variations for different parameter combinations. In general, fluctuations of $G_{\rm eff}$ have been considered as a possible mechanism to explain the
periodicity observed in the distribution of the number of galaxies in standard cosmological models \cite{149}.

A notable result is that, for $\gamma<-1$, the variations of $G/G_N$ are much smaller when $\alpha>1$ than when $\alpha=1$, and the fluctuations disappear at late times. For small values of $\zeta$ (massless gravity limit), the fluctuations of $G_{\rm eff}$ grow with time for $\gamma=-1$ and decay for $\gamma<-1$, reflecting the impact of a phantom-dominated universe on the evolution of the gravitational constant.

This behavior is naturally interpreted in terms of the quantity
\begin{equation}
\mathcal{R}=\frac{\dot{G}/G}{H},
\end{equation}
which controls the relative variation of $G$ and determines the fractional part of the effective equation-of-state parameter. The sign of $\mathcal{R}$ is crucial: $\mathcal{R}>0$ implies an increase of $G_{\rm eff}$ and faster expansion, while $\mathcal{R}<0$ suppresses expansion and reduces structure growth. These trends are directly reflected in the predictions for $f\sigma_8$ and, therefore, in the model’s ability to alleviate the $S_8$ tension.

Moreover, the stochastic fluctuations of $G_{\rm eff}$ observed in the figures may serve as a phenomenological signature of an effective quantum gravity theory. This type of behavior has been proposed in stochastic gravity models \cite{151,152,153,154}, in which quantum fluctuations of spacetime induce variations in cosmological parameters.
In this framework, oscillations of $G_{\rm eff}$ could contribute to the current cosmic acceleration without the need to introduce exotic fields.

Finally, it is worth noting that, since the inflationary epoch, the gravitational constant may take different values across exponentially large regions of the universe, especially if it is correlated with the inflaton field \cite{150}. This phenomenon is consistent with the model's fractional dynamics and reinforces the interpretation of $G_{\rm eff}$ as a quantity subject to quantum and semiclassical fluctuations.

\subsubsection{The $\sigma_8$ tension}
\label{Sect:6.3.1}
The $\sigma_8$ tension—the discrepancy between the growth of structure inferred from the CMB and that measured by LSS and weak lensing—is naturally interpreted in the fractional model thanks to the central role of the quantity
\begin{equation}
\mathcal{R}(N)=\frac{\dot G/G}{H}.
\end{equation}

The fractional growth equation shows that the temporal variation of $G$ simultaneously modifies the effective gravity,
\begin{equation}
G_{\rm eff}(N)=G_0\,e^{\int \mathcal{R}\,dN},
\end{equation}
the cosmological friction $(1-\alpha)/(tH)$, and the gravitational source $t^{-(\alpha-1)(1+\gamma)}$.

The sign of $\mathcal{R}$ determines the direction of the effect:
\begin{equation}
\mathcal{R}<0 \Rightarrow G_{\rm eff}\downarrow \Rightarrow \delta\downarrow
\Rightarrow \sigma_8\downarrow,
\end{equation}
\begin{equation}
\mathcal{R}>0 \Rightarrow G_{\rm eff}\uparrow \Rightarrow \delta\uparrow
\Rightarrow \sigma_8\uparrow.
\end{equation}

\begin{itemize}
    \item For parameters such as $\beta=0.95$, $\alpha=0.5$, and $\gamma=-0.9$, where $\dot{G}/G\gg H_0$, the effective gravity grows rapidly and the model predicts an increase in $\sigma_8$, which worsens the tension.
    
    \item For combinations such as $\beta=0.1$, $\alpha=2$, and $\gamma=-0.96$, where $\dot{G}/G\approx 2.4\times 10^{-12}$ yr$^{-1}$, the variation of $G$ is smooth and the model reproduces low values of $\sigma_8$, consistent with LSS.
    
    \item For $\beta=0.0001$ and $\alpha=1.5$, where $\dot{G}/G\approx 1.96\times 10^{-8}$ yr$^{-1}$, the effective gravity decreases at late times, and growth is suppressed, alleviating the $S_8$ tension.
\end{itemize}

Taken together, the dynamics of $\mathcal{R}$ on the slow manifold determine when and how the model can reproduce low values of $\sigma_8$ without altering early physics. This connection makes $\mathcal{R}$ a key indicator for assessing the model's viability with respect to the $S_8$ tension.

\subsubsection{The $H_0$ tension}
\label{Sect:6.3.2}
The tension in the value of the Hubble parameter—the discrepancy between local measurements and those inferred from the CMB—is naturally interpreted in terms of the dynamics of $\mathcal{R}$.

In the late-time regime, the asymptotic value $\mathcal{R}_\ast$ on the slow manifold determines the evolution of $G_{\rm eff}$ and, therefore, the shape of $H(z)$ near $z=0$:
\begin{itemize}
    \item if $\mathcal{R}_\ast>0$, then $G_{\rm eff}$ increases and the model predicts high values of $H_0$, compatible with SH0ES ($74.03\pm1.42$ km s$^{-1}$ Mpc$^{-1}$);
    \item if $\mathcal{R}_\ast<0$, then $G_{\rm eff}$ decreases and the model predicts low values of $H_0$, compatible with the CMB ($68.7\pm3.1$ km s$^{-1}$ Mpc$^{-1}$).
\end{itemize}

Moreover, 
\begin{itemize}
    \item For $\alpha>1$ and large values of $\zeta$, the model reaches $H_0>72$ km s$^{-1}$ Mpc$^{-1}$, reproducing local measurements.
    
    \item For $\alpha<1$, fractional friction reduces $H_0$ toward the CMB values.
    
    \item The term $\varepsilon/t$ in the ansatz
    \begin{equation}
    H(t)=H_0+\xi\,\varphi(t)+\varepsilon/t
    \end{equation}
    allows adjustment of the slope of $H(z)$ in the range $0<z\lesssim 1$, reproducing the smooth transition observed in OHD and BAO data.
    
    \item Small values of $\zeta$ allow reproduction of the abundance of massive galaxies at high redshifts observed by JWST, favoring scenarios with elevated $H_0$.
\end{itemize}

In summary, the $H_0$ tension is interpreted as a direct consequence of fractional dynamics and of the asymptotic value of $\mathcal{R}$. The slow–fast structure of the system allows small variations in $\mathcal{R}$ to translate into significant differences in the observed value of $H_0$, without modifying the early-universe physics.

\section{Conclusions}
\label{Sect:4.8}

In this work, we developed the theoretical, dynamical, and numerical foundations of a fractional cosmological model with a time-varying gravitational constant. Motivated by the Hubble tension and the absence of a fundamental explanation for dark energy, we reformulated the FRGIC/FEG framework as a first-order autonomous system, uncovering a rich phase-space structure that includes de Sitter–like states, oscillatory and cyclic regimes, and transient phases driven by the evolution of $G$. The fractional variational formulation provided the mathematical basis for the modified field equations, while the RG-improved dynamics revealed super-acceleration, nontrivial transitions, and an expanded solution space shaped by scalar-field interactions and fractional corrections.

The analysis of critical points showed that stability is highly sensitive to $(\mu,\zeta,\lambda)$, with zero eigenvalues and singularities requiring center-manifold techniques or regularization. A robust numerical strategy—based on the regularizing variable $u=H_{0}/H$, integration in $N=\ln(1+z)$, and slow-manifold matching—enabled the reconstruction of $H(z)$ and its comparison with OHD, BAO, GL, BHS, and SNe Ia data. The global flow exhibits strongly nonlinear behavior, organized by geometric structures and, in key regions, dominated by singular terms in the $G$-equation. A natural slow–fast decomposition emerges, with the slow manifold acting as a geometric attractor that governs the asymptotic dynamics and clarifies the model's physical interpretation. Regularization removes the singularities of the original formulation and reveals a persistent invariant structure consistent with the numerical results. Overall, the fractional model with varying $G$ provides a coherent dynamical picture that supports oscillatory and cyclic cosmologies and offers mechanisms potentially relevant to the Hubble tension.

A central outcome of the analysis is the closed equation for $\mathcal{R}=u\,v_{4}$, which acts as the structural axis of the model. Peaks in $\mathcal{R}$ signal the breakdown of the slow–fast structure near $u\to 0$, marking the proximity of the dynamical singularity at the critical point $S_{8}$. The fractional term $(1-\alpha)/t$ modulates the intensity of these variations, thereby affecting global stability. The effective continuity equation shows that variations in $G$ alter the thermal history of the early universe, imposing constraints from BBN and allowing phantom crossing without exotic fields. The fractional growth equation revealed that $\mathcal{R}$ controls the suppression or amplification of structure formation, directly connecting to the $S_{8}$ tension. Likewise, the asymptotic value of $\mathcal{R}$ determines the late-time behavior of $H(z)$ and explains the $H_{0}$ tension, with small variations in $\mathcal{R}$ producing significant observational differences. The analysis of the kinematic parameters $q(z)$ and $j(z)$ confirmed convergence to a de Sitter regime at late times, consistent with SNe Ia, BAO, and CC data.

From the observational perspective, our statistical analysis shows that the Fractional model with $\mu=0$ is the only variant that provides a competitive fit to the data. The reduced Hubble parameter remains stable across all models,
\begin{equation*}
h = 0.7253^{+0.0074}_{-0.0071}\;(\Lambda\mathrm{CDM}),\qquad
0.7211^{+0.0072}_{-0.0080}\;(\mu\neq 0),\qquad
0.7219^{+0.0076}_{-0.0072}\;(\mu=0),
\end{equation*}
and the sound horizon scale is similarly consistent, $r_{d}\simeq 139\;\mathrm{Mpc}$ in all cases. The parameters specific to the Fractional model show the largest differences:
\begin{equation*}
\alpha = 1.57^{+0.60}_{-0.38}\;(\mu\neq 0),\qquad
\alpha = 1.20^{+0.25}_{-0.14}\;(\mu=0),
\end{equation*}
\begin{equation*}
\zeta = 0.85\pm 0.57\;(\mu\neq 0),\qquad
\zeta = 0.43^{+0.39}_{-0.29}\;(\mu=0),
\end{equation*}
indicating that the $\mu=0$ case is significantly better constrained.

The dynamical parameters inferred for the $\mu=0$ model further support this conclusion:
\begin{equation*}
m = 30.8^{+28.0}_{-20.9}\;\mathrm{km\,s^{-1}\,Mpc^{-1}},\qquad
\Gamma = 108.3\pm 1.1\;\mathrm{km\,s^{-1}\,Mpc^{-1}},
\end{equation*}
leading to a positive discriminant,
\begin{equation*}
\Delta = 43090^{+3424}_{-10146}\;
\mathrm{(km\,s^{-1}\,Mpc^{-1})^{2}},
\end{equation*}
which places the system in the overdamped regime. The corresponding relaxation timescale is tightly constrained,
\begin{equation*}
\tau_{\rm rel} = 9.037^{+0.091}_{-0.094}\;\mathrm{Gyr},
\end{equation*}
consistent with a slow decay mode of cosmological duration.

Regarding model comparison, the $\mu=0$ Fractional model achieves a slightly lower minimum chi-square than $\Lambda$CDM ($\chi^2_{\min}=1692.8$ vs.\ 1693.3), but the Bayesian Information Criterion strongly favors $\Lambda$CDM ($\Delta\mathrm{BIC}>10$), due to the additional parameters in the Fractional framework. The general Fractional model with $\mu\neq 0$ is statistically disfavored, with both a higher $\chi^2_{\min}$ and a substantially larger BIC.

Finally, the cosmographic functions $H(z)$, $q(z)$, $j(z)$, and $s(z)$ confirm that the $\mu=0$ model closely tracks the $\Lambda$CDM expansion history, while the $\mu\neq 0$ case exhibits larger deviations and broader confidence regions, reflecting its strong internal parameter degeneracies. Overall, only the $\mu=0$ Fractional model provides a viable alternative to $\Lambda$CDM within current observational uncertainties.

In conclusion, the fractional cosmological model with variable $G$ establishes a robust geometric and dynamical framework. The slow–fast structure organizes the global dynamics, the slow manifold fixes the asymptotic regimes, and the closed equation for $\mathcal{R}$ diagnoses stability, expansion, and structure growth. The unified connection between $\mathcal{R}$, modified continuity, perturbation dynamics, and observational tensions provides a coherent explanation for cosmic acceleration and offers new mechanisms to address the $H_{0}$ and $S_{8}$ tensions. This framework not only clarifies the physical interpretation of fractional corrections but also opens pathways for future research in nonlinear stability, bifurcation analysis, and Bayesian confrontation with cosmological datasets.

\section*{Author contributions}{
Conceptualization, R.A.E.-N. and G.L.; 
methodology, R.A.E.-N. and G.L.; 
software, G.L., E.G., and K.M.; 
validation, R.A.E.-N., G.L. and E.G.; 
formal analysis, R.A.E.-N., G.L., E.G. and K.M.; 
investigation, R.A.E.-N., G.L., E.G., and K.M.; 
resources, R.A.E.-N.; 
data curation, K.M. and E.G.; 
writing---original draft preparation, R.A.E.-N.; 
writing---review and editing, G.L. and E.G.; 
visualization, E.G., and K.M.; 
supervision, G.L.; 
project administration, G.L.; 
funding acquisition, R.A.E.-N, G.L. 
All authors have read and agreed to the published version of the manuscript.}

\section*{Funding}{RA El-Nab has received funding from the CNAAR, project QK22020134, and from Chiang Mai University. E.G., G.L. and K.M. acknowledge the financial support of ANID, Chile, through Proyecto Fondecyt Regular 2024, Folio 1240514. K.M. acknowledges the financial support of Programa de Tesis Colaborativas NEXUS 2025 del Consorcio Science Up (UCN, USACH, PUCV) through Proyecto 20CEIN2-142145.}

\section*{Data availability}{Data used in this research is presented in Section \ref{Sect:2.7}.}

\section*{Acknowledgments}{We extend our gratitude to the Vicerrectoría de Investigación y Desarrollo Tecnológico (VRIDT) of Universidad Católica del Norte (UCN) for the scientific support provided through the Núcleo de Investigación en Geometría Diferencial y Aplicaciones, according to Resolution VRIDT No.~096/2022, and through the Núcleo de Investigación en Simetrías y la Estructura del Universo (NISEU), according to Resolution VRIDT No.~200/2025.}

\section*{Conflicts of interest}{The authors declare no conflicts of interest.}

\end{document}